\documentclass[10pt,3p,times,procedia]{elsarticle}

\usepackage{ecrc}
\usepackage{lipsum}
\usepackage{graphicx}
\usepackage{caption}
\usepackage{subcaption}
\usepackage{array}
\usepackage{amsmath}
\usepackage{color}
\usepackage[dvipsnames]{xcolor}
\usepackage[T1]{fontenc}
\usepackage{mathtools}
\usepackage{fullpage}
\usepackage{lineno}
\biboptions{numbers,sort&compress}

\volume{-}

\firstpage{1}

\journalname{-}

\runauth{}

\jid{egypro}

\usepackage{amssymb}

\usepackage{epstopdf}

\definecolor{darkblue}{RGB}{83,0,93}

\usepackage{float}

\def\d{\textrm d}
\def\i{\textrm i}

\def\ba#1\ea{\begin{align}#1\end{align}}
\def\!#1\!{}

\def\bsa#1\esa{\begin{subequations}
\begin{align}#1\end{align} \end{subequations}}

\def\lb{\left[}
\def\rb{\right]}

\def\F{\hat{F}_{exc}}
\makeatletter
\let\start@align@nopar\start@align
\let\start@gather@nopar\start@gather
\let\start@multline@nopar\start@multline
\long\def\start@align{\par\start@align@nopar}
\long\def\start@gather{\par\start@gather@nopar}
\long\def\start@multline{\par\start@multline@nopar}
\newcommand{\R}{\mathbb{R}}
\makeatother
\usepackage{soul}
\def\dd{0.25\textwidth}
\begin{document}

\begin{frontmatter}



\dochead{}

\title{Title\tnoteref{label1}}
\title{Shape Optimization of a Submerged Planar \\
Pressure Differential Wave Energy Converter}
\title{Shape Optimization of Wave Energy Converters for\\ Broadband Directional Incident Waves}

\author{Soheil Esmaeilzadeh\corref{cor1}\fnref{fn1}}
\address{Department of Energy Resources Engineering, Stanford University, CA, 94305}

\author{Mohammad-Reza Alam\fnref{fn2}}
\address{Department of Mechanical Engineering, University of California Berkeley, CA, 94720}

\cortext[cor1]{$~~$Corresponding author}
\fntext[fn1]{$~~$soes@stanford.edu}
\fntext[fn2]{$~~$reza.alam@berkeley.edu}


\begin{abstract}
Here, through a systematic methodology and the use of high performance computing, we calculate the optimum shape for a wave energy converter under the action of incident waves of (i) monochromatic unidirectional, (ii) monochromatic directional, (iii) polychromatic unidirectional and (iv) polychromatic directional (with both directional symmetry and asymmetry). As a benchmark for our study, without loss of generality,  we consider a submerged planar pressure differential wave energy converter, and use Genetic Algorithm to search through a wide range of shapes. A new parametric description of absorber shape based on Fourier decomposition of geometrical shapes is introduced, and for each shape hydrodynamic coefficients are calculated, optimum power take-off parameters are obtained, and overall efficiency is determined. We show that an optimum geometry of the absorber plate can absorb a significantly higher energy (in some cases \textit{few times} higher) when compared to a circular shape of the same area. Specifically, for a unidirectional incident wave, the optimum shape, as expected, is obtained to be the most elongated shape. For directional incident waves, a butterfly-shape is the optimum geometry whose details depend on not only the amplitude and direction of incident wave components, but also the relative phases of those components. For the latter effect, we find an optimally averaged profile through a statistical analysis.

\end{abstract}
\end{frontmatter}

\section{Introduction}

Oceanic surface waves carry a much higher energy density, sometimes by two orders of magnitude, than wind and solar. Estimates show that ocean wave energy can realistically provide about 10\% of world's electricity need \cite{Cruz2008,Panicker1976,Besio2016,Reguero2015,Voss1979}. This potential, along with proximity to load centers that are typically along coastal areas, as well as good predictability and power consistency makes wave energy an appealing solution to avoid archaic pollutant-rich energy production methods \cite{Lopez-Ruiz2016,Carballo2015}. Harnessing ocean wave energy, however, faces several major challenges that have, so far, barred wave energy industry from taking off.
\begin{center}
\begin{tabular}{ | m{1.5 em}  m{5cm} m{2.1em} m{8.2cm} |} 
\hline
\vspace{5pt}
\textbf{Nomenclature} &  &  & \\
PTO & power take-off unit &  $\lambda$  &  wave length [$m$]  \\ 
BEM & boundary element method &  $d$ & absorber plate's thickness [$m$]   \\ 
CWR & capture width ratio &  $\theta$ & wave's angle of incidence [$^\circ$]  \\ 
$\xi$ & shape difference factor &  $\omega_p$ & peak wave's angular frequency  \\ 
$h_{opr}$ & absorber plate's operating depth [\textit{m}] & $\gamma$ & peak enhancement factor  \\ 
$h_{btm}$ & sea bottom depth [\textit{m}] & $\alpha_p$ & Philips parameter\\ 
$z$ & vertical (heave) displacement [\textit{m}] & $H_s$ & significant wave height\\
$K$ & spring's coefficient [$N/m$] & $D\,(\theta)$ & spreading function \\ 
$C$ & damper's coefficient [$Ns/m$] & $\theta_m$ & wave's mean angle of incidence [$^\circ$]\\ 
$a$ & wave amplitude [$m$] & $\Theta$ & directional spectrum's angular span [$^\circ$]\\ 
$m$ & absorber plate's mass [$kg$] & $\mu$ & asymmetry parameter  \\ 
$\widetilde{A}$ & added mass [$kg$] &  $S\,(\omega,\theta)$ & wave's spectral density \\ 
$\widetilde{B}$ & added damping [$Ns/m$] & $r_0$ & radius of base circle  [$m$]\\
$\omega$ & wave's angular frequency [$rad/s$] & $A$ & area of a shape [$m^2$]\\
$F_{exc}$ & wave induced excitation force [$N$] & $a_n$ & coefficient of n$^{th}$ order Fourier mode [$m$] \\
$\phi_{F}$ & phase of excitation force [$rad$] & $\phi_n$ & phase of n$^{th}$ order Fourier mode [$rad$]\\
$\phi_{z}$ & phase of system response [$rad$] & $N_c$ & total number of Fourier modes\\	
$\widehat{\square}$ & amplitude of $\square$ & $p, q$ & elongation coefficients \\
$r_e$ & radius of equivalent circle & $N_{\theta}$ & number of directional spreading subdivisions  \\
$T$ & wave period [$s$] & $N_{\omega}$ & number of frequency spreading subdivisions  \\
$k_p$ & peak wave number [$rad/m$]  & $P_r$ & normalized time-averaged extracted power \\
$k$ & wave number [$rad/m$] & $N$ & Genetic Algorithm's number of generations\\
$S_f\,(w)$ & frequency spectrum & $P_{abs}$ & time-averaged extracted power by the optimum shape [$W$]\\
$T_p$ & peak wave period [s] & $P_{circ}$ & time-averaged extracted power by the equivalent circle [$W$]\\

\hline
\end{tabular}
\end{center}
 These include complexities associated with working in the harsh ocean environment such as corrosive nature of the ocean water, or extreme loads and impacts during storms. Even in a typical day, incoming ocean waves arrive almost always as a \textit{spectrum}, composed of waves of different frequencies and directions. For a wave energy device which is required to work at resonance, this poses an extra challenge; which is also new to the industry and research: conventionally offshore structures are designed to avoid resonance as much as possible, and to stay stable by minimizing all motions. A wave energy device on the other hand must target resonance and maximize its motion. 

Over the past century, more than one thousand patents have been filed on different ideas for harvesting ocean wave power. Over the past few decades, numerous clever theoretical, computational and experimental advancements have resulted in a much clearer picture of what can be achieved from ocean waves \cite[cf. e.g.][]{Falnes2014,mei2005theory,Liu2017}. Most of our today's knowledge, however, is for the case of monochromatic unidirectional incident waves, while ocean waves come in spectrum of frequencies and from many different directions. What makes the investigation of a frequency-spectrum difficult is frequency-dependent coefficients in the governing equation of a wave energy device that prevents closed-form solutions. High-performance computational tools are now allowing computational study of a wide range of scenarios that happen under the action of broadband frequency and directional waves.

To maximize the efficiency of a wave energy converter, a number of strategies are employed. For instance, we know today that (i) depending on the wave condition, depth of the water and the bathymetry, a different class of wave energy converters may be more suitable (i.e. more efficient and cost effective) \cite{Drew2009,BabaritAurelienJorgenHals2011,Lehman2016,Bouali2017,Rusu2015,Viet2016,Zhang2017}. (ii) A wide range of active control strategies have been proposed and their performance have been investigated \cite{Korde2002,Son2017,Kovaltchouk2013,Jama2015,Nielsen2013}. (iii) Nonlinear mechanisms have been shown to enhance the bandwidth of high performance of wave energy devices \cite{Younesian2017,Ramian2010}. (iv) Flexibility of the absorber may positively contribute (in some cases significantly) to the efficiency of  wave energy harnessing devices \cite{Desmars2016}, and (v) optimized \textit{geometry}, \textit{shape}, and \textit{power take-off mechanism} of a wave energy converter play an important role in the overall performance of the device \cite{Shadman2018,  Lopez2017, Xiao2017,Liu2018,Henriques2011}. Shape optimization is of relevance in all classes of wave energy conversion techniques such as oscillating water columns where it is found that immersion depth of 0.45 of water depth and chamber diameter of 0.92 of water depth yields the maximum efficiency \cite{Bouali2013}; and overtopping devices where it is found that the optimum length to opening ratio has to be $\sim 2.5-3$ \cite{Kramer2002}. Most efforts so far, nevertheless, have been limited to comparing few specific shapes. As mentioned above, this has been mainly due to the computational costs of finer searches among all possibilities. 


For a heaving point absorber wave energy converter under wave conditions of Belgian coastal area of the North Sea (shallow water), Vantorre et al. \citep{Vantorre2004} compared performance of a set of geometries including a hemisphere and a few conical geometries. They found that a $90^\circ$ cone with cylindrical extension that pierces the water surface yields the largest efficiency among their set. 
 For a point absorber working in wave conditions off the west coast of Ireland, Goggins and Finnegan \cite{Goggins2014} considered variations of a vertical cylinder (e.g. truncated cylinder with a hemispherical base, with a $45^\circ$ linear chamfer, or with a quarter encircle chamfer), and found that a truncated cylinder with a hemisphere attached to its base with the total draft to radius ratio of 2.5 is the optimum shape having the largest significant heave velocity response. Through experimentations, Hager et al. \citep{Hager2012} showed that a buoy with concave faces performs better than a buoy with flat or convex faces, and that the capture width ratio can reach up to $94.5\%$. For a surge motion of a wave energy device, McCabe \cite{McCabe2013} parametrized the device geometry by bi-cubic B-spline surfaces and showed that the optimum shape has asymmetry in the direction of wave propagation with distinct, pointed prows and sterns.

Here through a systematic approach, we find optimum shape of a single wave energy converter under different wave conditions including monochromatic, polychromatic, unidirectional and directional seas. While the methodology proposed is general and applies to any wave energy converter design, we focus our attention on the optimization of a submerged planar pressure differential wave energy converter called ``\textit{Wave Carpet}" \citep{Alam2012,Boerner2015,Lehmann2014} for deep waters. {The original idea of the wave carpet was to mimic the natural phenomenon of surface waves getting damped by muddy seabeds, and to place a synthetic wave absorber near the seabed that responds to the action of overpassing waves similar to how mud responds. If water is deep, the device may be elevated in the water column so that the device stays close enough to the surface, hence retaining required performance.} A submerged wave energy converter is more survivable under storm conditions (e.g. resulting slamming impacts \cite{Baker2001}) as the water column buffers surface forces, does not interfere with surface vessels and fishing boats, has no visual pollution, and does not impede atmosphere-sea surface interactions such as Oxygen and CO2 exchange, which is of major concerns from environmental point of view particularly when large-scale deployments (e.g. \textit{wave energy farms} \cite{Greenwood2016, Bozzi2017,Borgarino2012,Erselcan2017}) are sought. 

\begin{figure}
\centering
\includegraphics[width=0.5\textwidth]{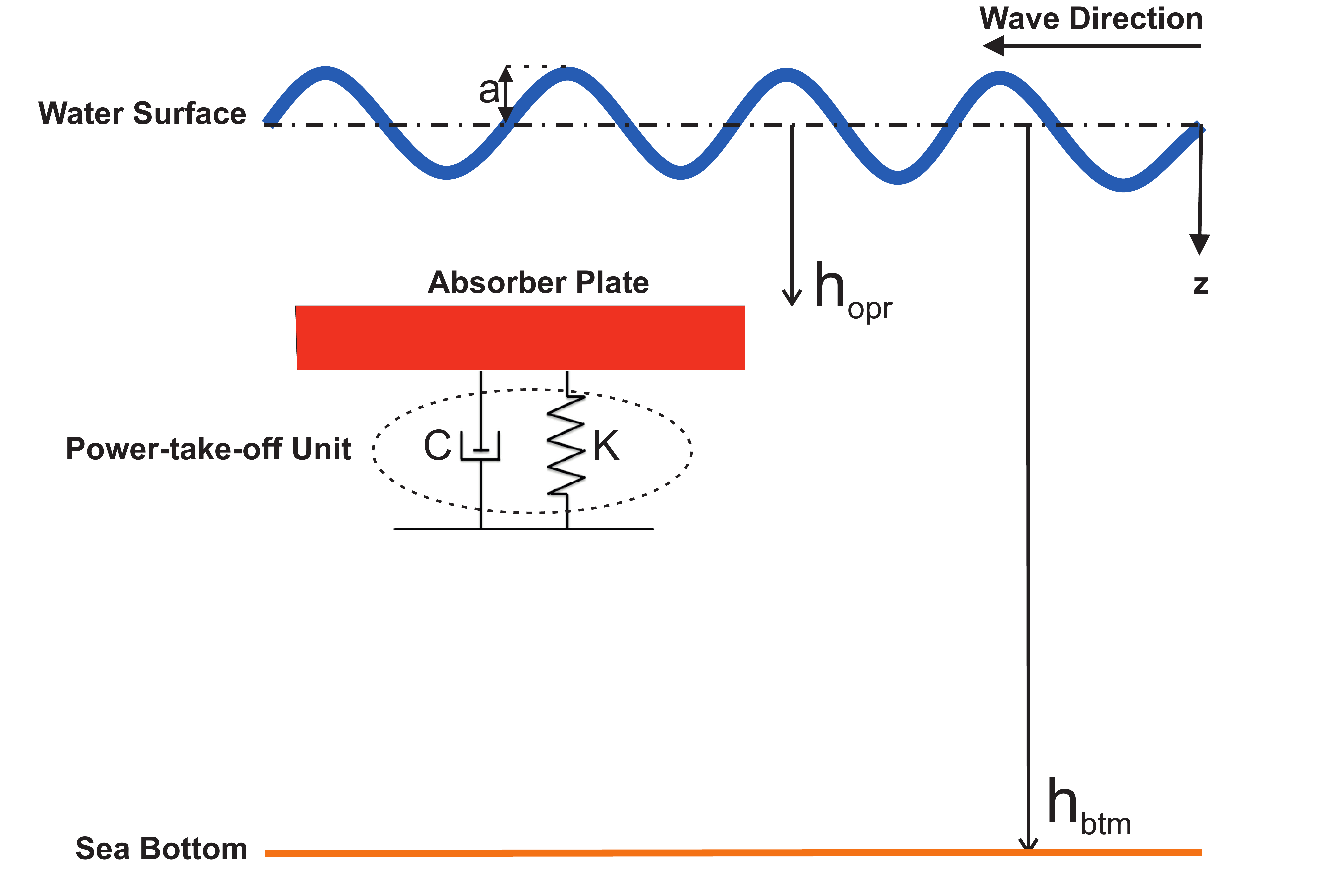}
	\caption{Schematic of a submerged planar pressure differential wave energy converter in sea; $h_{opr}$, $a$, $h_{btm}$, $K$, $C$ are absorber plate's operating depth, wave amplitude, sea bottom depth, and power take-off's spring and damper coefficients respectively}
	\label{fig_schematic}
\end{figure}

{In our investigation, without loss of generality, we consider a single deep water version of the aforementioned wave energy converter, a rigid planar absorber (figure \ref{fig_schematic}), and restrict our analysis to its heave-only motion.}
The objective is to find the optimum shape that produces maximum power under the action of specific form of incident waves. In our optimization, we compare performance of shapes of \textit{the same area}. This is because the overall cost of ships and offshore structures is roughly proportional to their weights. Therefore, for a wave energy device to have a comparable price with another of the same type (say Wave Carpet in this study), they must have, roughly speaking, the same weight and therefore the same area. Clearly this is not the only factor determining the cost \cite[cf.][]{Kurniawan2012}, and it is not difficult to find variations of a shape with the same area whose production leads to costs orders of magnitude higher than other shapes (e.g. infinitely long and thin plate). To compensate for such effects, we consider that all permissible shapes are variations of a fixed base ellipse plus a few Fourier modes, but put a limit on how much elongated the ellipse can be, and also on the amplitude of Fourier modes to avoid sharp corners. For each given wave condition, we optimize (i) spring stiffness, (ii) damping coefficient of generator, (iii) major and minor axes of the ellipse, and (iv) amplitude and phase of each Fourier mode (the last two items iii and iv under the constraint that the total area is the same). It's worth to mention that an allowable range of values for the spring and damping coefficients are considered that are practical in real world applications, and always the WEC is kept submerged.

As optimization algorithm, we tried different strategies including gradient-based methods, direct search methods, stochastic methods, and Genetic Algorithm. It turns out that for the problem in hand Genetic Algorithm has the fastest convergence rate, and we confirmed that it in fact converges to the true solution via cross-validating its results with those obtained via the direct search methods. Genetic Algorithm is a heuristic search method that has gained attention due to its efficient capability in finding optimum solutions; particularly in cases where the objective function cannot be expressed analytically in terms of its constitutive parameters/variables and where there exists many local extrema in the search area \citep{Amaran2014}. Genetic algorithm is not foreign to offshore industry, and has been used for optimization of offshore structures \cite[e.g.][]{McCabe2013,Kurniawan2012,Birk2009,Giassi2018}, and also for optimizing wave energy devices  \cite[e.g.][]{McCabe2013,Kurniawan2012}.

Here, we start with problem statement and presentation of the governing equations (\S2), followed by details of our shape optimization methodology and definition of the objective function (\S3). We then present and discuss results of optimization for the absorber's shape under four different wave conditions (\S4): (i) monochromatic unidirectional, (ii) monochromatic directional, (iii) polychromatic unidirectional and (iv) polychromatic directional waves (with both directional symmetry and asymmetry).



\section{Fundamentals}

We consider a single submerged flat plate with uniform thickness of $d$ in $z$-direction at the depth $h_{opr}$ connected to a power take-off (PTO) system and under the action of surface gravity waves (see figure \ref{fig_schematic}). The power take-off system is composed of a linear spring with the stiffness $K$, and a generator modeled as a linear damper with a damping coefficient $C$. The idea is that the net pressure difference between the top and the bottom of the plate due to the passage of surface gravity waves can move the plate that in turn moves the end of the power take-off system and results in the generation of power. Therefore, energy of surface waves is harnessed by the power take-off unit. 

In our analysis, we assume that water is homogeneous, inviscid, incompressible and irrotational such that potential flow theory applies. We also neglect the effect of surface tension, and assume that the water depth is constant and equal to $h_{btm}$. Restricting the motion of the plate to the heave-direction only (i.e.  $z$-direction in figure \ref{fig_schematic} ) the governing equation for the motion of absorber plate reads
\ba\label{Eq_EqOfMotion}
(m+\widetilde{A})\,\ddot{z}+(C+\widetilde{B})\,\dot{z}+K\,z=F_{exc}
\ea
in which $m$ is the mass of the absorber plate, $\widetilde{A}$ and $\widetilde{B}$ are the hydrodynamic added mass, and added damping coefficients respectively, and $F_{exc}$ is the wave excitation force in the $z$-direction. Parameters $\widetilde{A},\widetilde{B}$ and $F_{exc}$ are functions of the absorber plate's geometry as well as the incident wave frequency ($F_{exc}$ is also linearly proportional to incident wave amplitude). In order to calculate $\widetilde{A},\widetilde{B}$ and $F_{exc}$ we use an open source Boundary Element Method-based solver, \emph{NEMOH}, developed at \textit{Ecole Centrale de Nante} by Babarit and Delhommeau \cite{Babarit2015}
\footnote{Alternative tools include ANSYS AQWA (\cite{Vantorre2004,Pastor2014}) and WAMIT (\cite{McCabe2013,Kurniawan2012,Birk2009,Li2015,Chen2016,Tay2017}).}
.

 If the incident wave is monochromatic with the frequency $\omega$ then  $F_{exc}=\F\,e^{\i \omega t}$ where $\F$ is the amplitude of the excitation. The absorber's response therefore is obtained readily as $z=\hat{z}\,e^{\i\,(\omega t+\phi_z)}$ where
\bsa
&\hat{z}=\F/\sqrt{\bigl[K-(m+\widetilde{A})\,\omega^2\bigr]^2+\omega^2\,\bigl[C+\widetilde{B} \bigr]^2}, \\
&\phi_z=\arctan \bigg[ \frac{\omega\,(C+\widetilde{B})}{K-(m+\widetilde{A})\,\omega^2} \bigg].
\esa
The absorbed power is therefore 
\ba \label{Eq_powerr}
P_{abs}=\frac{1}{T}\int_{0}^{T}c ~ \dot z^2\,dt=\frac{1}{2}c\omega^2\hat{z}^2.
\ea

\begin{figure}
\centering
\includegraphics[width=0.5\textwidth]{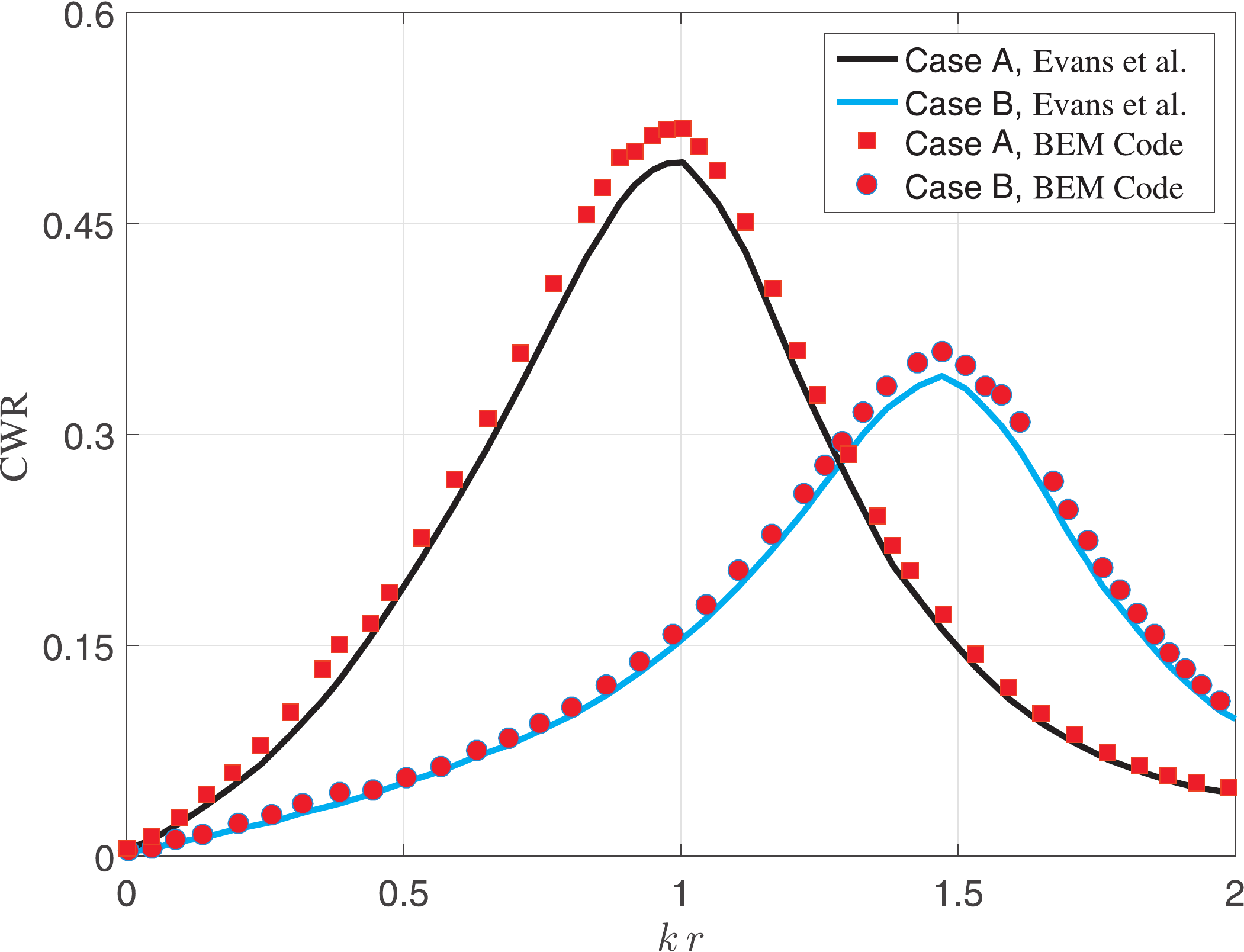}
	\caption{Capture width ratio (CWR) of a floating hemisphere (with relative density one) in heave as a function of dimensionless wavenumber of the incident wave $kr$ ($r$ is the radius of the hemisphere). Spring coefficient is chosen such that the undamped system is at resonance at respectively $kr=1.0$ (Case A), and $kr=1.5$ (Case B). Likewise damping coefficient of the power take-off is chosen to be the same as the radiation damping for $kr=1$, 1.5. Solid lines show CWR calculated analytically \cite{Evansss}, and markers show results obtained from NEMOH.}
	\label{fig_aspectratio}
\end{figure}

For validation of numerical calculation of hydrodynamic coefficients (added mass, added damping, and excitation force), we compare the Capture Width Ratio (CWR)\footnote{A measure of a wave energy converter's performance is the Capture Width Ratio (CWR) defined as the power extracted by the wave energy converter divided by the amount of power in the incident wave of crest length equal to the width of the device. For instance, if a device of width $w_d$ captures $p_d$ Watt under incident waves of power flux $p_w$ Watt per meter of the wave crest, then $CWR=p_d/(p_ww_d)$.} of a floating hemisphere with hydrodynamic coefficients obtained (i) analytically \cite{Evansss} and (ii) numerically through NEMOH. Figure \ref{fig_aspectratio} shows the comparison of results from analytical expressions (solid lines) and those obtained numerically (markers), for two cases of parameters tuned such that resonance occurs at $kr=1$ (case A) and $kr=1.5$ (case B).

%
%
%

\begin{figure}
\centering
                \begin{subfigure}[b]{0.33\textwidth}
                \includegraphics[width=\linewidth]{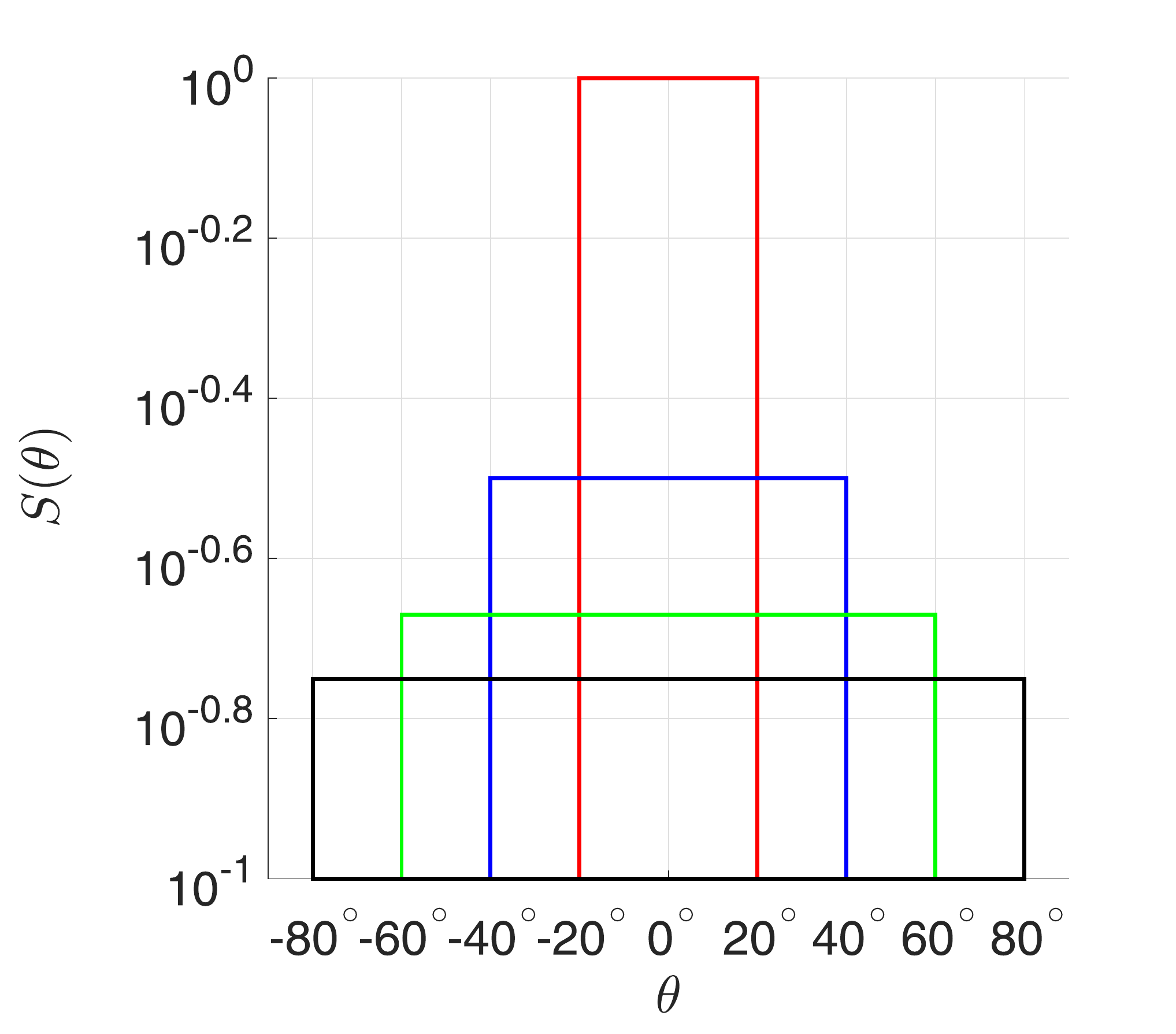}
                \caption{}
                \label{fig_amp_dis_mono_multidir}
        \end{subfigure}%
                \begin{subfigure}[b]{0.33\textwidth}
                \includegraphics[width=\linewidth]{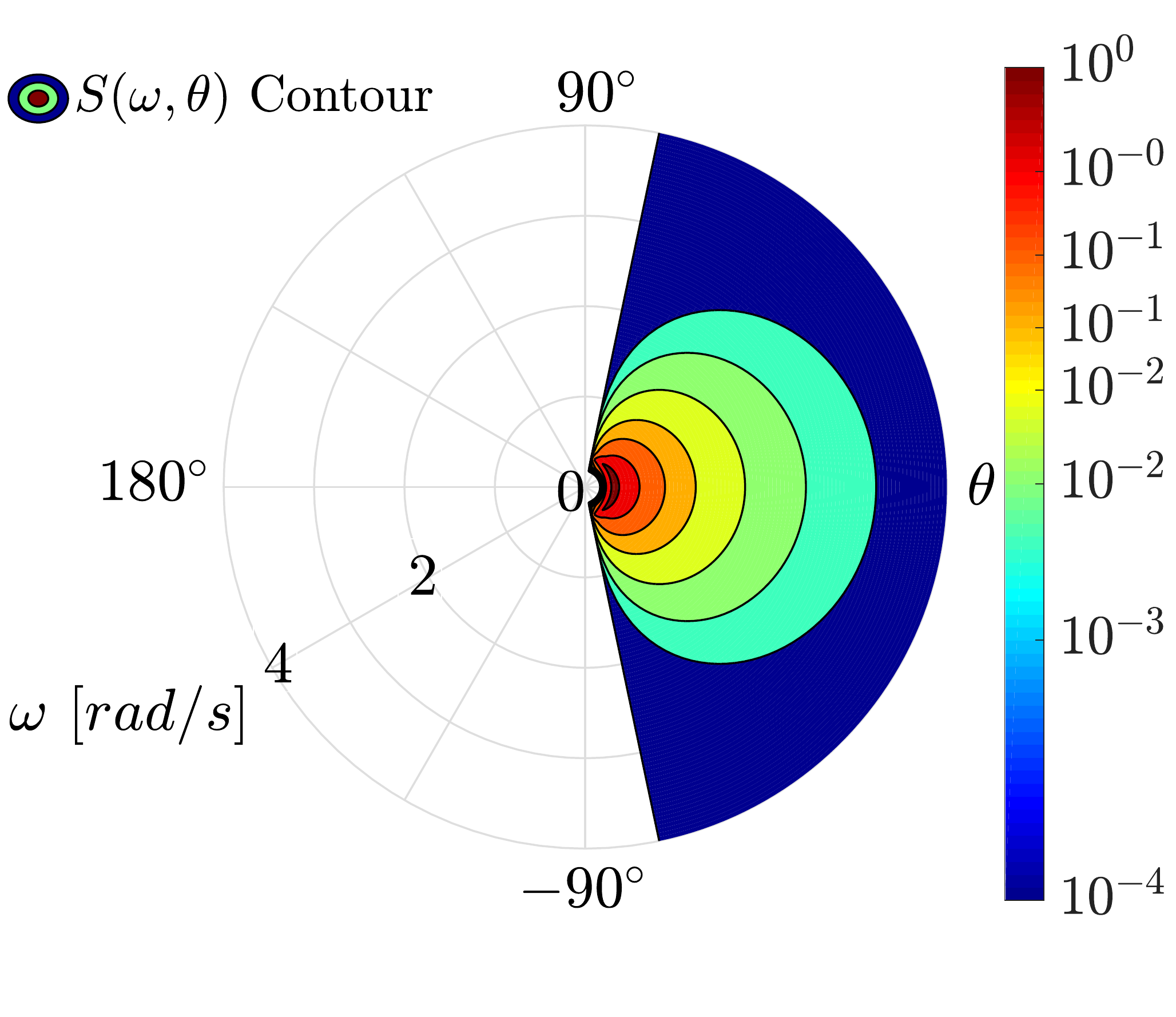}
                \caption{}
                \label{fig_amp_dis_sym_dir_spec}
        \end{subfigure}
        \begin{subfigure}[b]{0.33\textwidth}
                \includegraphics[width=\linewidth]{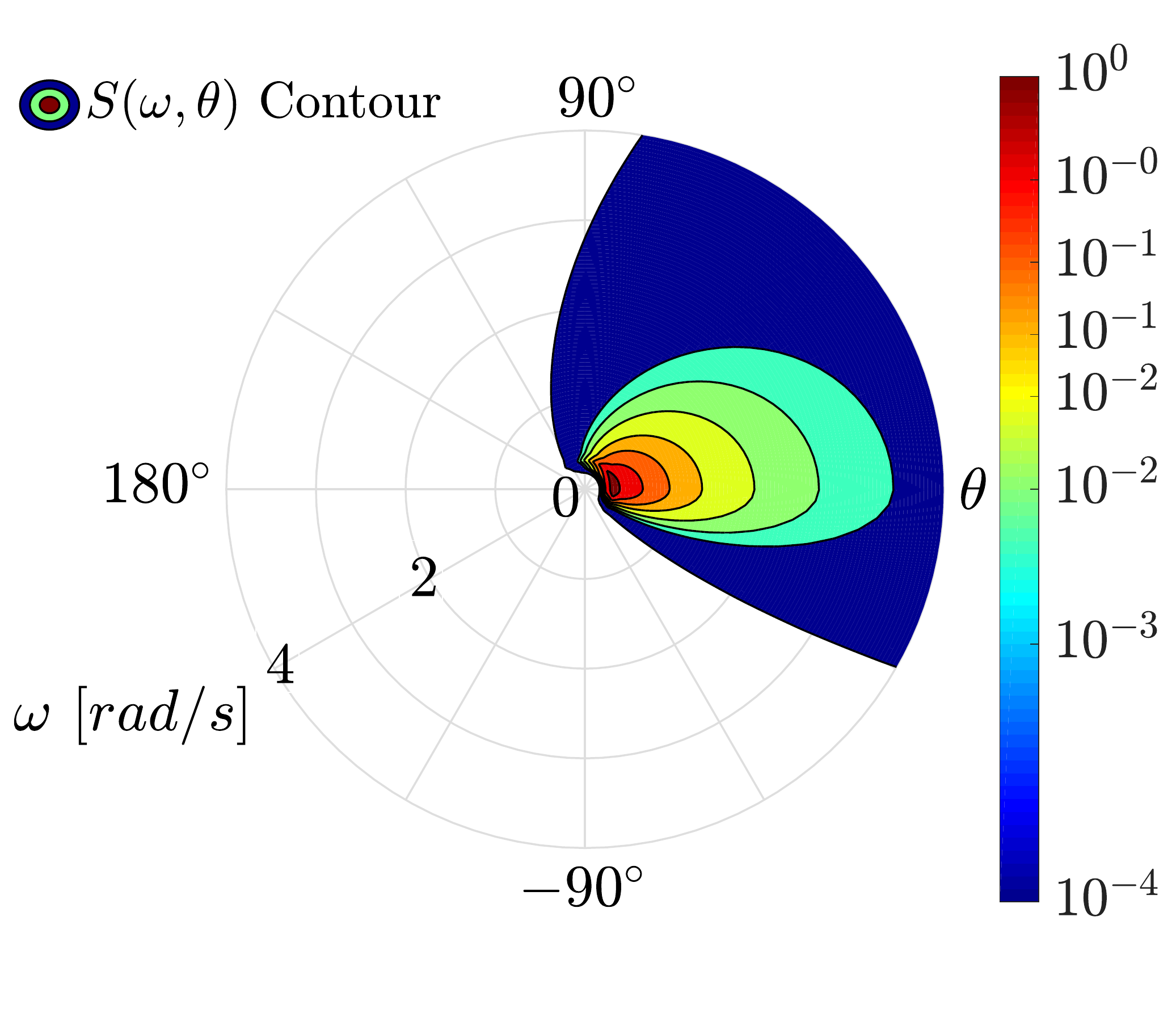}
                \caption{}
                \label{fig_amp_dis_nonsym_dir_spec}
        \end{subfigure}%
        \caption{Spectral density function ($S$) distribution of (a) monochromatic waves with directional spreading for four angular spans of $\pm20^\circ$, $\pm40^\circ$, $\pm60^\circ$ and $\pm80^\circ$, 
        (b) a JONSWAP broadband directional spectrum with symmetric angular distribution, $\mu=0,s=2,\Theta=160^\circ,H_s=3.25\,[m],T_p=9.7\,[s]$, (c) a JONSWAP directional spectrum with an asymmetric angular distribution, $\mu=0.5,s=75,\Theta=160^\circ,H_s=3.25\,[m],T_p=9.7\,[s]$}\label{figspr}
\end{figure}

The goal of the current research is to look for an optimized shape of a wave energy converter under four different sea conditions: $i.$ normal incidence of monochromatic unidirectional waves, $ii.$ monochromatic waves with directional spreading (figure \ref{figspr}.a), $iii.$ normal incidence of polychromatic unidirectional waves, and $iv.$ polychromatic directional spectrum of waves with a symmetric directional distribution (figure \ref{figspr}.b) and with a general (i.e. asymmetric) directional distribution (figure \ref{figspr}.c). For the directional spectra (wave climates $ii$, $iv$) the basic premise is that the spectral density function $S(\omega,\theta$) is the product of a frequency spectrum $S_f(\omega)$ and a spreading function $D(\theta)$. For the frequency spectrum, we choose the Joint North Sea Wave Project (JONSWAP) spectrum in the form (cf. \cite{Hasselmann1973})
\ba\label{Eq_Somega}
    S_f(\omega)=\frac{{\alpha}_p g^2}{\omega^5} \exp \bigg[-\frac{5}{4}\bigg(\frac{\omega}{\omega_p}\bigg)^{-4}\bigg]\gamma ^{\exp\,[-(\omega-{\omega}_p)^2/(2\,{\sigma}^2\,{{\omega}_p}^2)]}~,
\ea
where $\omega$ is the wave frequency, ${\omega}_p$ the peak wave frequency, and $\gamma$ the peak enhancement factor specifying the spectral bandwidth typically ranging between 1<$\gamma$<9 for which we choose the mean value of $\gamma=3.3$ \cite{Hasselmann1973}. $\alpha_p=H^2_s\,\omega^4_p/[16I_0(\gamma)g^2]$ is the Phillips parameter, $H_s$ is significant wave height, and the Zeroth-order moment $I_0(\gamma)$ which varies in the range of 0.2<$I_0(\gamma)$<0.5 is usually calculated numerically \cite{Carter1982} being equal to $I_0(3.3)=0.3$ for our application. In this manuscript, without losing generality we focus on the sea state five (rough sea conditions) with $H_s=3.25$ meters and $T_p=9.7$ seconds \citep{Alam2014}.
For the directional spreading component of the directional spectra (wave climates $ii$, $iv$), without loss of generality, we assume zero degree mean for the wave direction (i.e. ${\theta}_m=0^\circ$) and use the following directional spreading function $D(\theta)$ (c.f. e.g. \cite{Lee2010}) 
\begin{subequations}
\begin{align}
    D(\theta)=\left\{
\begin{array}{c l}     
    \frac{2}{\Theta}\cos^s(\frac{\pi\theta}{2\Theta}\zeta) & \lvert~\theta\frac{\zeta}{2}~\rvert\leqslant{\Theta}/{2}\\
    0 & \lvert~\theta\frac{\zeta}{2}~\rvert>{\Theta}/{2}
\end{array}\right. 
\\
\zeta=\left\{
\begin{array}{c l}     
    \exp\,(-\mu) & \theta\geq0\\
    \exp\,(+\mu) & \theta<0
\end{array}\right. ~~~~~~~~~~~~~~\hspace{5pt}
\end{align}\label{Eq_S_theta}
\end{subequations}
where $\Theta$ is angular span, and $\mu$ is the asymmetry parameter. Physically speaking, a positive $\mu$ means a broader angular distribution of energy for $\theta>\theta_m=0$ (see e.g. figure \ref{figspr}.c). Here, for a symmetric directional spreading we choose $\mu=0$, $s=2$, $\Theta=160^\circ$, and for asymmetric directional spreading we choose $\mu=0.5$, $\Theta=160^\circ$, and $s=75$ to limit the spectrum in a finite angular span. These cases are shown in figure \ref{figspr}.  For a monochromatic directional spectrum, amplitude of individual waves are $a_j=\sqrt{2\,S(\theta_j)\,\d\,\theta}$, and for a broadband directional spectrum $a_j=\sqrt{2\,S(\omega_j,\theta_j)\,\d\,\omega\,\d\,\theta}$.

\section{Shape Optimization Methodology}
\label{sec_ShapeoptimizationMethodology}
%
%
%

The objective of optimization is to maximize the normalized time-averaged power generation $P_r$ defined by
\ba\label{pr}
P_r=\frac{P_{abs}}{P_{circ}}
\ea
where $P_{abs}$ is the time-averaged extracted power by a wave energy converter, and $P_{circ}$ is the time-averaged extracted power by a \textit{circular-shaped} device of the same area (equivalent circle). Note that both ${P_{abs}},{P_{circ}}$ are obtained for optimum power take-off parameters $C,K$. 

The power extracted for the case of unidirectional monochromatic incident waves (case $i$) is very simple and given by \eqref{Eq_powerr}. But care must be taken in dealing with incident waves with directional spreading. The reason is that when we have directional spreading two waves of the same frequency may superimpose at the location of the device, but the extracted energy cannot be linearly superimposed as the contribution of phase-differences become of leading order importance.%
\footnote{This is similar in nature to the paradoxical classic problem of two same-frequency waves of amplitudes $a$ superposition. Let's assume we have two identical waves of amplitude $a$, each carry energy of $ca^2$ with $c$ being a constant. Therefore the two have $2ca^2$ energy. If we superpose these two waves, there is a new wave of amplitude $2a$ that must have, according to the wave theory, the energy of $4ca^2$ which is double the initial energy of the two initial waves. Alternatively, if we superpose the two waves with a $pi$ radian phase difference, then the outcome is a wave of amplitude zero with zero energy! The solution lies in the importance of consideration of phases.}
Specifically, power absorption under monochromatic directional incident waves (wave climate $ii$) is given by (see Appendix I)
\ba\label{Eq_objFunc_waveClimateII}
P_{abs}=\frac{1}{2}c\omega^2\,{\sum^{N_{\theta}}_{i=1}\sum^{N_{\theta}}_{j=1}}\,\hat{z}_i\,\hat{z}_j\,\cos(\phi_{z,i}-\phi_{z,j})~,
\ea
and power absorption under an incident directional spectrum (wave climate $iv$) is given by
\ba\label{Eq_objFunc_waveClimateIII}
P_{abs}=\frac{1}{2}c\sum_{k=1}^{N_{\omega}} \omega^2_k\,{\sum^{N_{\theta}}_{i=1}\sum^{N_{\theta}}_{j=1}}\,\hat{z}_i\,\hat{z}_j\,\cos(\phi_{z,i}-\phi_{z,j}),
\ea
where $\hat{z}$ is the system response amplitude, $\phi_z$ is the response phase, and $N_{\theta},N_{\omega}$ are the total number of directional and frequency spreading subdivisions (discretizations).

For shape parameterization, we consider a horizontal circle of radius $r_0$ (the base circle), and add $n=N_c$ Fourier components with amplitude $a_n$ and phase $\phi_n$ according to
\begin{equation}
r(\theta)=r_0+\sum_{n=1}^{N_c}a_n\,\sin\,(n\,\theta+\phi_n).
\label{Eq_radius_fouriermodes}
\end{equation}
This representation enables us to construct any arbitrary shape if $N_c$ is large enough. Note that with the introduction of phases the obtained shapes can be symmetric or asymmetric (cf. exempli gratia \cite{Newman2014,Newman2015}). For the problem in hand, in some cases, the shape has to be elongated in the $x$ or $y$-direction, therefore to make the optimization more efficient we define elongation coefficients $p,q$ according to 
\ba \label{Eq_xy}
x=p\,r\,\cos\,(\theta), \qquad y=q\,r\,\sin\,(\theta). 
\ea
These elongation coefficients $p,q$ will be optimized along with amplitudes $a_n$ and phases $\phi_n$ to obtain the optimum shape. We also impose the constraint that the total area of all shapes must be constant. Mathematically this can be expressed, using Green's theorem \citep{Odzijewicz2013} {(for derivation see Appendix III)}, as
\ba\label{Eq_sinModes_Area}
pq\lb\pi{r_0}^2+\frac{1}{2}\left(\sum_{n=1}^{N_c} {\pi\,a_n}^2\right)\rb=\mbox{Constant}.
\ea

For optimization, we use \textit{Genetic Algorithm (GA)} which is an evolutionary optimization scheme that mimics the process of natural selection (see e.g. \cite{Mitchell1998}). {The GA solver of MATLAB optimization toolbox, \textit{Optimtool}, is used as the optimization module in our work.} An optimization through Genetic Algorithm starts with an initial random set of points (i.e. candidate solutions) in the phase space called \textit{initial population}. Fittest members of the initial population, i.e. data points that better fulfill the optimization objective, are given preferences to breed and form the next generation. To form the next generation, each pair of survived data points, i.e. \textit{parents}, give birth, through a process called \textit{crossover} to one or more children that inherit properties from both parents, though not being identical to either. To make sure that the artificial evolution does not converge to a local extremum, \textit{mutation} is induced here and there through \textit{randomly} modifying a percentage of newborn children. There are many other details that have shown to increase the convergence efficiency of genetic algorithm, e.g. how many parents survive to the next generation, or putting a limit on the life of each member of the population. But eventually, after enough number of generations (that may range from few to billions or higher) in many cases a global optimum is reached. Genetic algorithm is a heuristic search that starts from a random search, but employs an evolutionary logic to find the optimum solution. Genetic algorithm is typically computationally expensive, but in many cases its performance can significantly exceed that of gradient-based methods, particularly when the objective function is not provided in an analytical form. 
\begin{figure}
\centering
\includegraphics[width=0.47\textwidth]{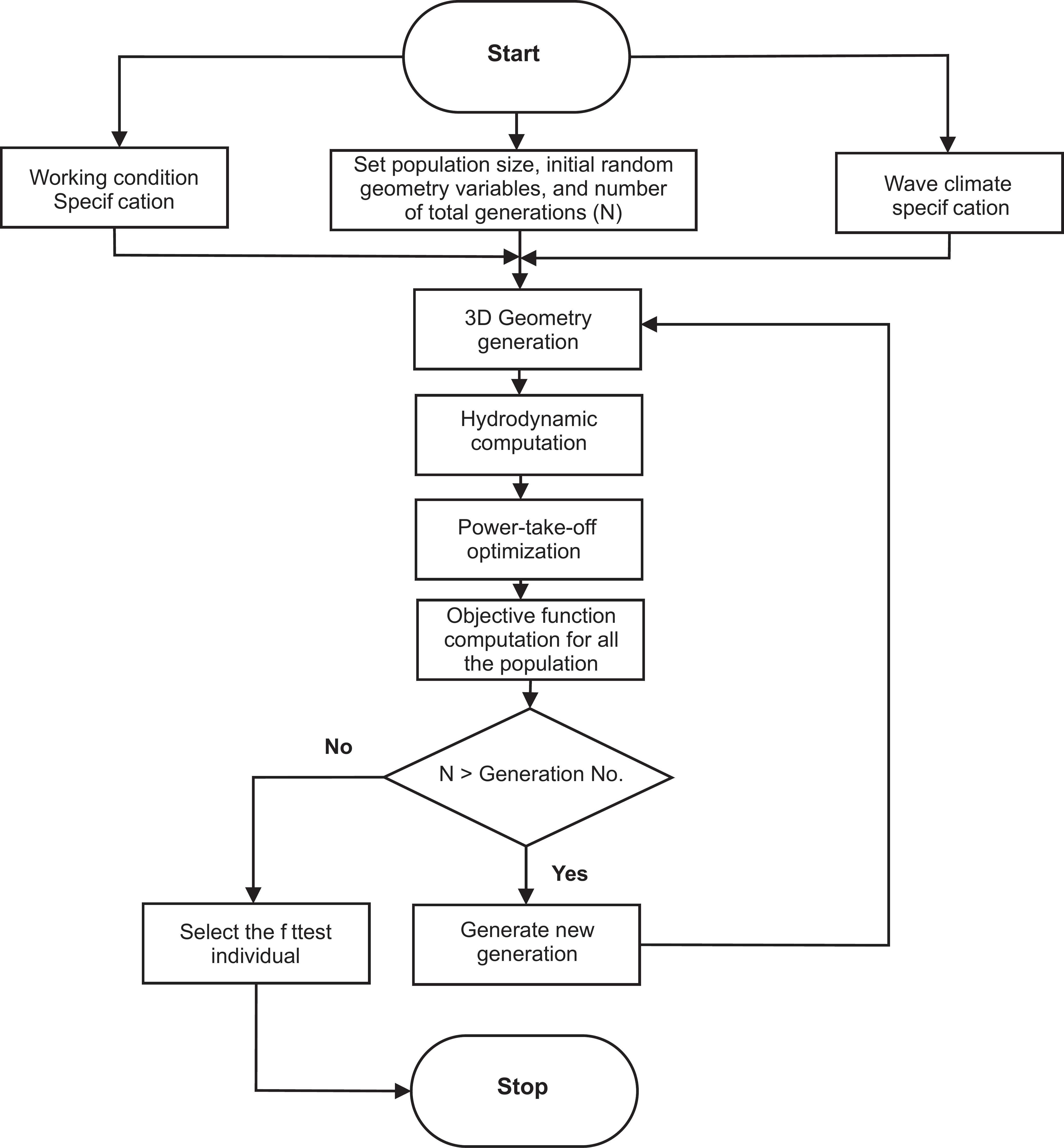}
	\caption{{Flowchart of shape optimization process. We start by specifying the working conditions of the device e.g. operating depth, sea bottom depth and incident spectrum, as well as Genetic Algorithm (GA) parameters. Next, by using our shape generation method,  shapes are generated and meshed. Then through Boundary Element Method (BEM) solver \textit{NEMOH}, hydrodynamic coefficients of each shape are calculated. Next, the optimum power take-off parameters for each shape are found using a separate optimization subroutine. At the next step, the objective function value, i.e. the power that can be absorbed by each shape, is calculated. Then the evolutionary process of GA optimization continues until reaching to the maximum number of generations where the optimization process stops and the shape with maximum absorbed power is chosen as the optimum shape. For each optimization case the maximum number of generations is tuned in a way that the increment in absorbed power by the fittest individual in the consecutive generations be small and when there is no significant improvement in the values of fitness of the population from one generation to the next. On average an optimization case for a broadband directional wave climate required 21.71 hours of computation on 8 CPU cores.}}
	\label{fig_flowchart}
\end{figure}

Here the optimization objective function is to maximize $P_r$ defined according to \eqref{pr}, and variables to be optimized are $a_n,\phi_n,p$ and $q$ given in equations \eqref{Eq_radius_fouriermodes} and \eqref{Eq_xy} under discussed constraints. The optimization procedure is as follows: we initialize our Genetic Algorithm with random selections of $a_n,\phi_n,p$ and $q$. Each set corresponds to a unique shape whose hydrodynamic coefficients can be calculated via NEMOH after proper meshing is generated. Since a wide range of different geometries are tested, a careful convergence test for each shape is performed through refining the mesh and re-calculating hydrodynamic coefficients. We then using Genetic Algorithm optimize power take-off parameters $c,k$ for the shape to yield the maximum power. This is done for every individual shape in the population. Then objective function is evaluated and subsequently the next generation is created. The procedure continues until convergence is reached.

\section{Results and Discussion}

\subsection{Monochromatic unidirectional incident wave}
\begin{figure}[h!]
\centering
        \begin{subfigure}[b]{0.25\textwidth}
        \centering
                \includegraphics[width=\linewidth]{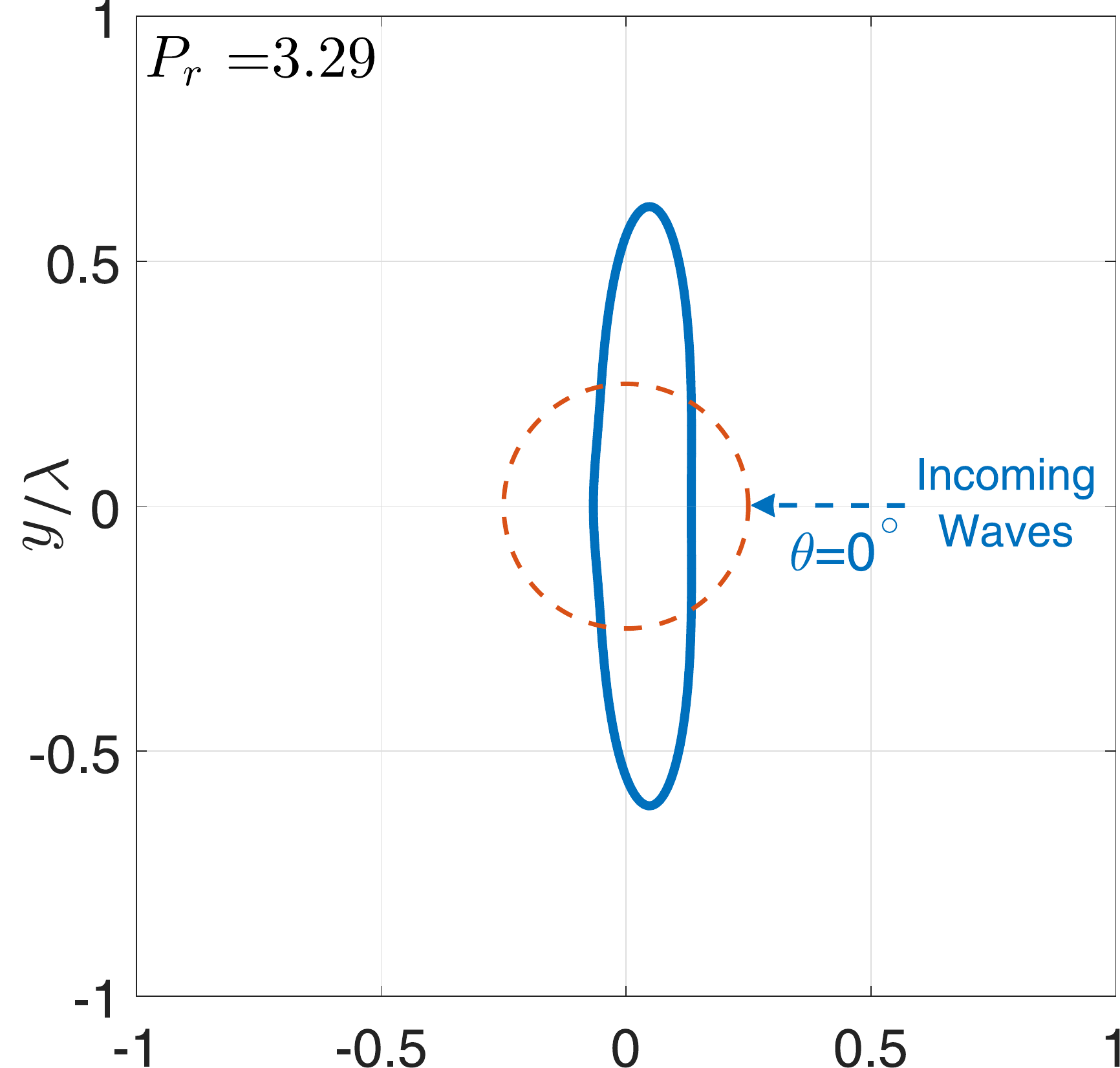}
                \caption{}
                \label{fig:MUW_a}
        \end{subfigure}%
        \begin{subfigure}[b]{0.25\textwidth}
        \centering
                \includegraphics[width=\linewidth]{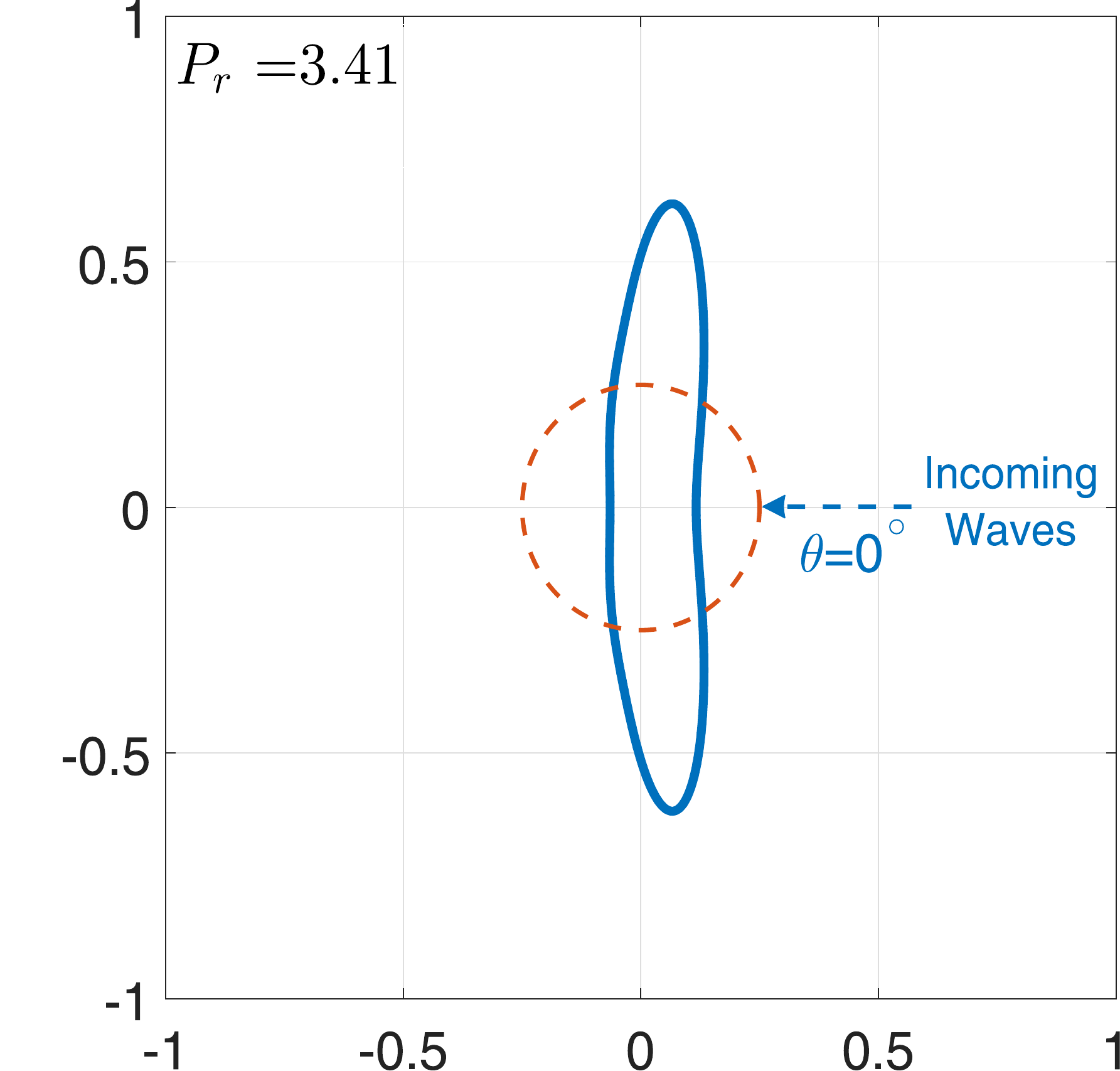}
                \caption{}
                \label{fig:MUW_b}
        \end{subfigure}%
        \begin{subfigure}[b]{0.25\textwidth}
        \centering
                \includegraphics[width=\linewidth]{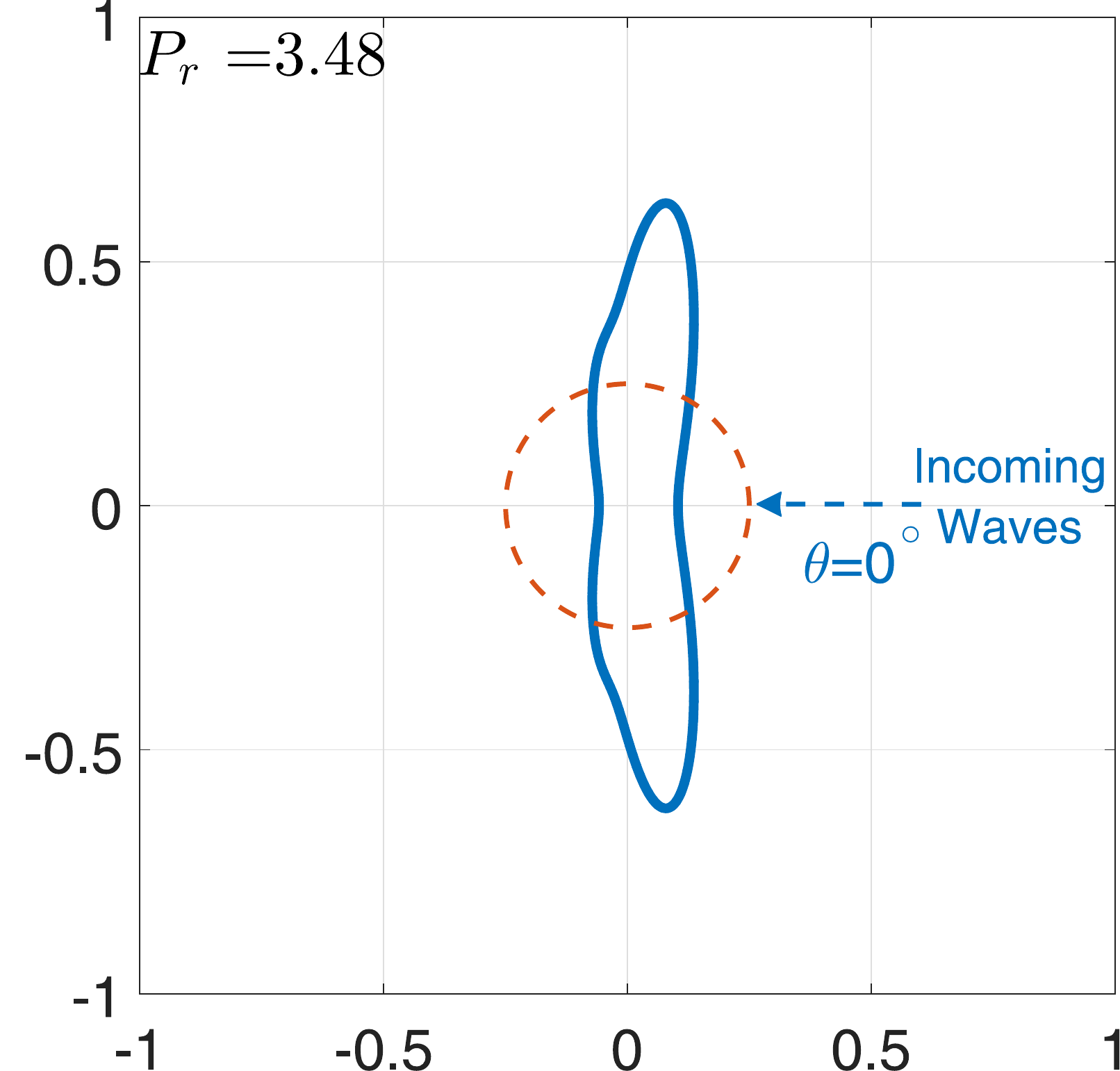}
                \caption{}
                \label{fig:MUW_c}
        \end{subfigure}%
        \\
        \begin{subfigure}[b]{0.25\textwidth}
        \centering
                \includegraphics[width=\linewidth]{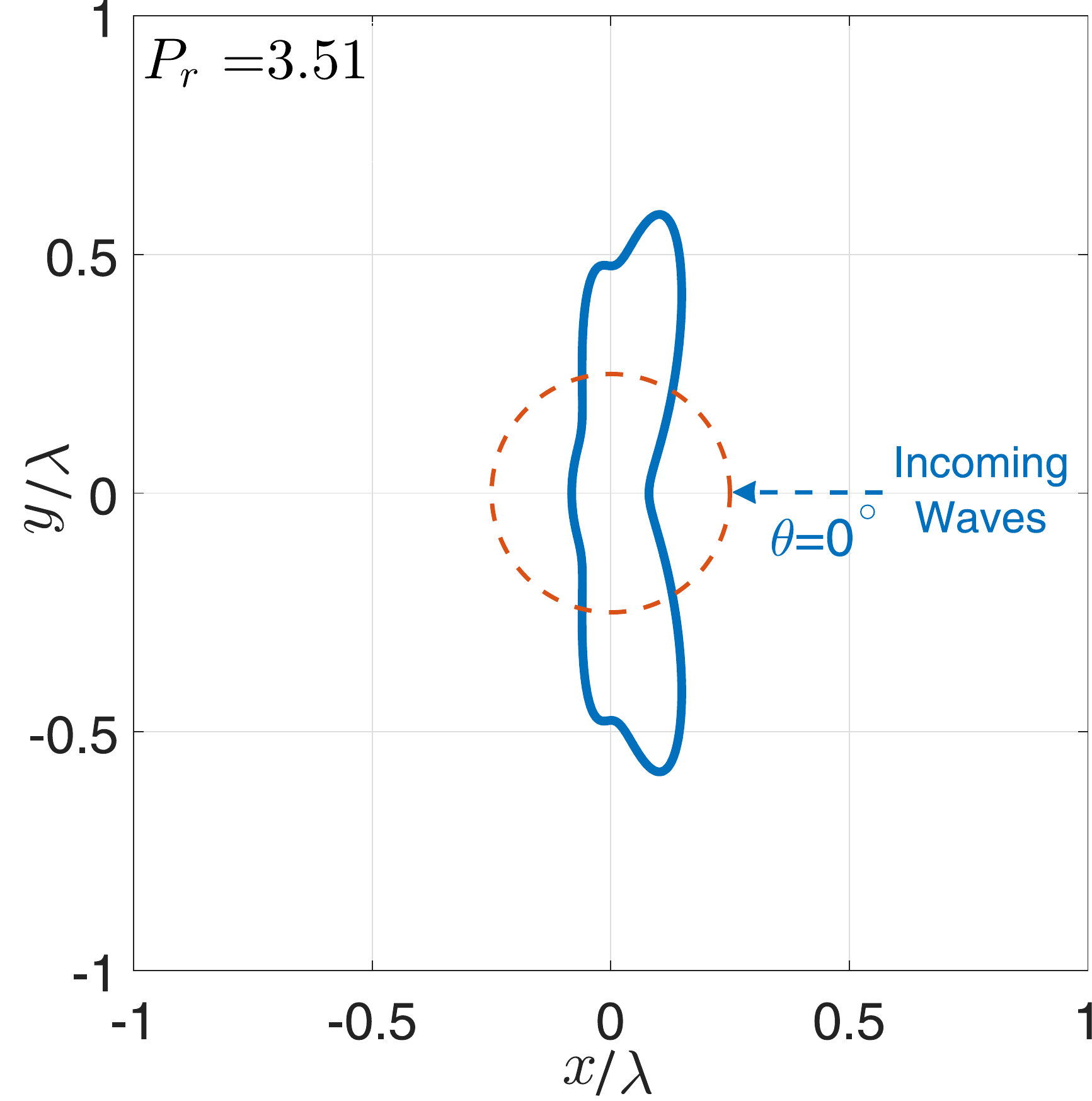}
                \caption{}
                \label{fig:MUW_d}
        \end{subfigure}%
        \begin{subfigure}[b]{0.25\textwidth}
        \centering
                \includegraphics[width=\linewidth]{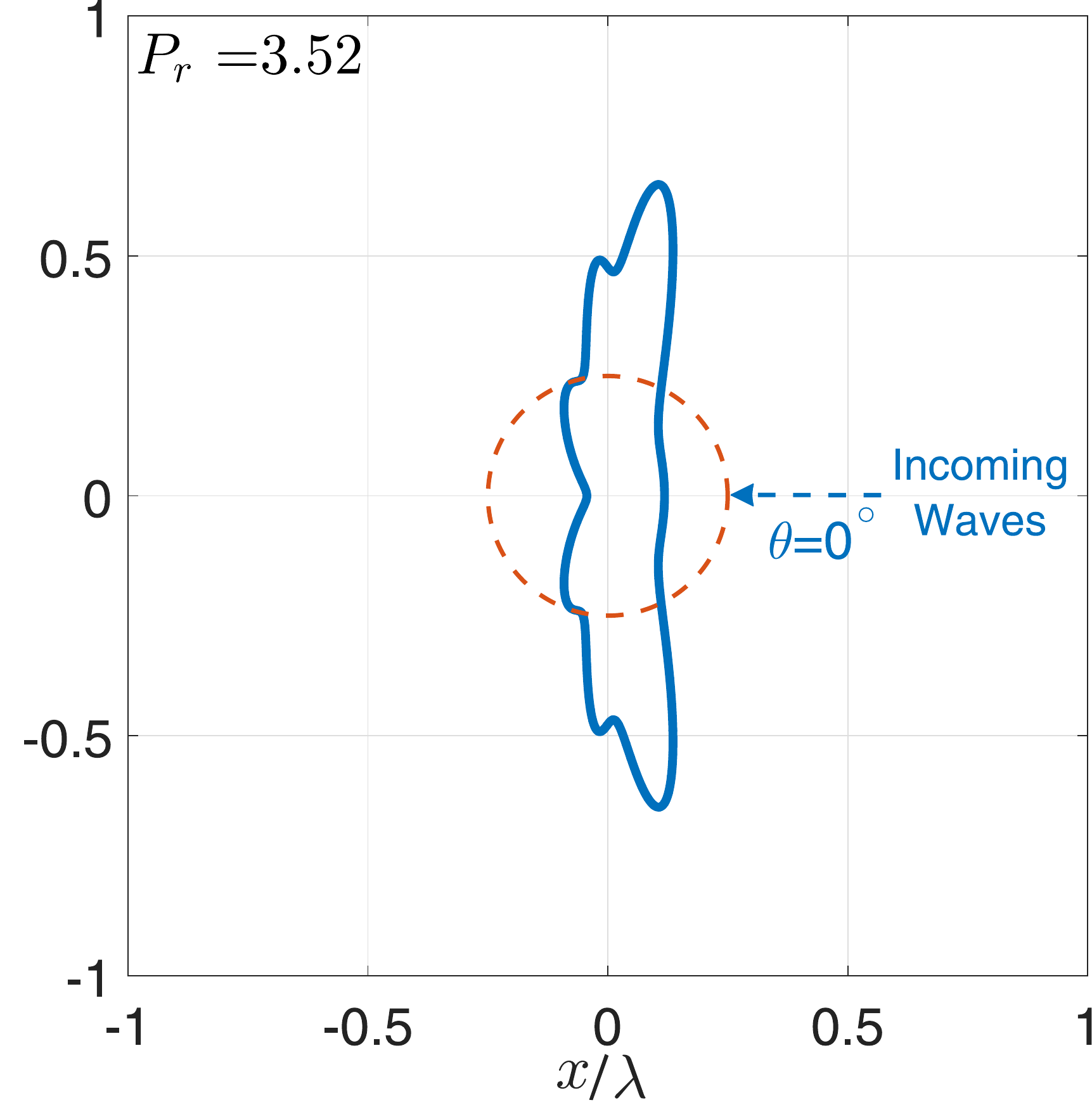}
                \caption{}
                \label{fig:MUW_e}
        \end{subfigure}%
        \begin{subfigure}[b]{0.25\textwidth}
        \centering
                \includegraphics[width=\linewidth]{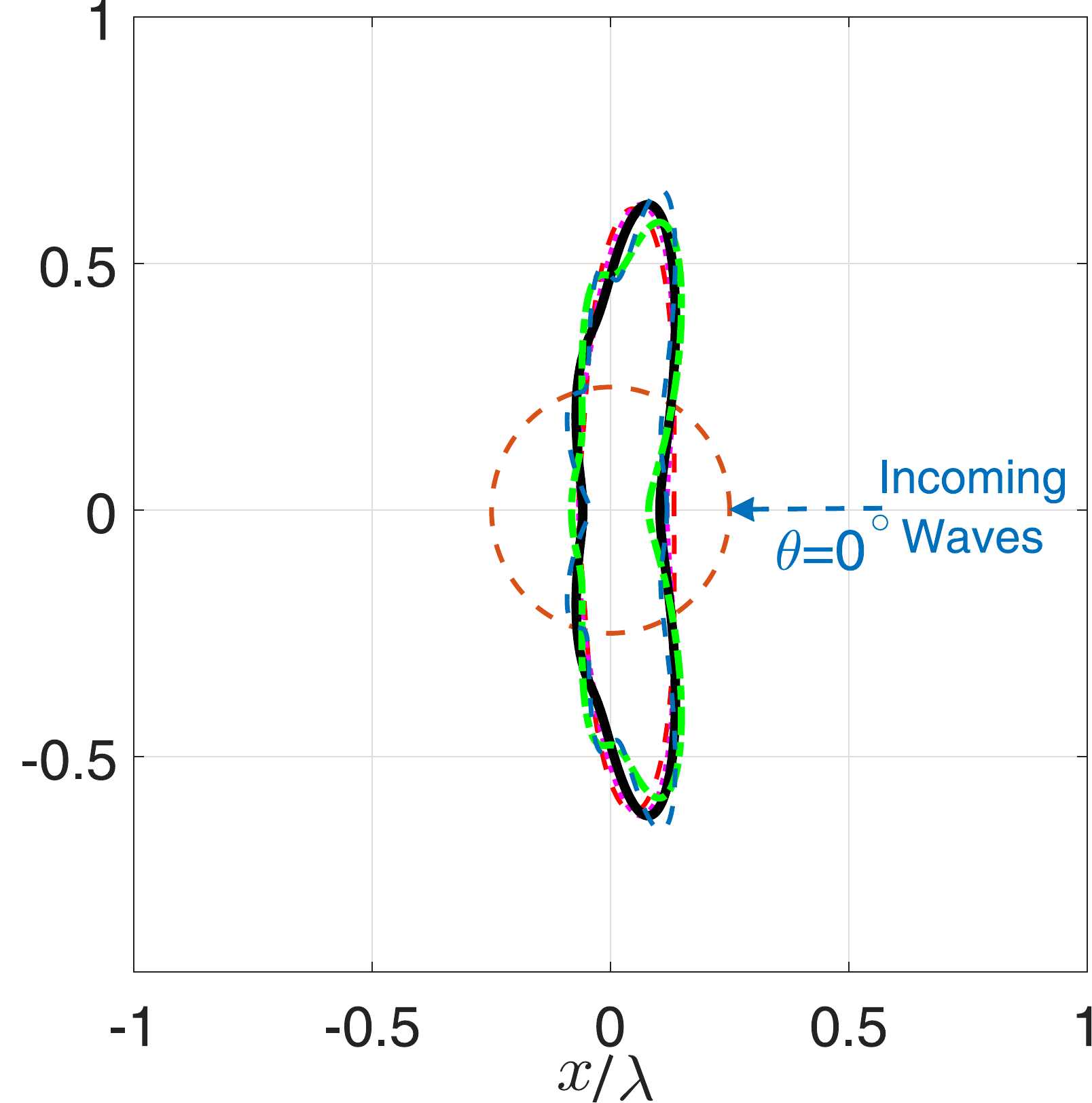}
                \caption{}
                \label{fig:MUW_f}
        \end{subfigure}%
        \caption{Optimum absorber plate shapes for monochromatic unidirectional incident waves arriving from the right (i.e. $\theta=0^{\circ}$). $N_c$ = 3(a), 4(b), 5(c), 6(d) and 7(e), for normalized area $A/\lambda^2=0.2$,  $kh_{opr}=1.13$ and $kd=0.11$. For comparison of shapes, we show a superposition of all optimum shapes in figure f. Shown in this figure are the equivalent circle of the same area (red dashed-line) and optimum shapes (blue solid line) and coefficients of shape generation Fourier modes are given in Appendix II.}
        \label{mono}
\end{figure}

We first consider the case of monochromatic unidirectional incident wave, for which we already expect that the optimum shape is long and narrow (i.e. stretched along $y$-axis). Therefore, this step also serves as a validation of our approach. Since typically the cost of ocean objects are approximated proportional to the material used, in our optimizations we assume that the total area of the device $A$ is constant. If the geometry is a simple circle, we call it the equivalent circle whose radius is $r_e=\sqrt{A/\pi}$. 

In practice, the range of variables to be optimized is always limited by design constraints that include for example fabrication limitations and force distribution and concentration. Here we assume $0.5<p,q<2$ which physically means that the optimum shape cannot be too narrow or long. Also for the equivalent circle radius $r_e$ and base circle radius $r_0$ (cf. equation 9)  we assume $r_0/r_e=0.95$, $a_n/r_0<0.40$ in order to keep the shape near a circle and prevent formation of complex geometries with sharp corners. Clearly the methodology is very general, and we have checked that the qualitative nature of results presented here do not change by varying these constraints within a reasonable range. 

For a total area of $A/\lambda^2=0.2$, where $\lambda$ is the wavelength of incident unidirectional monochromatic waves, we compare the optimum shape (shown in blue solid lines) that gives the maximum normalized average power $P_r$, along with the equivalent circle (shown in red dashed-line) in figures \ref{mono}a-e. We consider, respectively, the number of modes $N_c$=3-7 in figures \ref{mono}a-e, and an overlap of all shapes in figure \ref{mono}f. Clearly the optimum absorber plate shape tends toward an elongated shape perpendicular to the direction of incident wave (i.e. $\theta$=0), and is symmetric about that axis. With $N_c=$3 we obtain a power production $P_r$=3.29 with the optimum shape depicted in fig. \ref{mono}a, that is, the optimum shape has more than 300\% higher energy capture capability than a circular absorber of the same area. By increasing the number of modes $N_c$, at the cost of more curves and corners the performance consistently increases to $P_r=$ 3.41, 3.48, 3.51 and 3.52 for respectively $N_c=$ 4, 5, 6 and 7. Clearly the performance is asymptotically approaching an upper limit of $P_r \sim 3.5$ with the increase of $N_c$. 

We can also show that shapes converge as $N_c$ increases. To do so, we define a shape difference factor $\xi_{A,B}$, as a metric to measure the difference between two absorber plate shapes $A,B$, according to
\ba\label{sdf}
 {\xi_{A,B}}=\frac{1}{2\pi}\int_{\theta=0}^{2\pi} \lb\frac{r_B(\theta)-r_A(\theta)}{r_A(\theta)}\rb^2 \d \theta.
 \ea
We show the values of $\xi$ for (i) successive shapes, and (ii) each shape compared to the one with highest $N_c$(=7) in table \ref{tbl}. The difference between shapes decreases from 0.47\% for $N_c=$ 3, 4 to 0.32\% for $N_c=$ 6, 7, and also the difference with each shape and $N_c=$ 7 shape also decreases monotonically. Therefore, the optimization is convergent with respect to $N_c$, though as it is seen the rate of convergence is relatively slow.
\begin{table}
\centering
\begin{tabular}{c c c c }
 $\xi_{3,4}$ & $\xi_{4,5}$ & $\xi_{5,6}$ & $\xi_{6,7}$ \\
 \hline
 0.47 \% & 0.46\% & 0.39 \% & 0.32\%\\
 $\xi_{3,7}$ & $\xi_{4,7}$ & $\xi_{5,7}$ & $\xi_{6,7}$ \\
 \hline
3.72\% & 2.44 \% & 1.77 \% & 0.32 \%
\end{tabular}
\caption{The shape difference factor $\xi$, defined according to \eqref{sdf}, for comparison of relative absorber plate shapes presented in figure \ref{mono}. Differences in shape decrease as $N_c$ increases that indicates a convergence of our scheme with $N_c$.}
 \label{tbl}
\end{table}

It can be shown that optimum shapes in figure \ref{mono} under mentioned constraints have minimum or very low second moment of area about their $y$-axis (i.e. an axis parallel to the $y$-axis that goes through each shape's center of mass). This is equivalent of having the narrowest shape within the pool of shapes under above assumptions, which is expected since for monochromatic unidirectional waves under potential flow theory in fact the narrowest shape has the highest capture width ratio. We would like to comment that in some cases, for instance in the case of figure \ref{mono}-c, the optimum shape has a slightly ($<3\%$) higher second moment of area than the shape with lowest moment of inertia with $N_c=$ 5. The reason is that under considered constraints and for a given $N_c$ (for the case of figure \ref{mono}-c, $N_c=5$), the optimum shape for $P_r$ as well as the shape with the minimum second moment of area have (smooth) dents and bulges (see e.g. figure \ref{mono}-e). Once these dents and bulges are introduced, second moment of area may not suffice to describe the optimum shape and a full wave-analysis is needed to investigate wave-body interaction. That's the reason why in the presence of dents and bulges that appear for higher $N_c$'s, optimum $P_r$ shape may be slightly different than the shape with the minimum second moment of area.

\subsection{Monochromatic directional incident wave}

\begin{figure}[h!]
\centering

        \begin{subfigure}[b]{0.245\textwidth}
                \includegraphics[width=\linewidth]{0deg_opt5}
                \caption{$\theta=0^\circ$}
                \label{figgull}
        \end{subfigure}%
        \begin{subfigure}[b]{0.25\textwidth}
                \includegraphics[width=\linewidth]{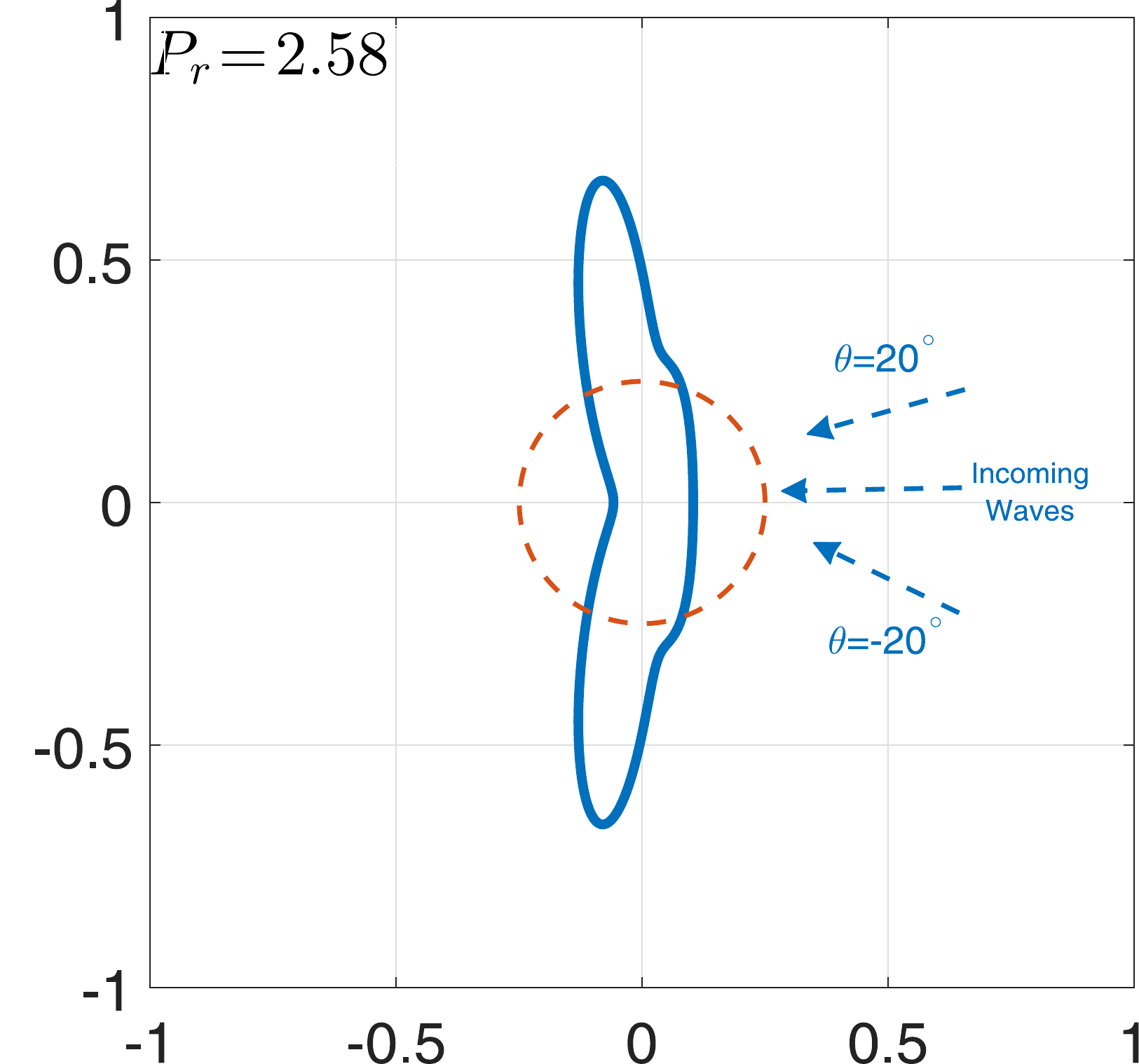}
                \caption{$|\theta|<20^\circ$}
                \label{figgull2}
        \end{subfigure}%
        \begin{subfigure}[b]{0.25\textwidth}
                \includegraphics[width=\linewidth]{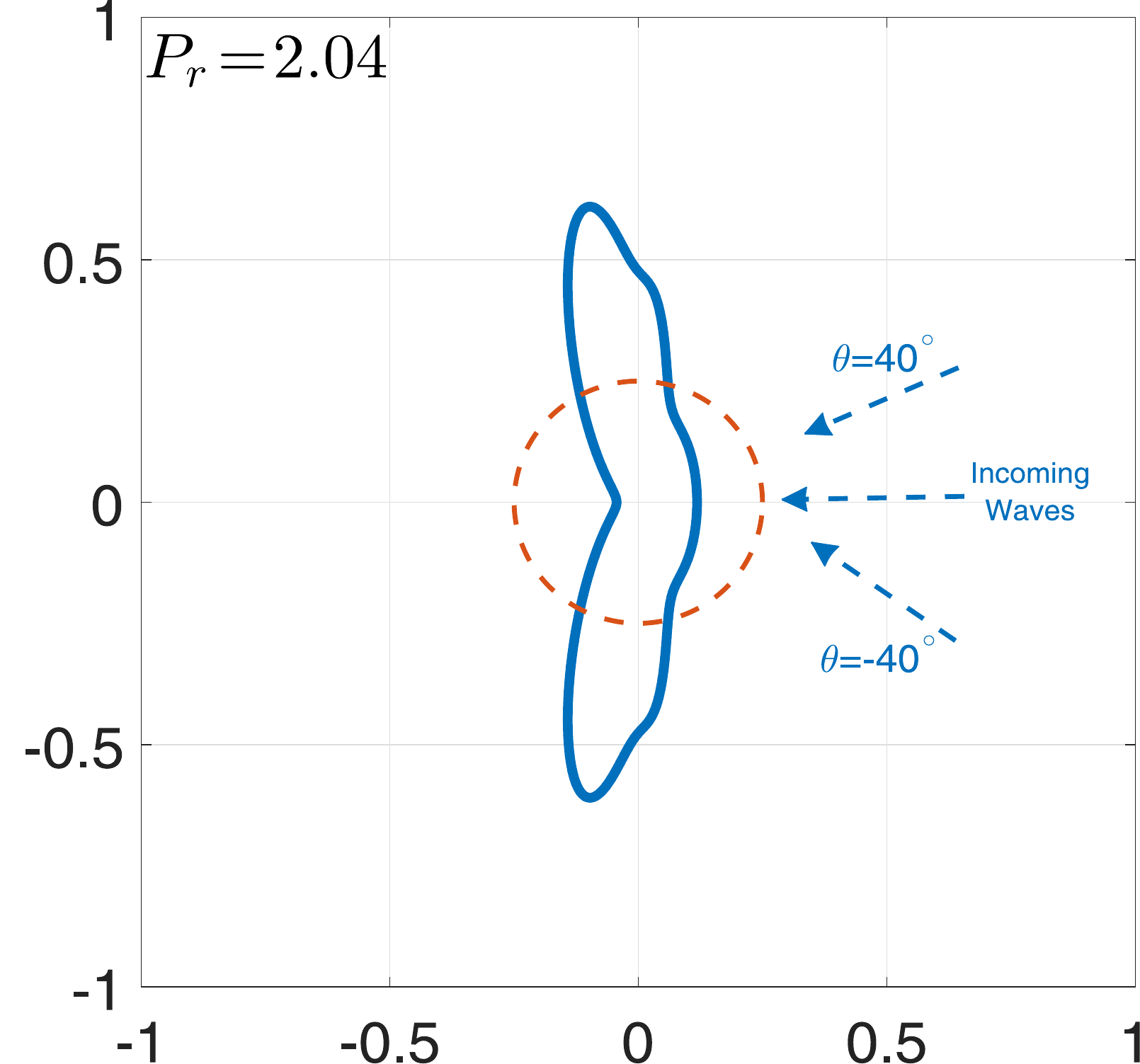}
                \caption{$|\theta|<40^\circ$}
                \label{figtiger}
        \end{subfigure}%
        \\
        \begin{subfigure}[b]{0.25\textwidth}
                \includegraphics[width=\linewidth]{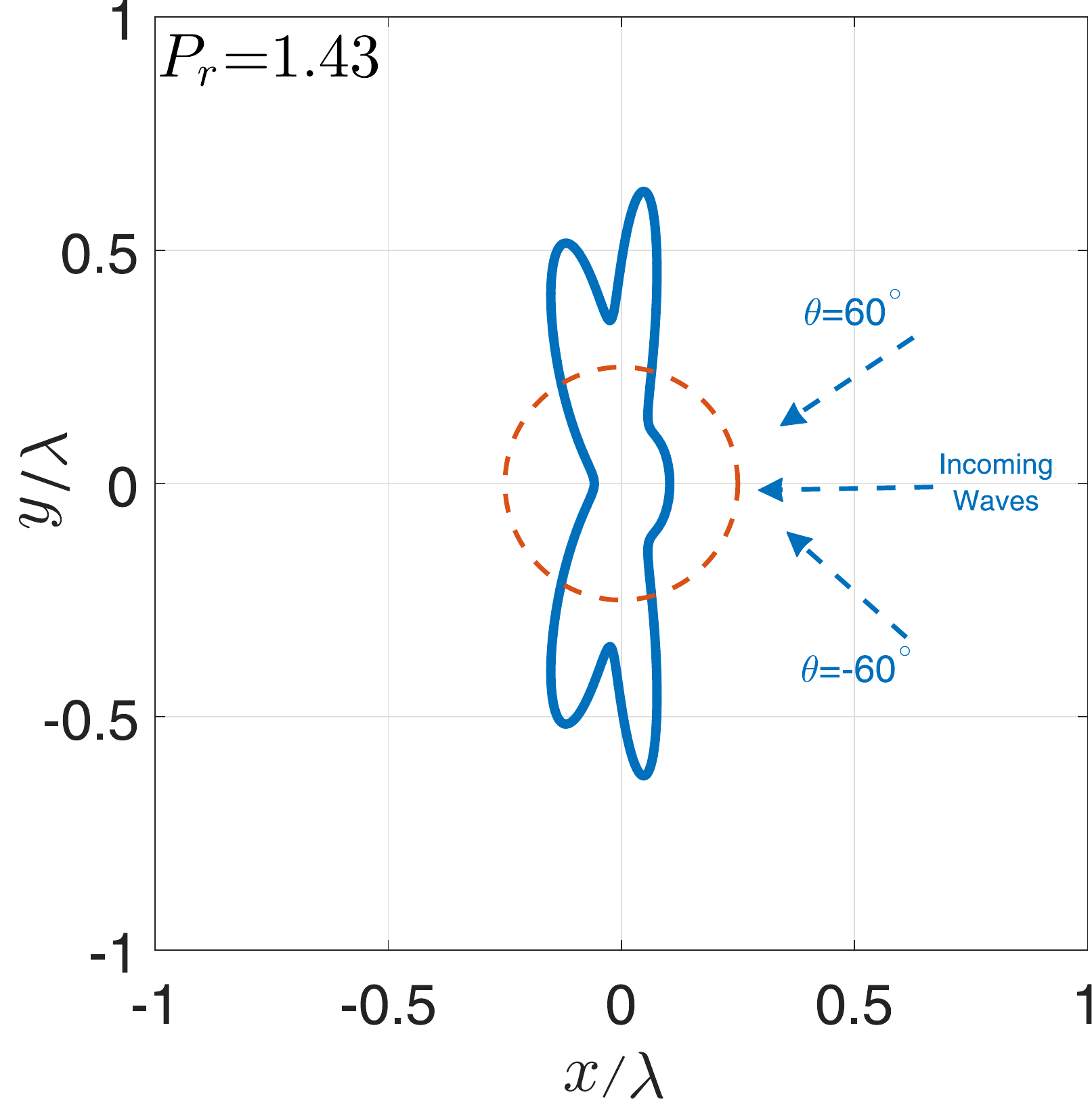}
                \caption{$|\theta|<60^\circ$}
                \label{figgull22}
        \end{subfigure}%
        \begin{subfigure}[b]{0.25\textwidth}
                \includegraphics[width=\linewidth]{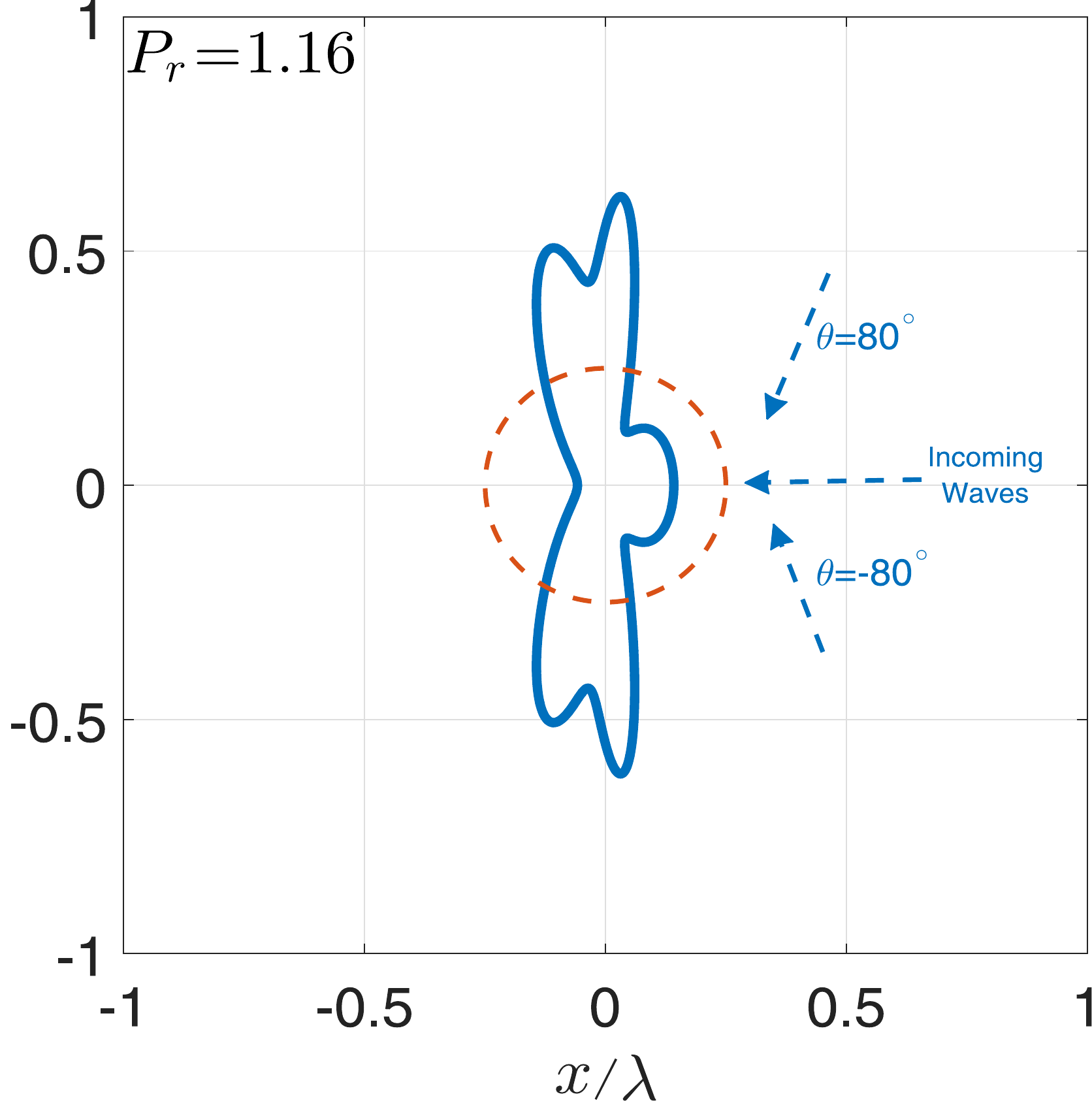}
                \caption{$|\theta|<80^\circ$}
                \label{figgull222}
        \end{subfigure}%
        \begin{subfigure}[b]{0.25\textwidth}
                \includegraphics[width=\linewidth]{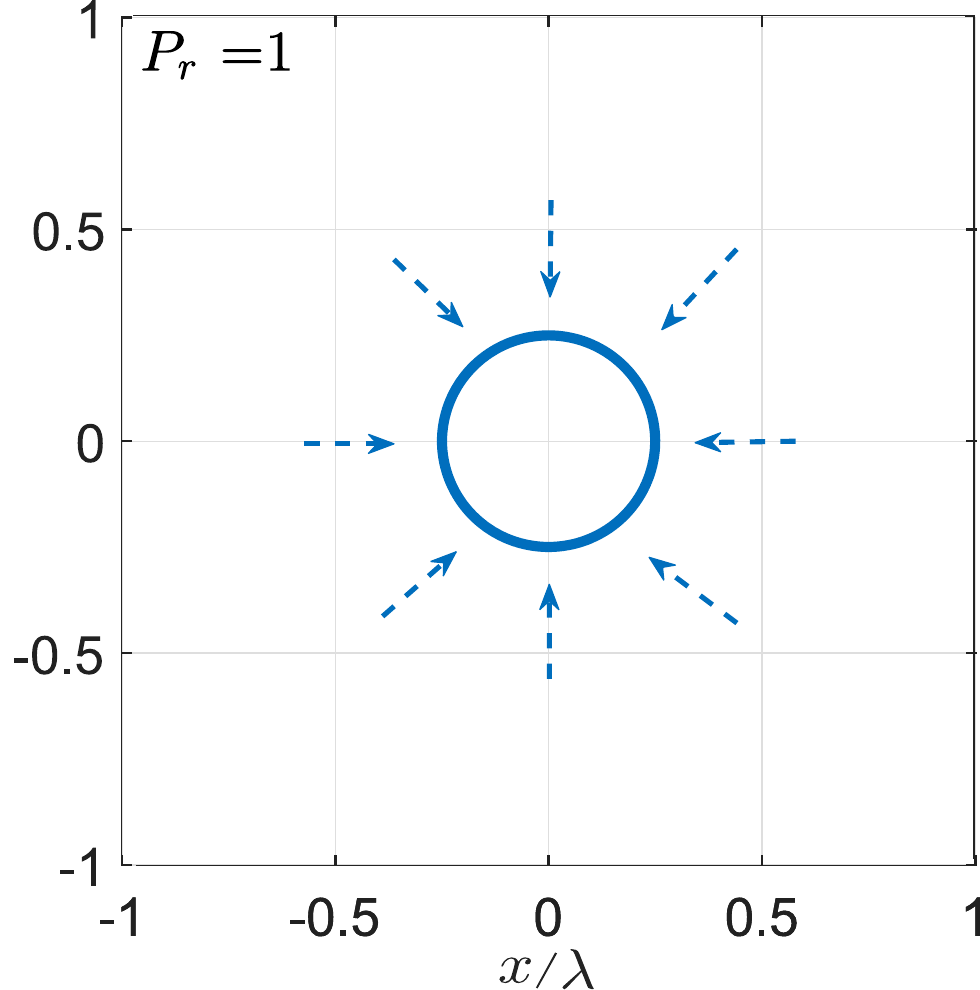}
                 \caption{$|\theta|<180^\circ$}
                \label{fig_circle}
        \end{subfigure}%
        \caption{Optimum absorber plate's shapes for monochromatic directional incident waves whose energy is uniformly distributed across the provided range of $\theta$ (cf. figure\ref{figspr}.a). As the spreading angle increases from $\theta=0^\circ$, the optimum shape leans symmetrically backward to face maximum area of itself normal to the incident wave energy. With a further increase in the directorial range, i.e. for $|\theta|<60^\circ,80^\circ$, a double-wing topology is obtained, as well as a bulge-shaped area in the middle that captures energy of high-angle incident waves. In the limit of $|\theta|<180^\circ$ (figure \ref{fig_circle}), the optimum shape is a circle. Parameters and line conversion is the same as in figure \ref{mono} and coefficients of shape generation Fourier modes are given in Appendix II.}
        \label{monodir}
\end{figure}

Here we consider the case of proposed absorber working in a monochromatic directional sea state, i.e. when all waves have the same frequency, but come from different directions. As discussed before (see also Appendix I), in this case the relative phases of waves arriving from different direction play an important role. Let's first consider the case in which all waves are in phase with respect to the origin of the coordinate system. We show in figure \ref{monodir} the optimum absorber shape for $P_r$, as calculated from \eqref{Eq_objFunc_waveClimateII}, for zero directionality of incident waves (figure \ref{monodir}a, which is the same as the case in figure \ref{mono}c), and cases of a directional incident wave with energy evenly distributed across  $|\theta|<20^\circ$ (figure \ref{monodir}b),  $|\theta|<40^\circ$ (figure \ref{monodir}c),   $|\theta|<60^\circ$ (figure \ref{monodir}d), and $|\theta|<80^\circ$ (figure \ref{monodir}e). Based on convergence behavior for the case of unidirectional waves (cf. figure \ref{mono}), in the following simulations we choose $N_c=$ 5.

In the absence of directionality (figure \ref{monodir}a) as discussed before, the optimum absorber shape has a narrow geometry. As the spreading angle departs from zero, say for $|\theta|<20^\circ$ in figure \ref{monodir}b, the optimum shape develops two tilted \textit{wings} in order to capture energy from all major incident directions. In other words, optimum shape takes a symmetric inclined form in order to increase the length of the plate perpendicular to the incoming waves. This trend is further highlighted in the case of $|\theta|<40^\circ$ (figure \ref{monodir}c). 

For a higher directionality angle of $|\theta|<60^\circ$ (figure \ref{monodir}d), the absorber develops two wings on each side (total of four wings). It appears that development of these four wings is to capture energy from even more directions: each wing captures the incoming energy normal to its axis, hence adds to the absorber's capturing orientational diversity. As the directionality angle further increases (figure \ref{monodir}e), the optimum absorber shape deviates further from the slender shape and tends more toward a circle. This is in fact expected since in the limit of $|\theta|<180^\circ$, waves coming symmetrically from all directions, the optimum shape becomes a circle, as is seen in figure \ref{monodir}f. The trend toward a circle is also seen from the development of a bulge near the center that grows with the increase in the spreading angle  (see figures \ref{monodir}d, e).

The normalized absorbed power, as expected, decreases as the spreading angle increases. This is also expected since for the case of unidirectional incident waves crests are aligned and optimum shape is simply the narrowest shape, whereas for directional spreading case this is not the case and therefore overall efficiency is negatively affected. The overall power absorption decreases from $P_r=$ 3.48, to respectively $P_r=$ 2.58, 2.04, 1.43 and 1.16 as spreading angle increases from 0 to 80 degrees (figures \ref{monodir}). It is worth nothing that for incident waves with directional symmetry investigated here, as expected all optimum shapes are also symmetric with respect to the mean wave direction (i.e. $\theta_m=0^\circ$). 
\begin{figure}[h!]
\centering
\captionsetup[subfigure]{justification=centering}
        \begin{subfigure}[b]{\dd}
                \includegraphics[width=\linewidth]{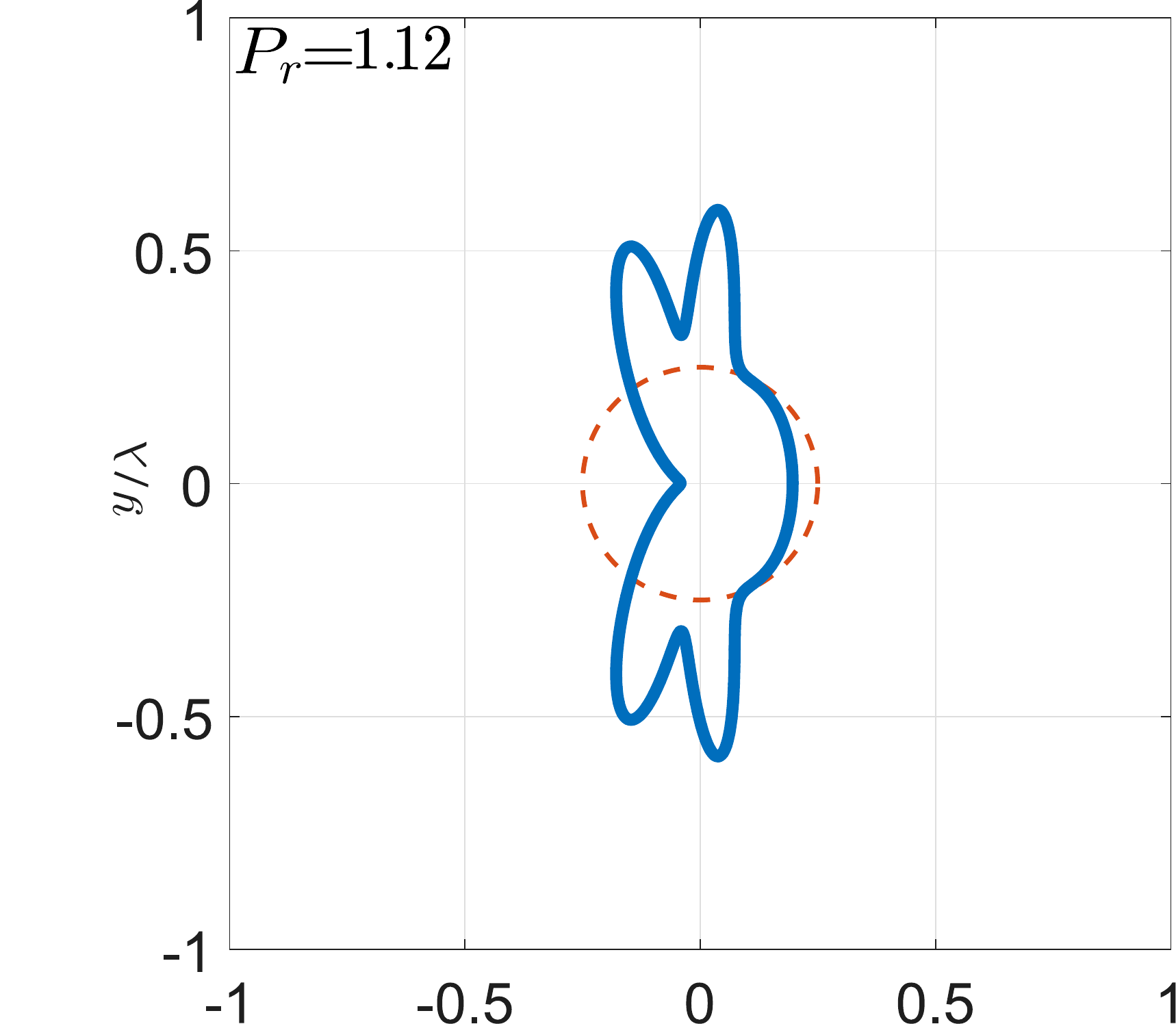}
                \caption{}
                \label{fig_sym_dir_all_a}
        \end{subfigure}%
        \begin{subfigure}[b]{\dd}
                \includegraphics[width=\linewidth]{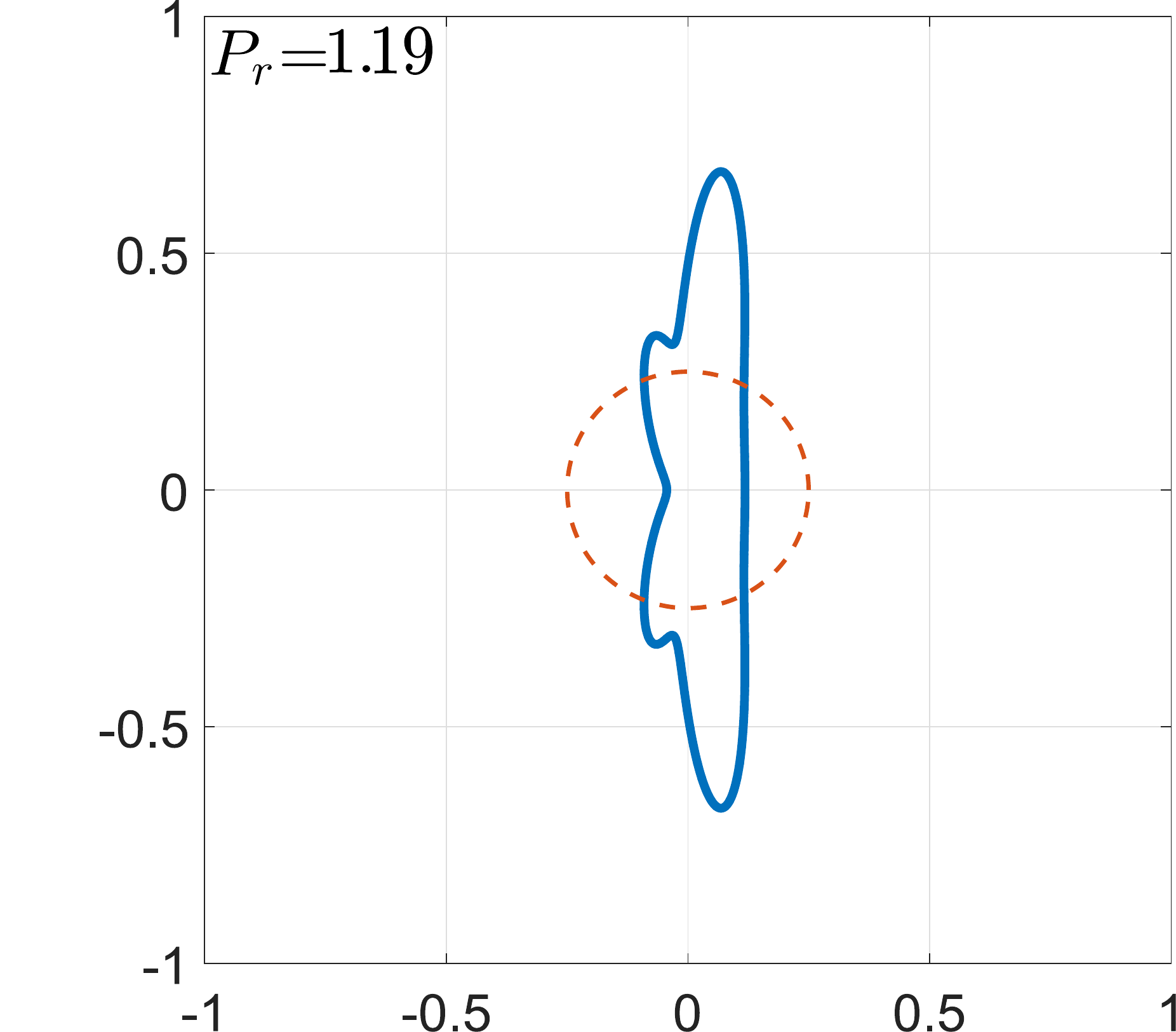}
                \caption{}
                \label{fig_sym_dir_all_b}
        \end{subfigure}%
        \begin{subfigure}[b]{\dd}
                \includegraphics[width=\linewidth]{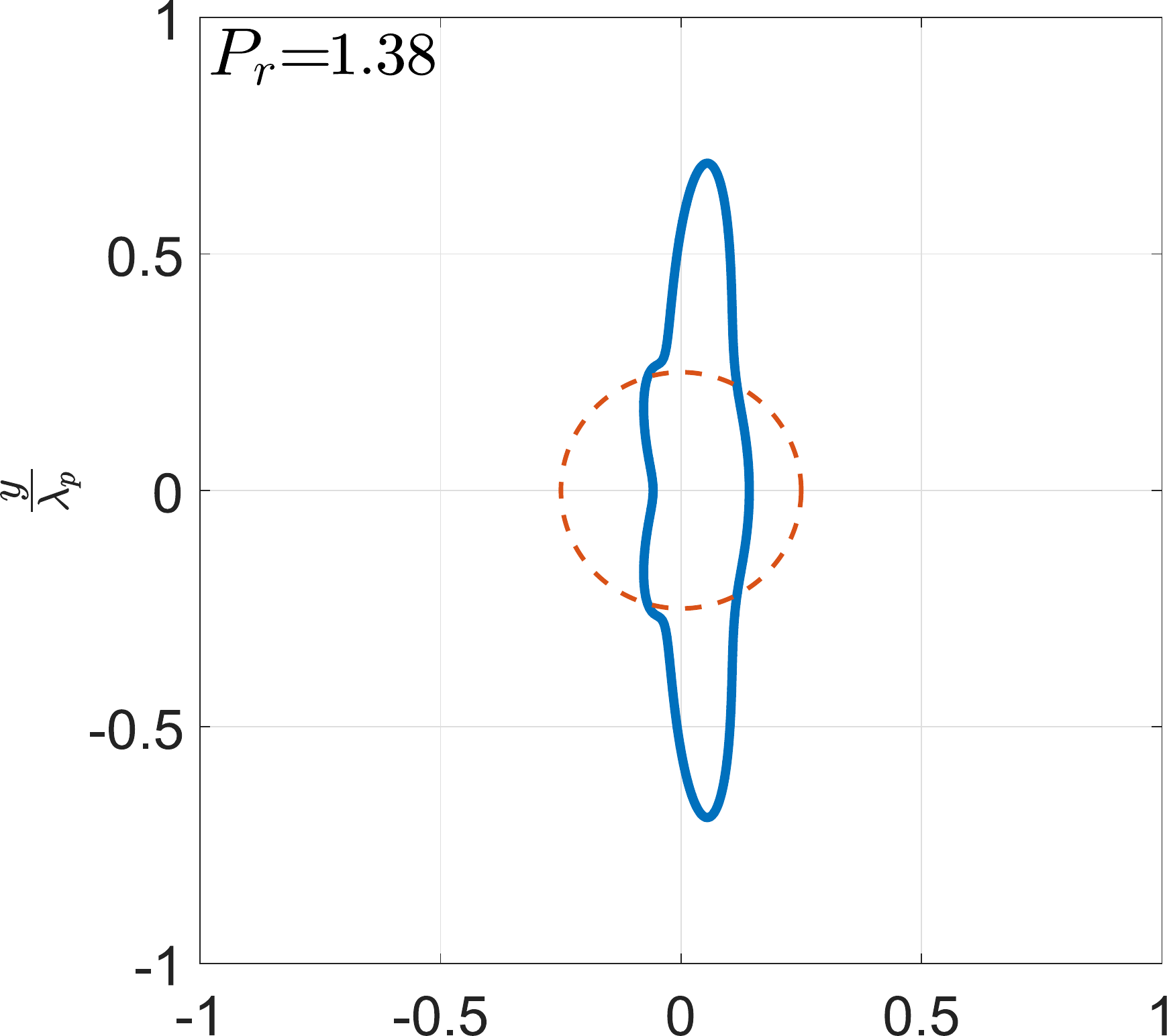}
                \caption{}
                \label{fig_sym_dir_all_c}
        \end{subfigure}%
        \begin{subfigure}[b]{\dd}
                \includegraphics[width=\linewidth]{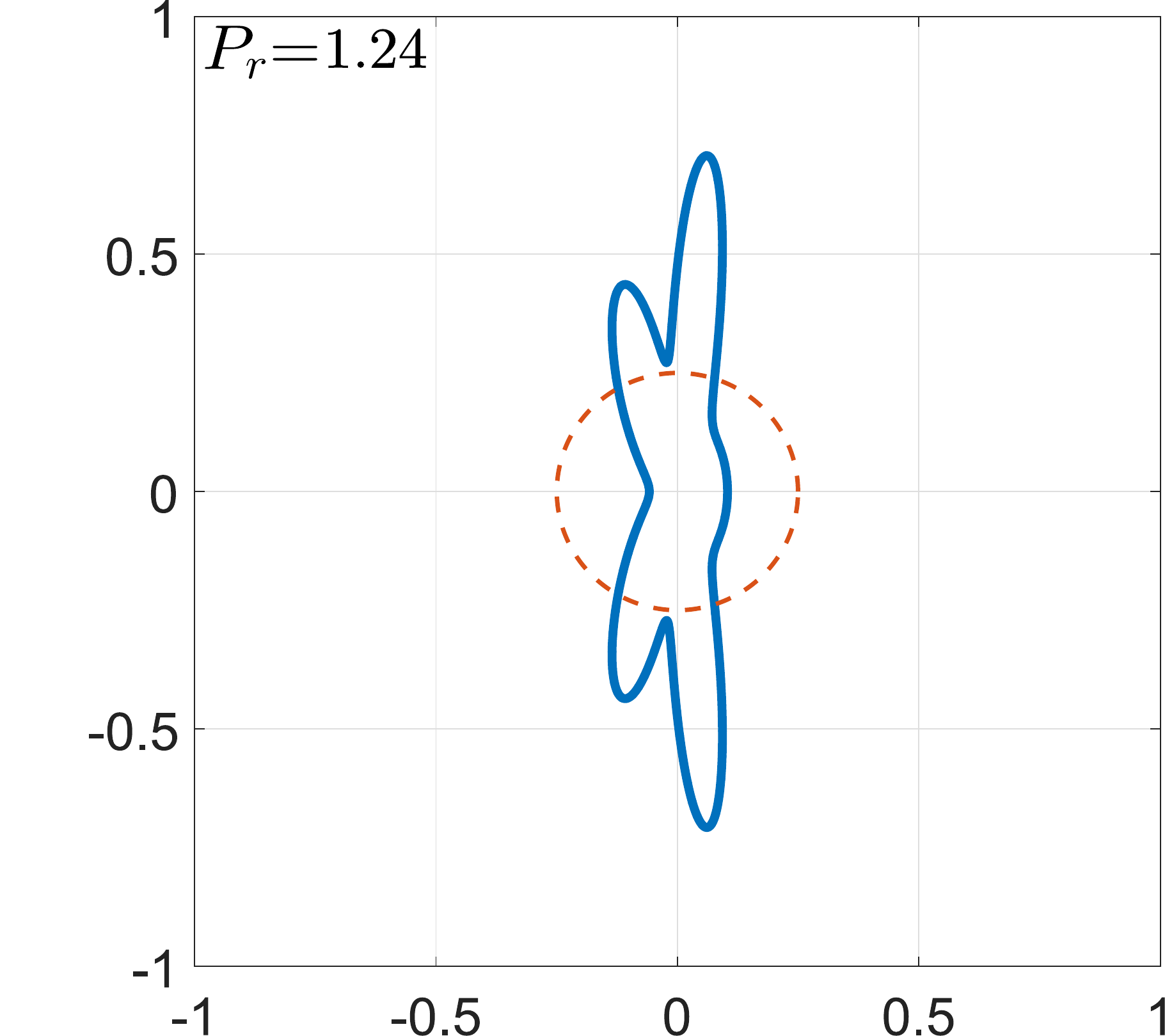}
                \caption{}
                \label{fig_sym_dir_all_d1}
        \end{subfigure}%
	\\
        \begin{subfigure}[b]{\dd}
                \includegraphics[width=\linewidth]{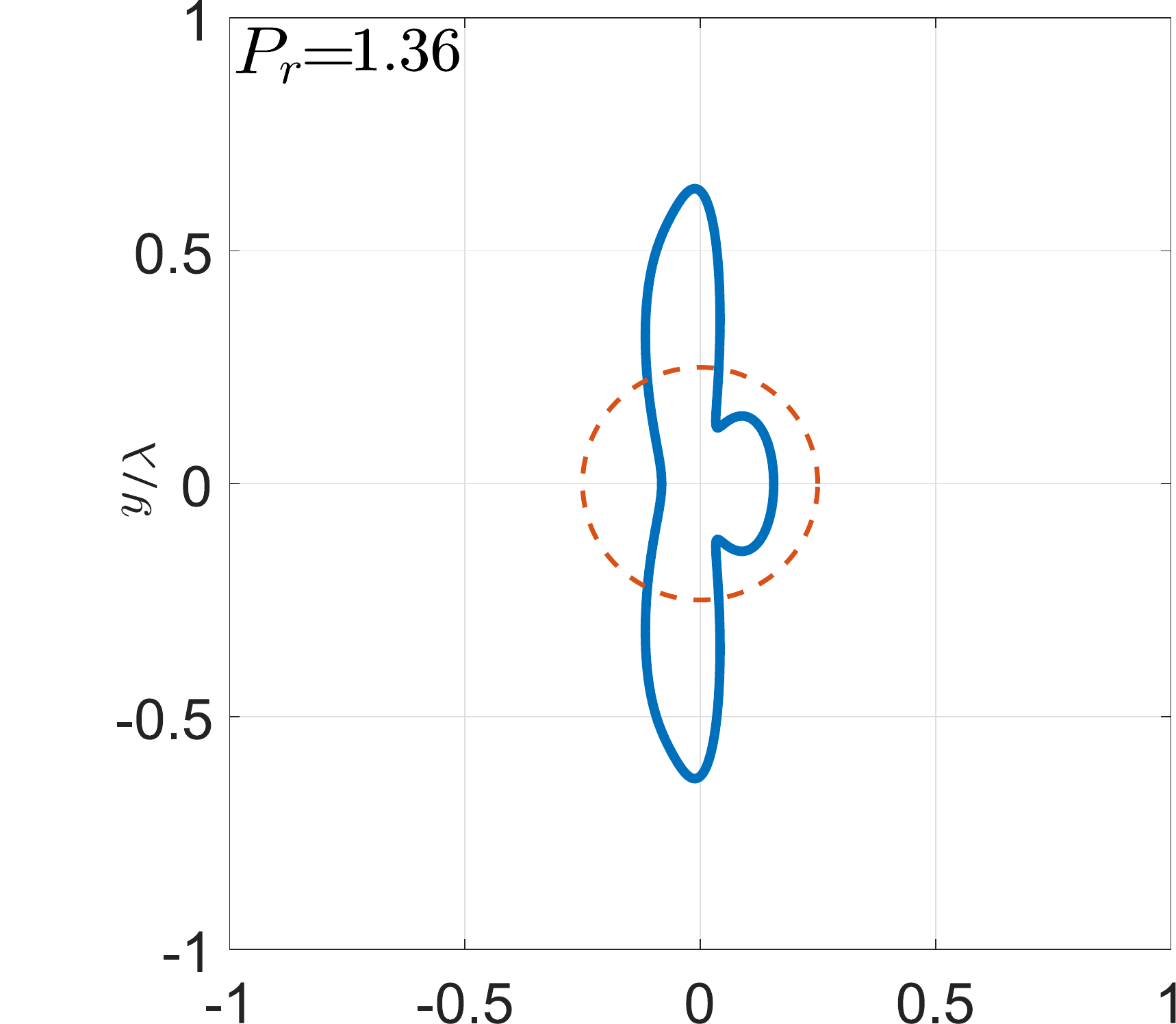}
                \caption{}
                \label{fig_sym_dir_all_d2}
        \end{subfigure}%
        \begin{subfigure}[b]{\dd}
                \includegraphics[width=\linewidth]{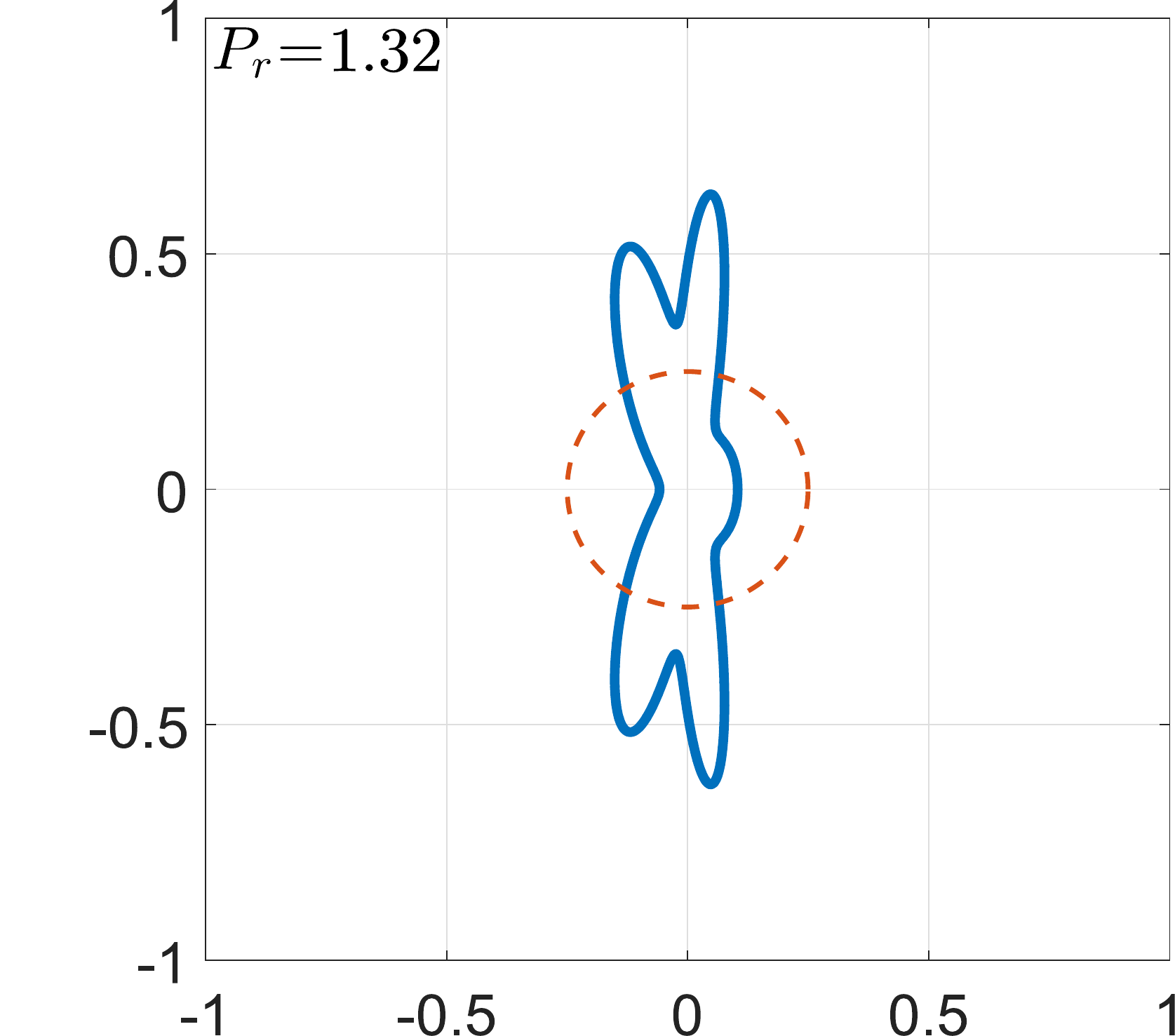}
                \caption{}
                \label{fig_sym_dir_all_d3}
        \end{subfigure}%
      \begin{subfigure}[b]{\dd}
                \includegraphics[width=\linewidth]{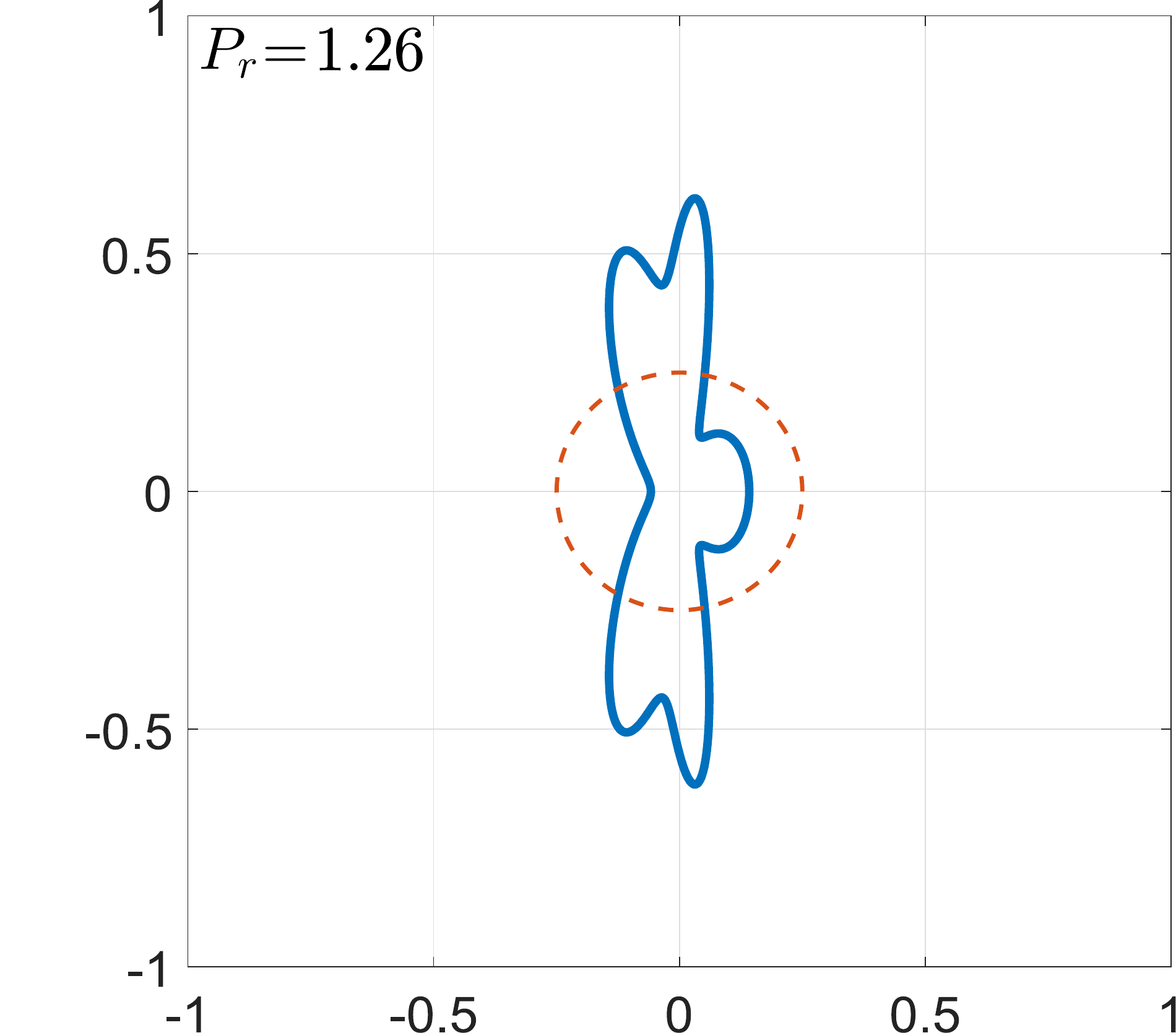}
                \caption{}
                \label{fig_sym_dir_all_d4}
        \end{subfigure}%
        \begin{subfigure}[b]{\dd}
                \includegraphics[width=\linewidth]{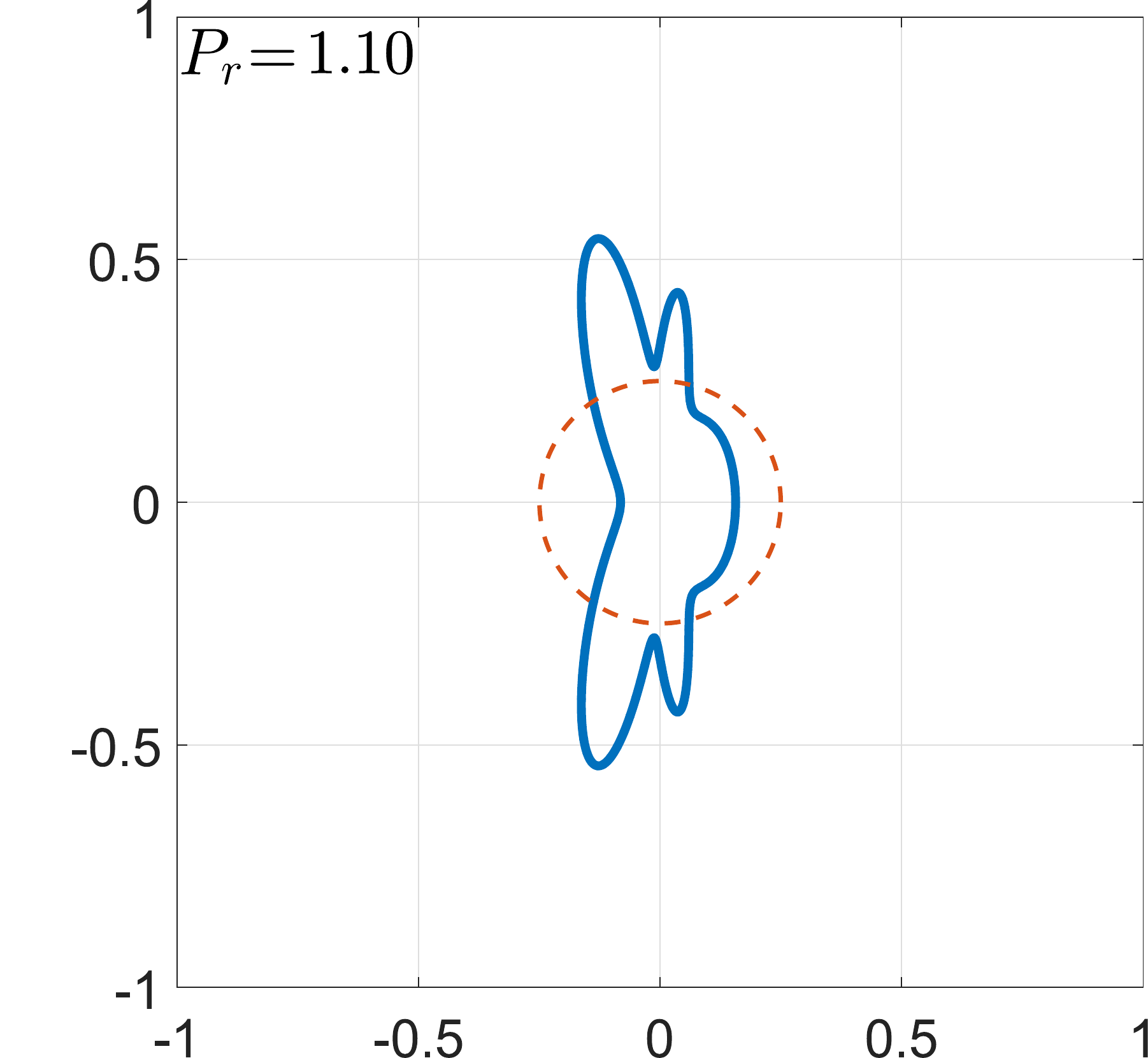}
                \caption{}
                \label{fig_sym_dir_all_d5}
        \end{subfigure}\\
        \begin{subfigure}[b]{\dd}
                \includegraphics[width=\linewidth]{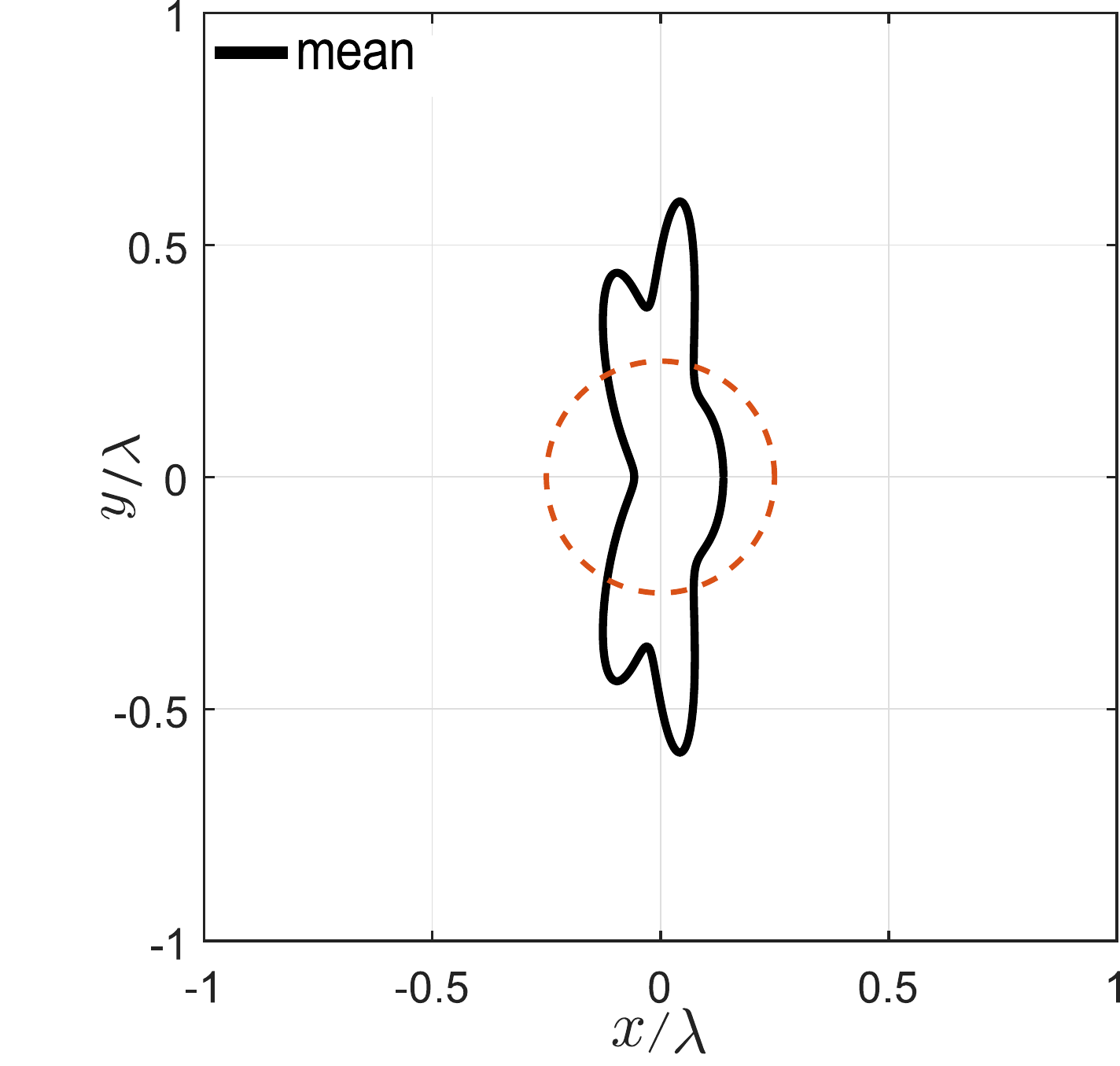}
                \caption{}
                \label{fig_sym_dir_all_a}
        \end{subfigure}%
        \begin{subfigure}[b]{\dd}
                \includegraphics[width=\linewidth]{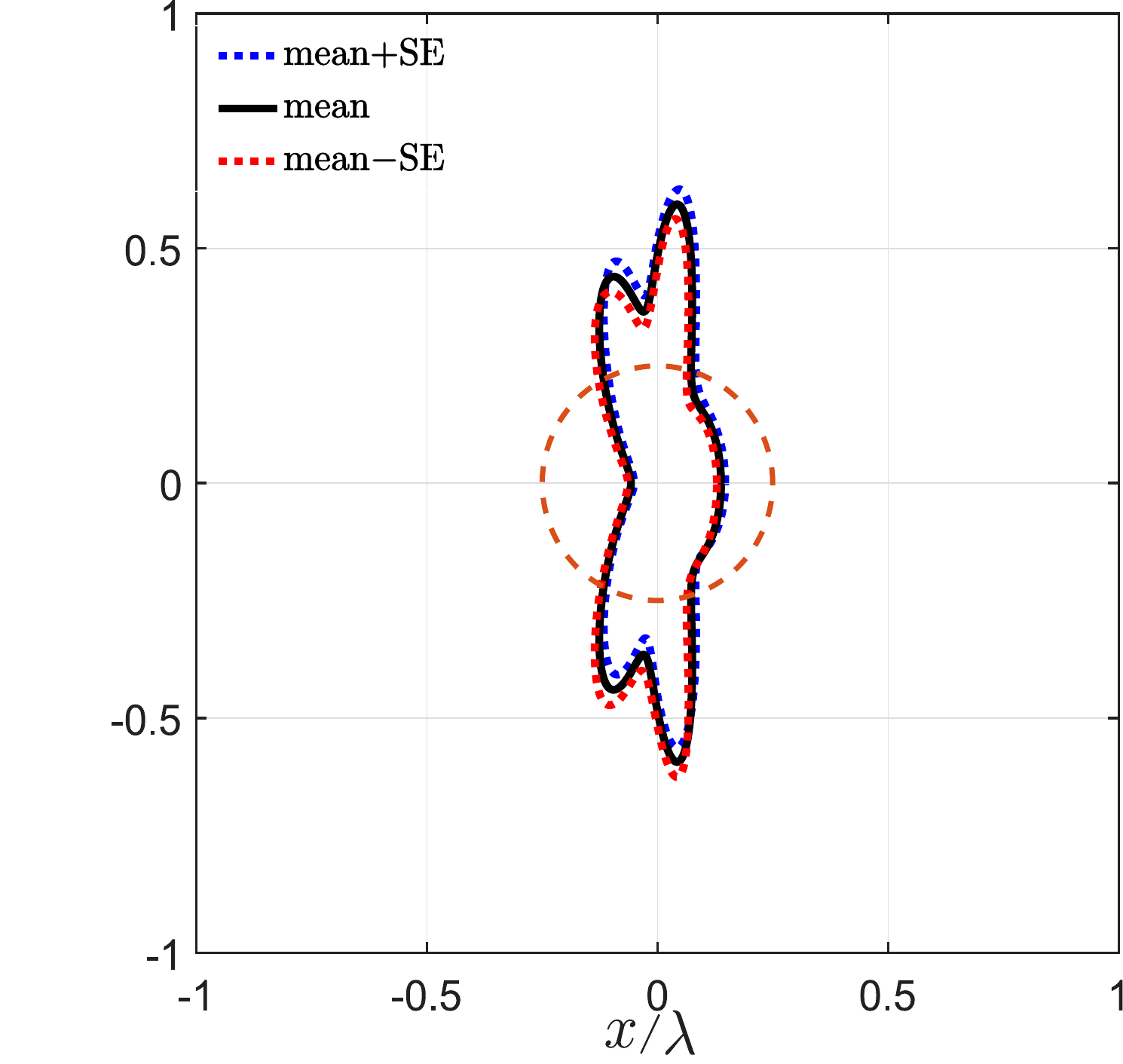}
                \caption{}
                \label{fig_sym_dir_all_bb}
        \end{subfigure}%
        \begin{subfigure}[b]{\dd}
\includegraphics[width=\linewidth]{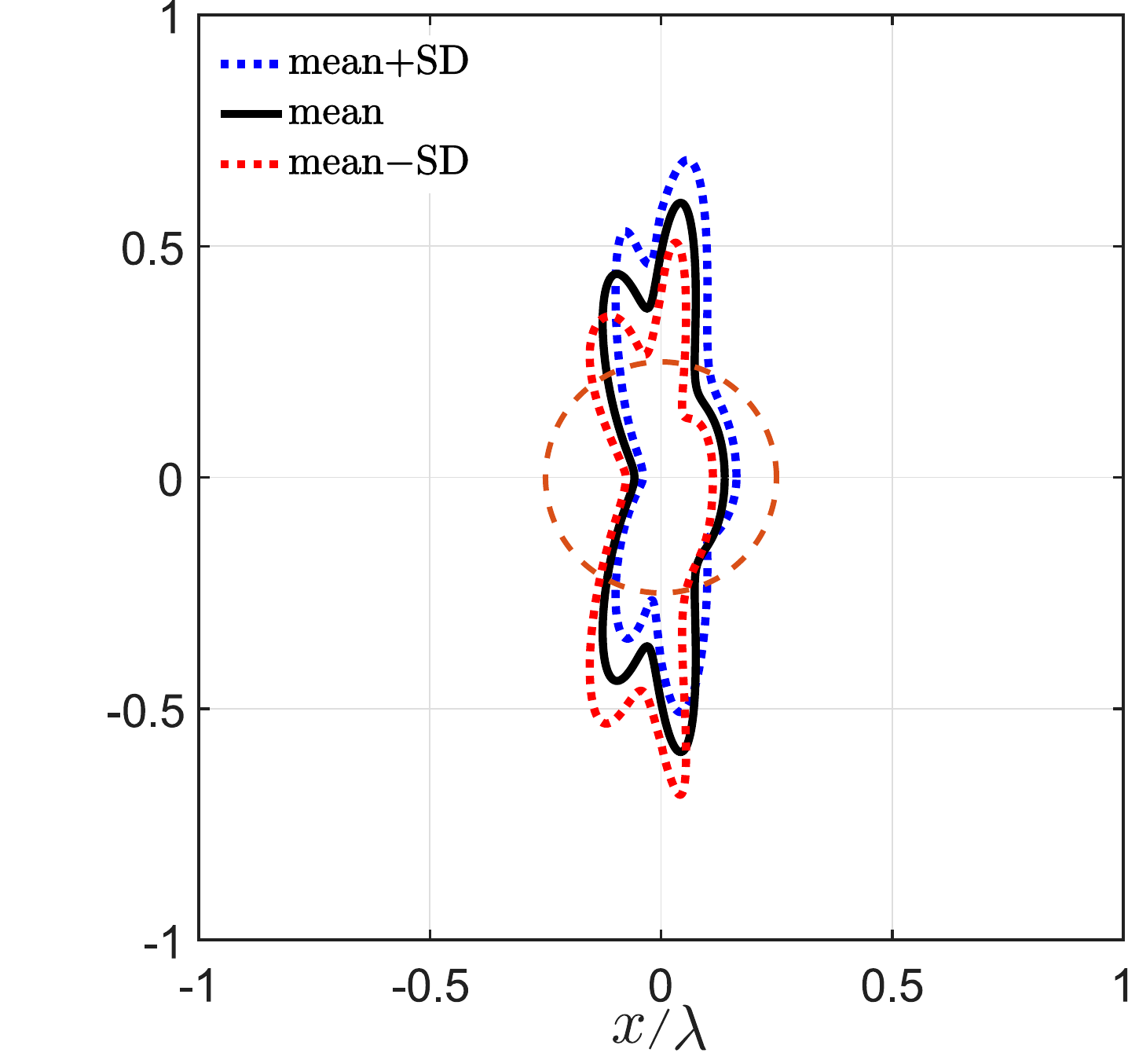}
                                \caption{}
                \label{fig_sym_dir_all_cc}
        \end{subfigure}%
        \begin{subfigure}[b]{\dd}
                \includegraphics[width=\linewidth]{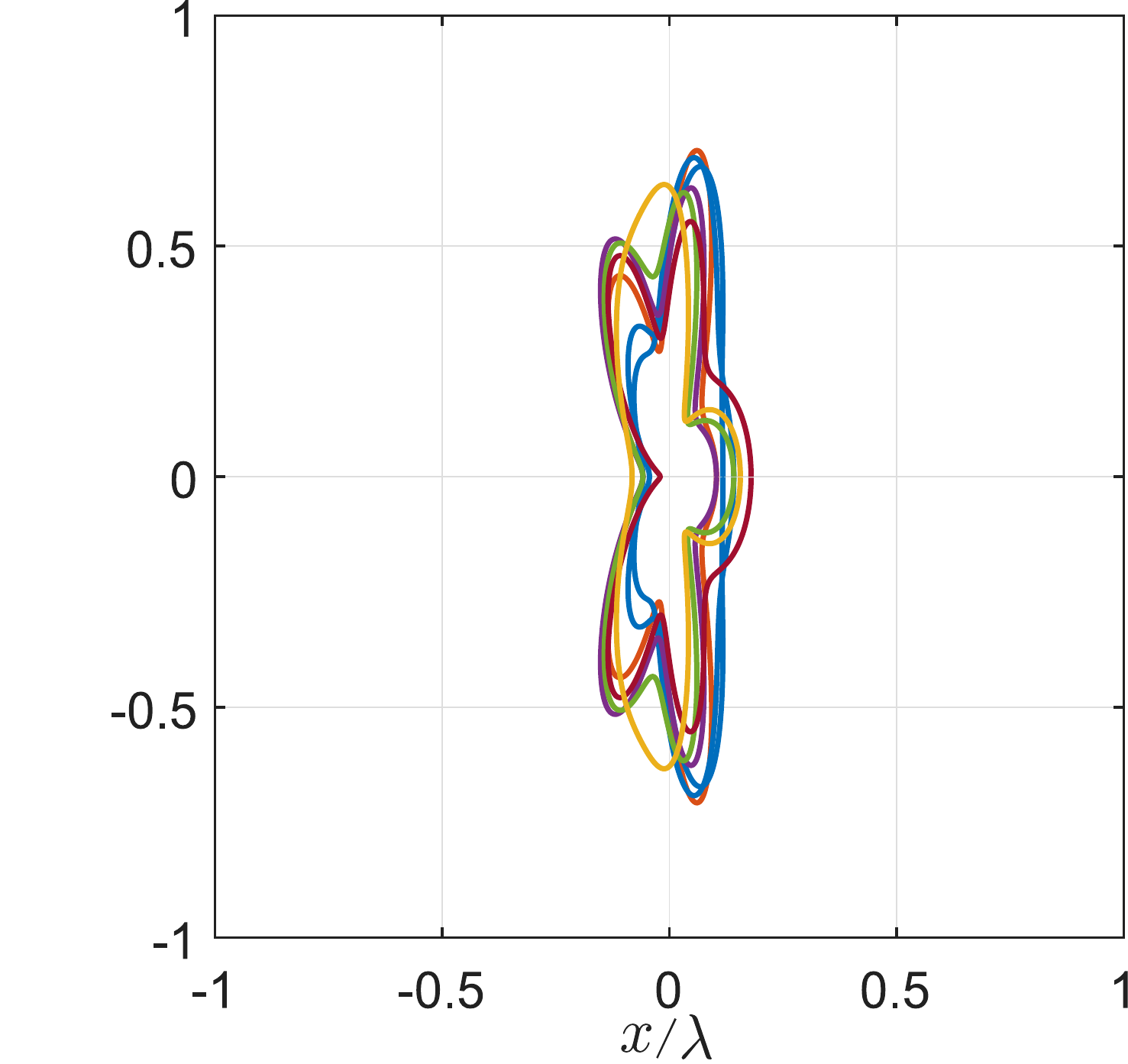}
                \caption{}
                \label{fig_sym_dir_all_d6}
        \end{subfigure}%
        \caption{Shape optimization of the absorber plate under monochromatic directional incident waves ($|\theta|<60^\circ$) with random phases (c.f. figure \ref{monodir}d which is the same optimization but with zero phases). Figures a-h show the optimum shape and normalized power $P_r$ for eight different sets of random phases (uniformly distributed random numbers from $[0,2\pi]$). Clearly, for each set of random phases a different optimized shape is obtained. The average enhancement in the power extraction is $\sim 25\%$. The average of these shapes is shown in figure (i), along with standard error (figure j) and standard deviation (figure k). The average enhancement in the power extraction using the average shape is $\sim 14\%$. Finally, we show the superposition of all shapes and the average in figure (l). Parameters and variables are the same as in figure \ref{monodir} and coefficients of shape generation Fourier modes are given in Appendix II.}
\label{monodirphase}
\end{figure}

If the monochromatic directional spectrum arrive with random phases, then the optimum shape depends on those specific phases. That is, for every set of random phases, a different optimum shape is obtained. Let's consider the case of figure \ref{monodir}d, in which waves are arriving with directional spread of $|\theta|<60^\circ$. The optimum shape for eight set of random phases (with uniform distribution) is shown in figures \ref{monodirphase}a-h. The absorbed power in all cases are higher than a circular-shape absorber from 10\% to 38\%, with an average of 25\% increase in the efficiency. The optimum shape for an unknown set of random phases (which is the case in real ocean), therefore, will be the average of the obtained geometries (figure \ref{monodirphase}i). Calculation of standard error shows a relatively good convergence to the mean (figure \ref{monodirphase}j), though shapes for different random phases may deviate from this mean by an average standard deviation of less than five percent (see figure \ref{monodirphase}k). We finally show in figure \ref{monodirphase}l superposition of all eight geometries and the average shape. It is certainly of interest to see how the average geometry (figure \ref{monodirphase}i) performs under different random phases of figure \ref{monodirphase}a-h. Our simulations show that if the average profile is chosen, under the eight wave conditions discussed above, we obtain respectively $P_r=$ 1.09, 1.08, 1.17, 1.19, 1.12, 1.25, 1.19 and 1.02, which is on average 14\% increase in the efficiency compared to a circular-shape geometry. 

Similar results are obtained for other angular distributions. For instance, for uniformly distributed monochromatic waves across $|\theta|\le 40^\circ$ (cf. figure\ref{monodir}c) but with random phases, case-by-case optimization yields an average increase in power of 60\%, and the average profile obtains an average increase of 42\%. Note that the increase in the average absorbed power is expected as the angular distribution decreases due to a similar mechanism discussed before.

\subsection{Polychromatic (broadband) unidirectional incident wave}

If the incident wave is polychromatic and unidirectional, similar to the case of a monochromatic unidirectional wave, the optimum shape must be the most elongated shape. Our optimization obtains identical shapes to those reported in figure \ref{mono}, for each $N_c$. The absorbed power, however, as expected will be much less. This is due to the fact that our linear power take-off parameters can only be tuned to one frequency, and therefore the rest of the spectrum is \textit{detuned} to the power take-off (It should be noted that the optimum power take-off parameters are found through optimization as well to ensure maximum extracted power for each shape). For an incident JONSWAP spectrum of  sea state five (rough sea conditions) with $H_s=3.25$ meters and $T_p=9.7$ seconds and for $N_c=3-7$ we obtain, respectively, $P_r=$ 1.41, 1.52, 1.56, 1.58 and 1.61. The pattern is similar to the case of figure \ref{mono}, but it is converging to $P_r\sim$1.6, i.e. our optimum shape performs about 60\% better than a circular absorber. 

\subsection{Polychromatic (broadband) directional incident wave}

For the case of polychromatic directional incident waves, similar to monochromatic directional waves, phases of waves play an important role both in the optimum shape and the absorbed power. This is due to the fact that when phases are randomly distributed, it is possible that some waves at specific directions undergo a destructive interference with other waves of the same frequency coming from another direction, and therefore their effective contributions to power absorption decreases. Alternatively, some may undergo constructive interference whereby contribute positively to the power absorption.

\begin{figure}[h!]
\centering
        \begin{subfigure}[b]{\dd}
                \includegraphics[width=\linewidth]{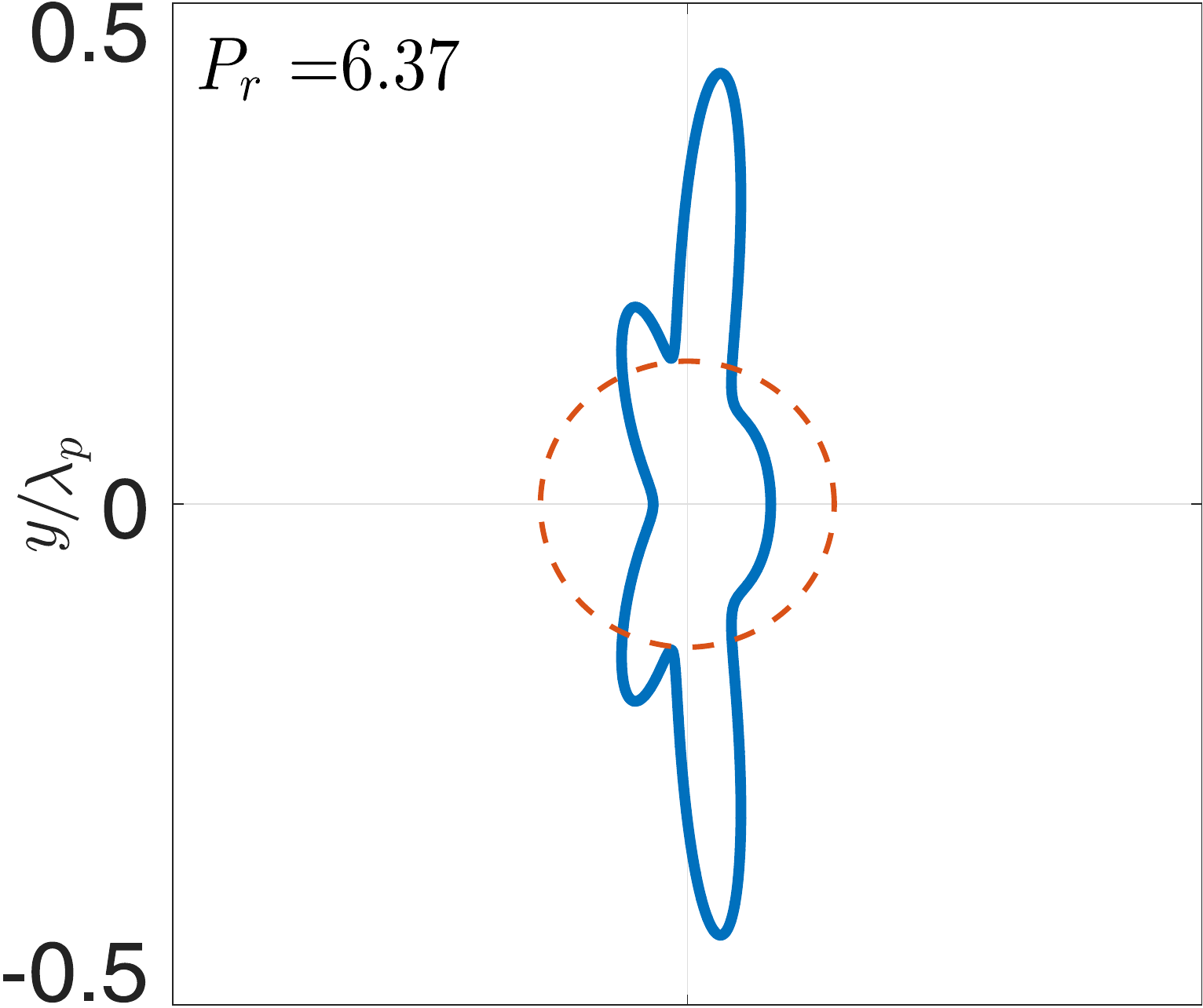}
                \caption{}
                \label{fig:gull}
        \end{subfigure}%
        \begin{subfigure}[b]{\dd}
                \includegraphics[width=\linewidth]{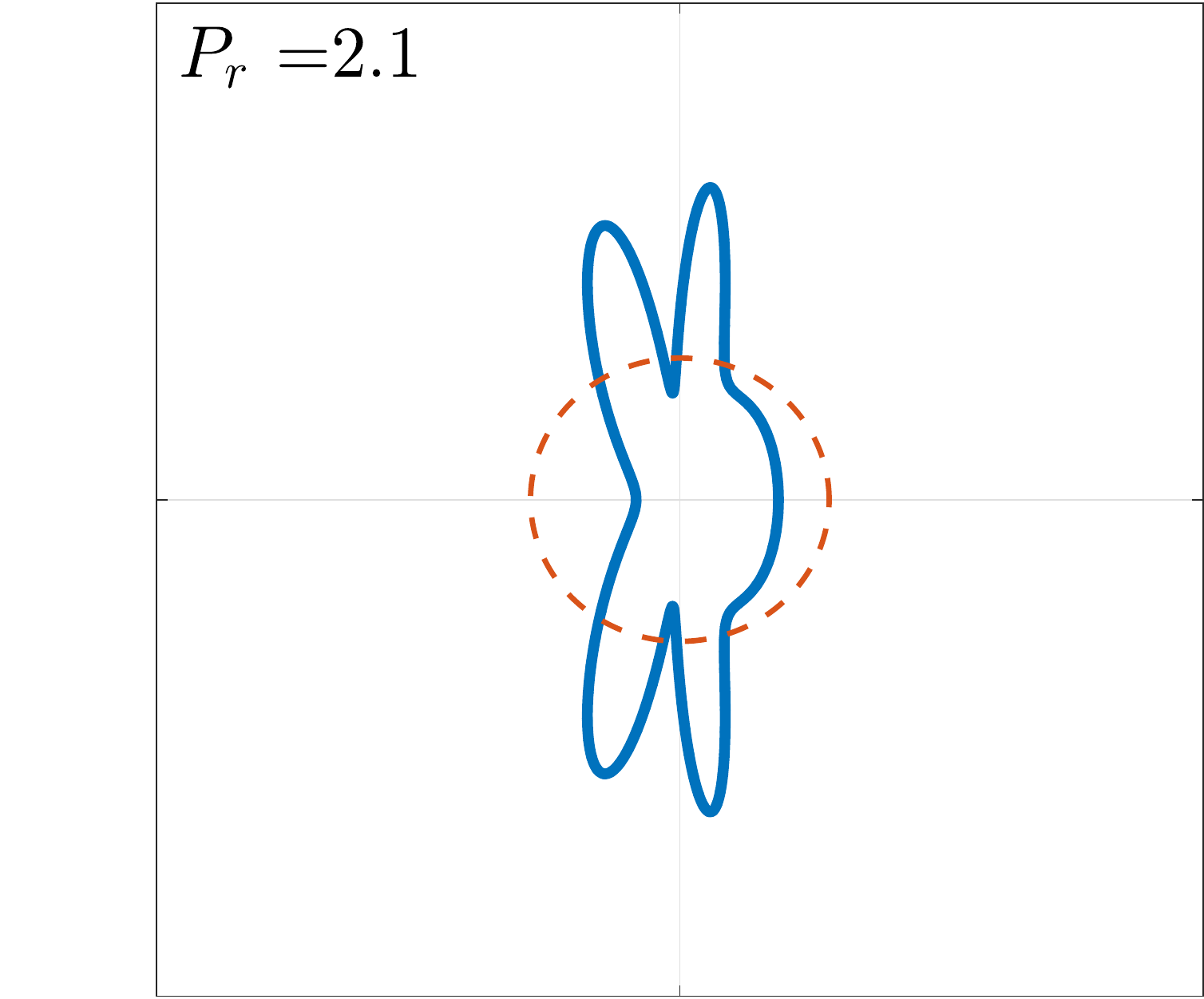}
                \caption{}
                \label{fig:gull2}
        \end{subfigure}%
        \begin{subfigure}[b]{\dd}
                \includegraphics[width=\linewidth]{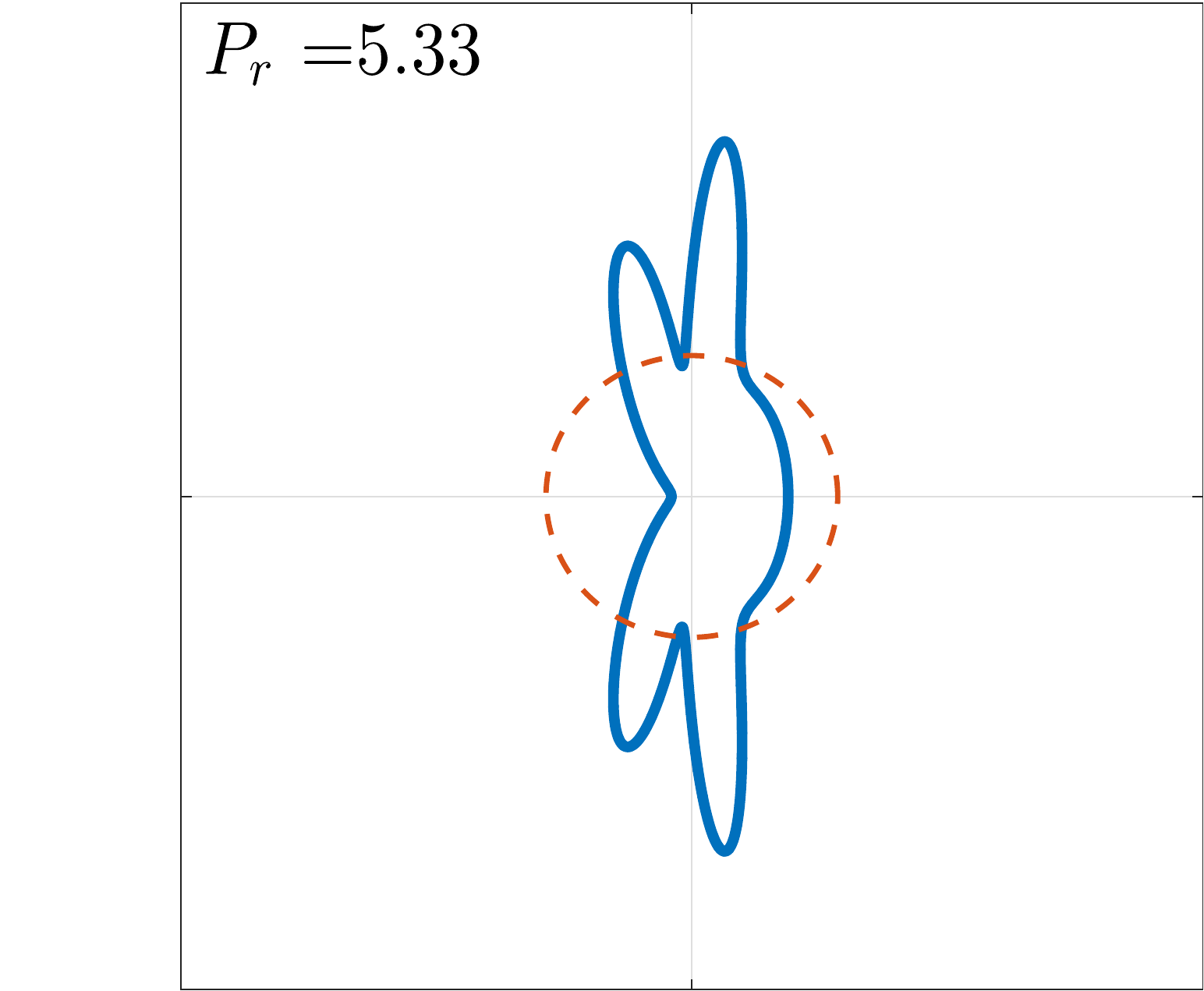}
                \caption{}
                \label{fig:tiger}
        \end{subfigure}%
        \begin{subfigure}[b]{\dd}
                \includegraphics[width=\linewidth]{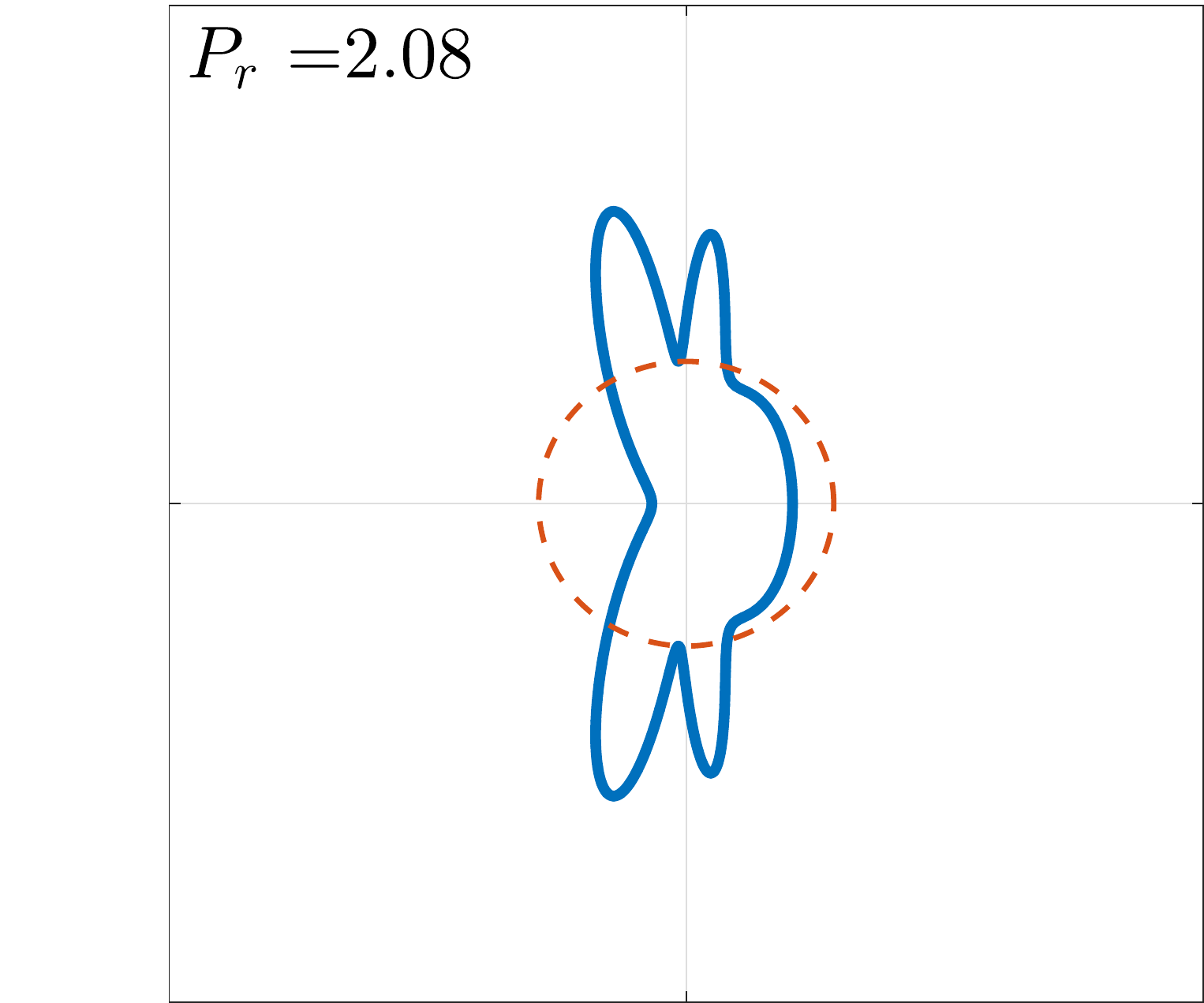}
                \caption{}
                \label{fig:gull}
        \end{subfigure}%
        \\
        \begin{subfigure}[b]{\dd}
                \includegraphics[width=\linewidth]{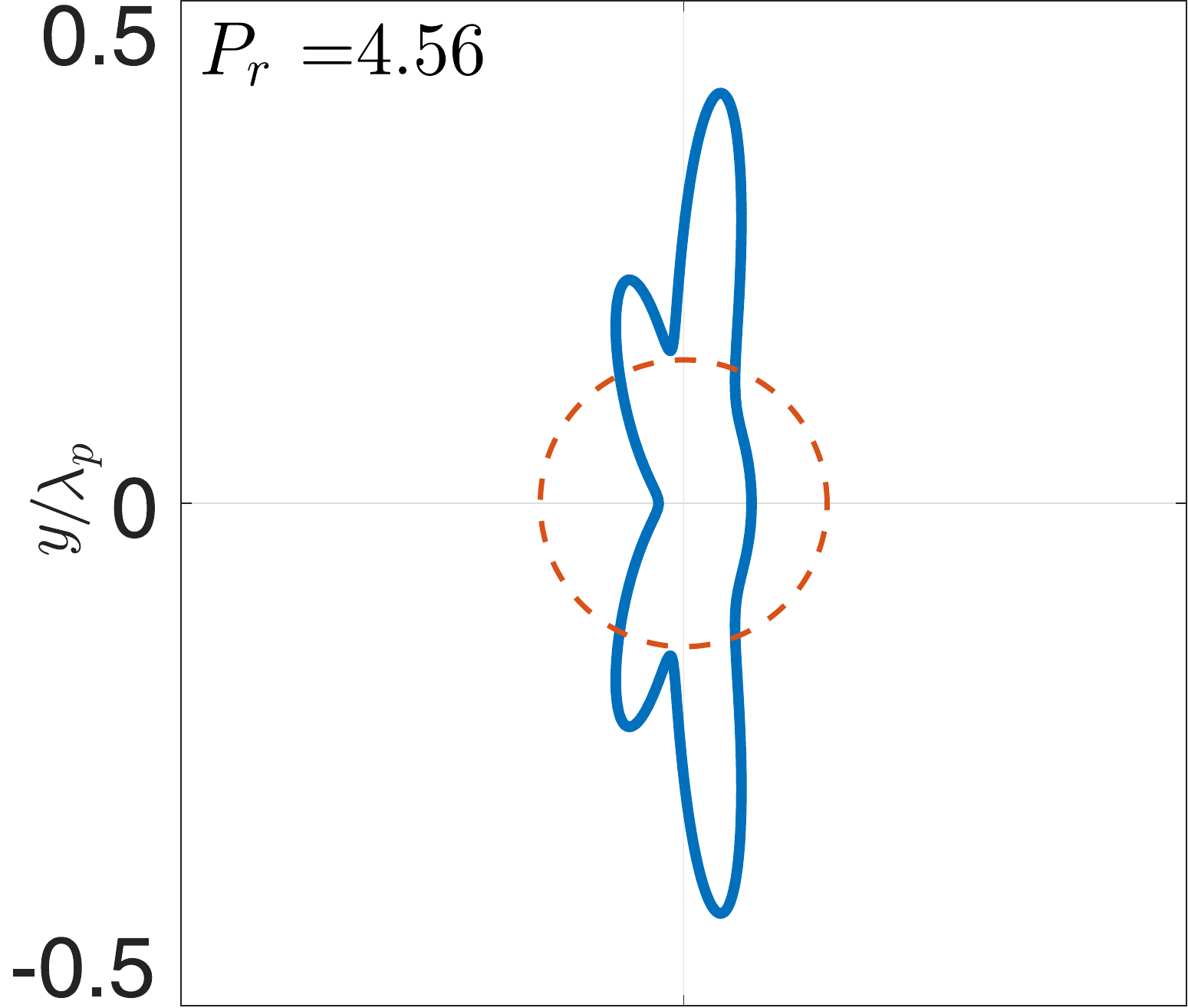}
                \caption{}
                \label{fig:gull222}
        \end{subfigure}%
                \begin{subfigure}[b]{\dd}
                \includegraphics[width=\linewidth]{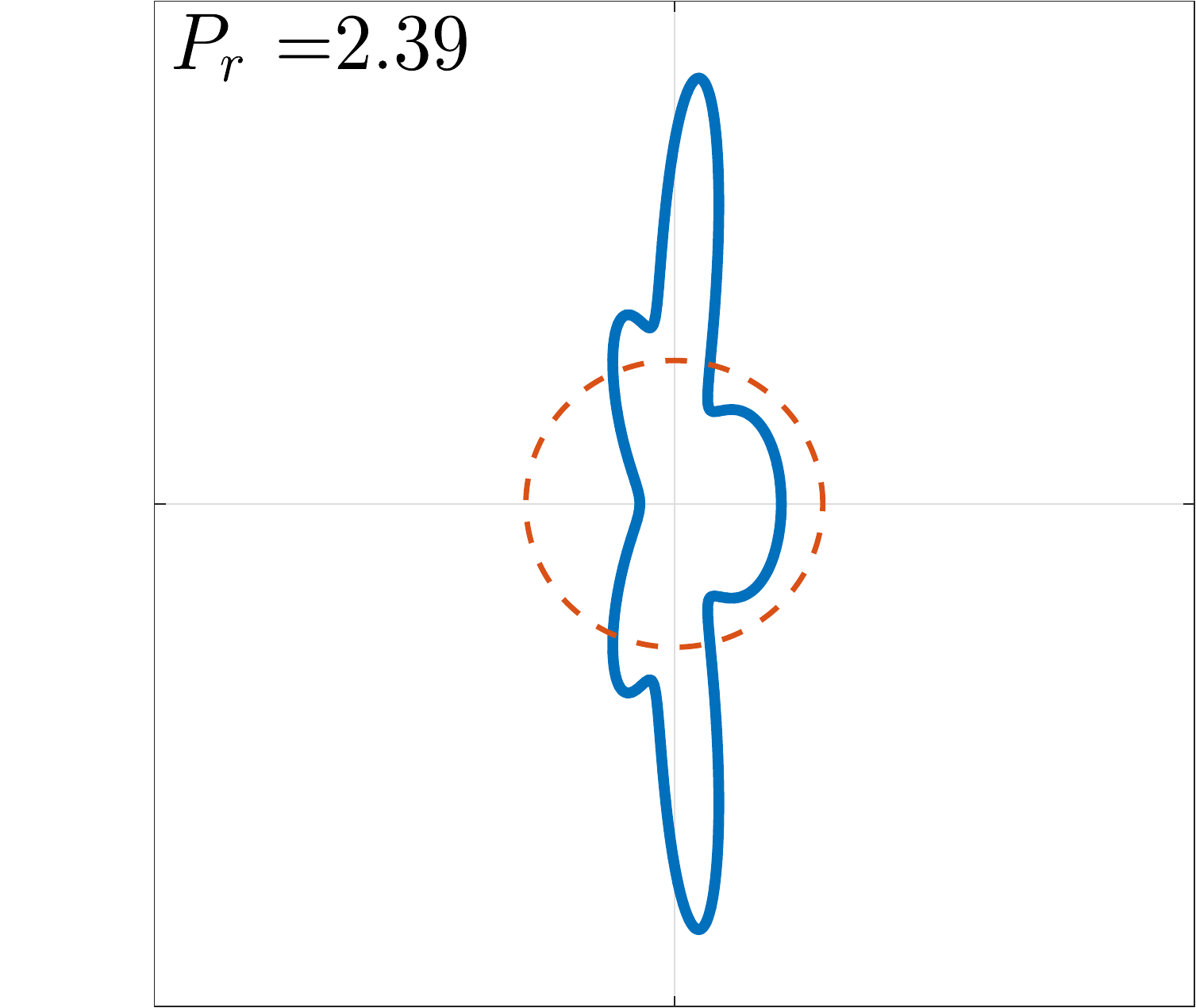}
                \caption{}
                \label{fig:gull22}
        \end{subfigure}%
                        \begin{subfigure}[b]{\dd}
                \includegraphics[width=\linewidth]{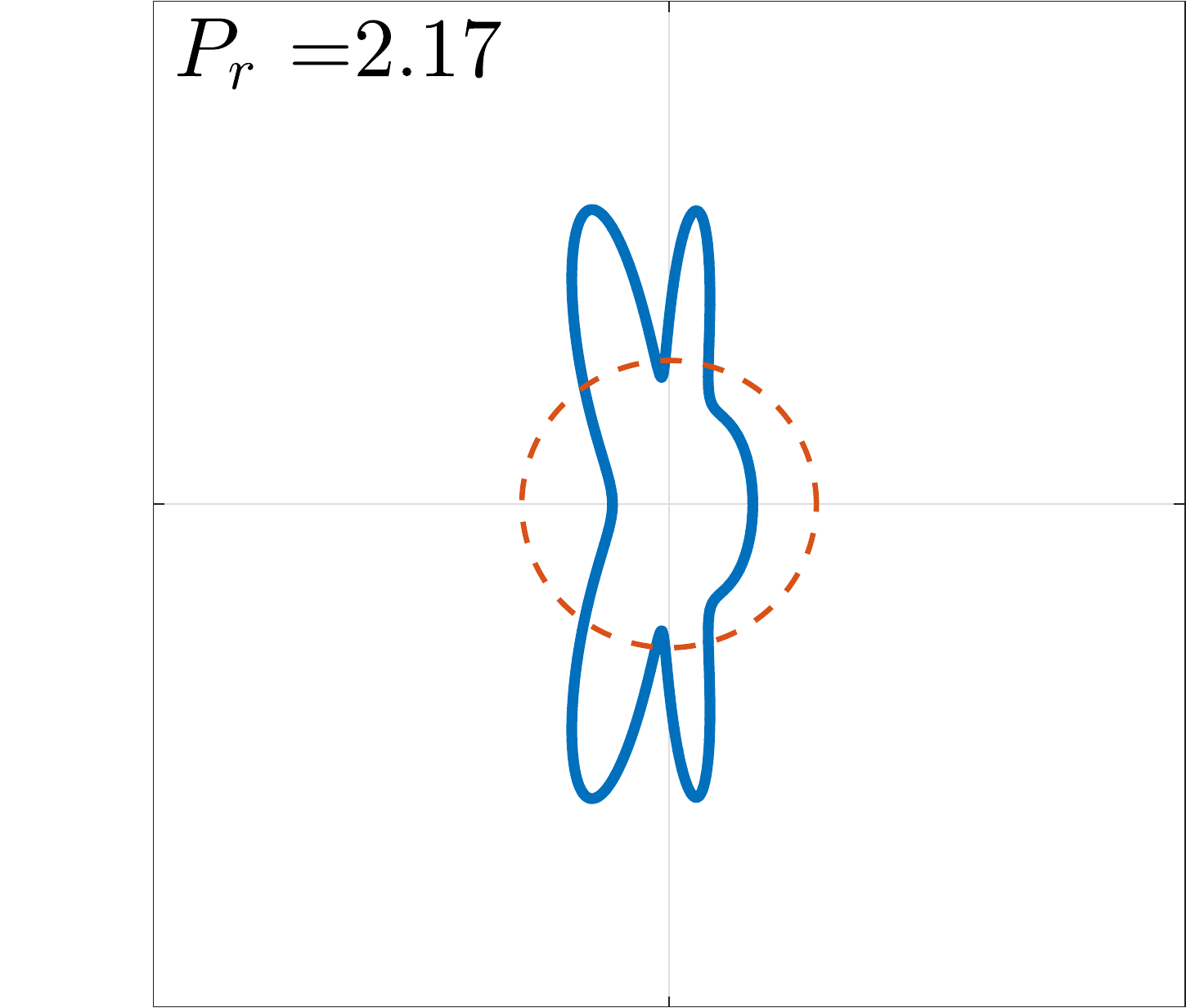}
                \caption{}
                \label{fig:gull22}
        \end{subfigure}%
                        \begin{subfigure}[b]{\dd}
                \includegraphics[width=\linewidth]{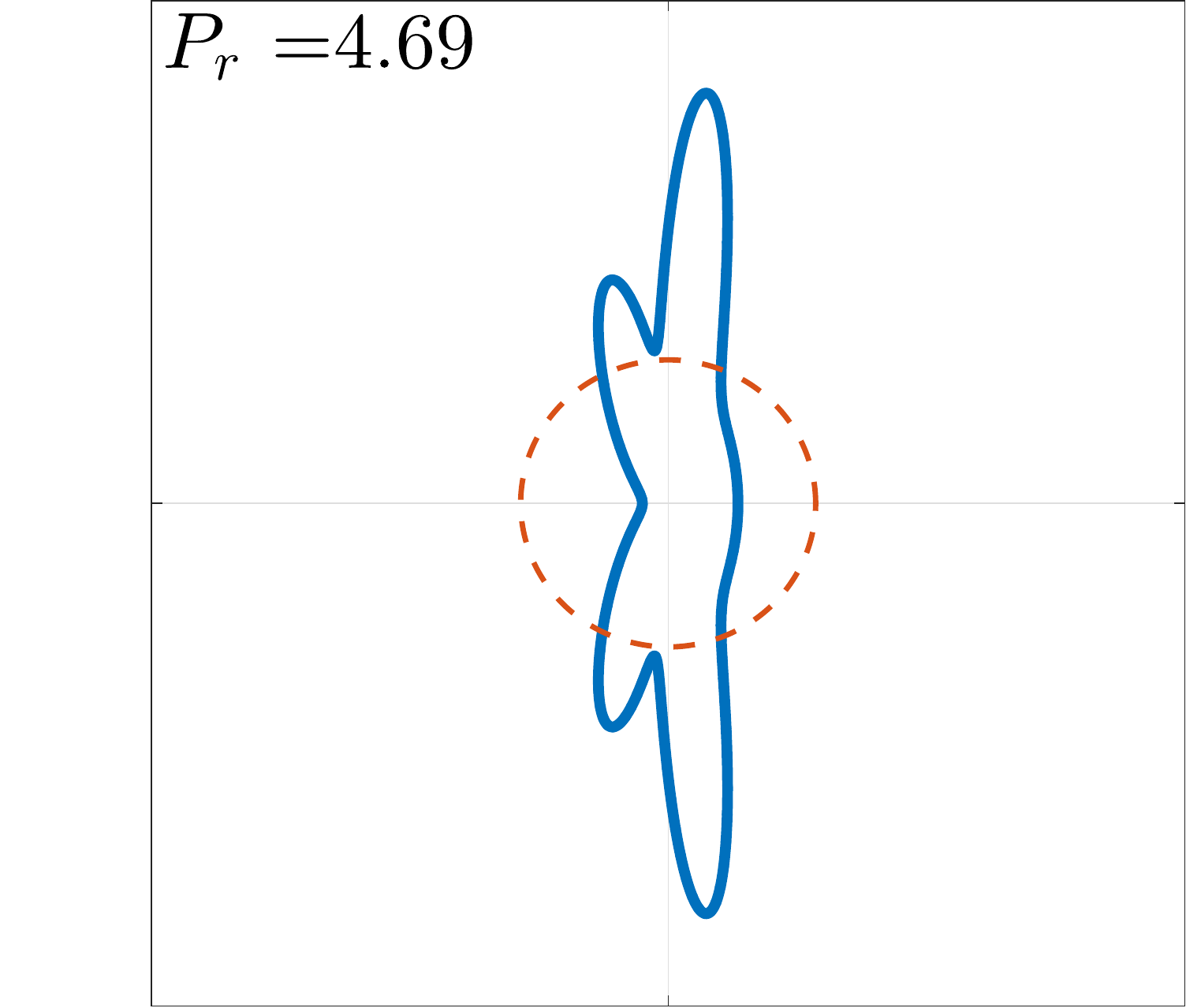}
                \caption{}
                \label{fig:gull22}
        \end{subfigure}%
				\\
				
        \begin{subfigure}[b]{\dd}
                \includegraphics[width=\linewidth]{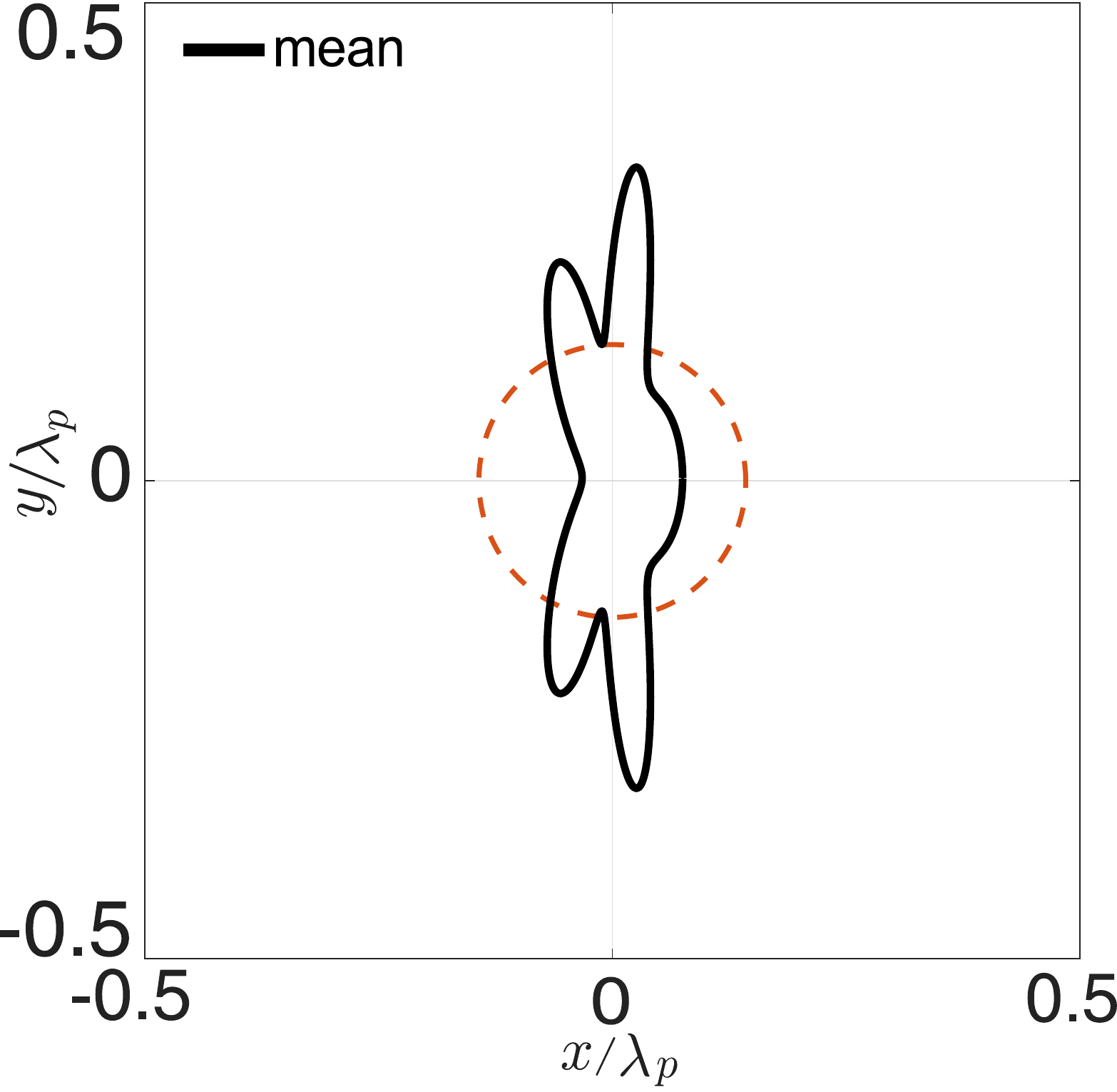}
                \caption{}
                \label{fig_sym_dir_all_bbb}
        \end{subfigure}%
        \begin{subfigure}[b]{\dd}
                \includegraphics[width=\linewidth]{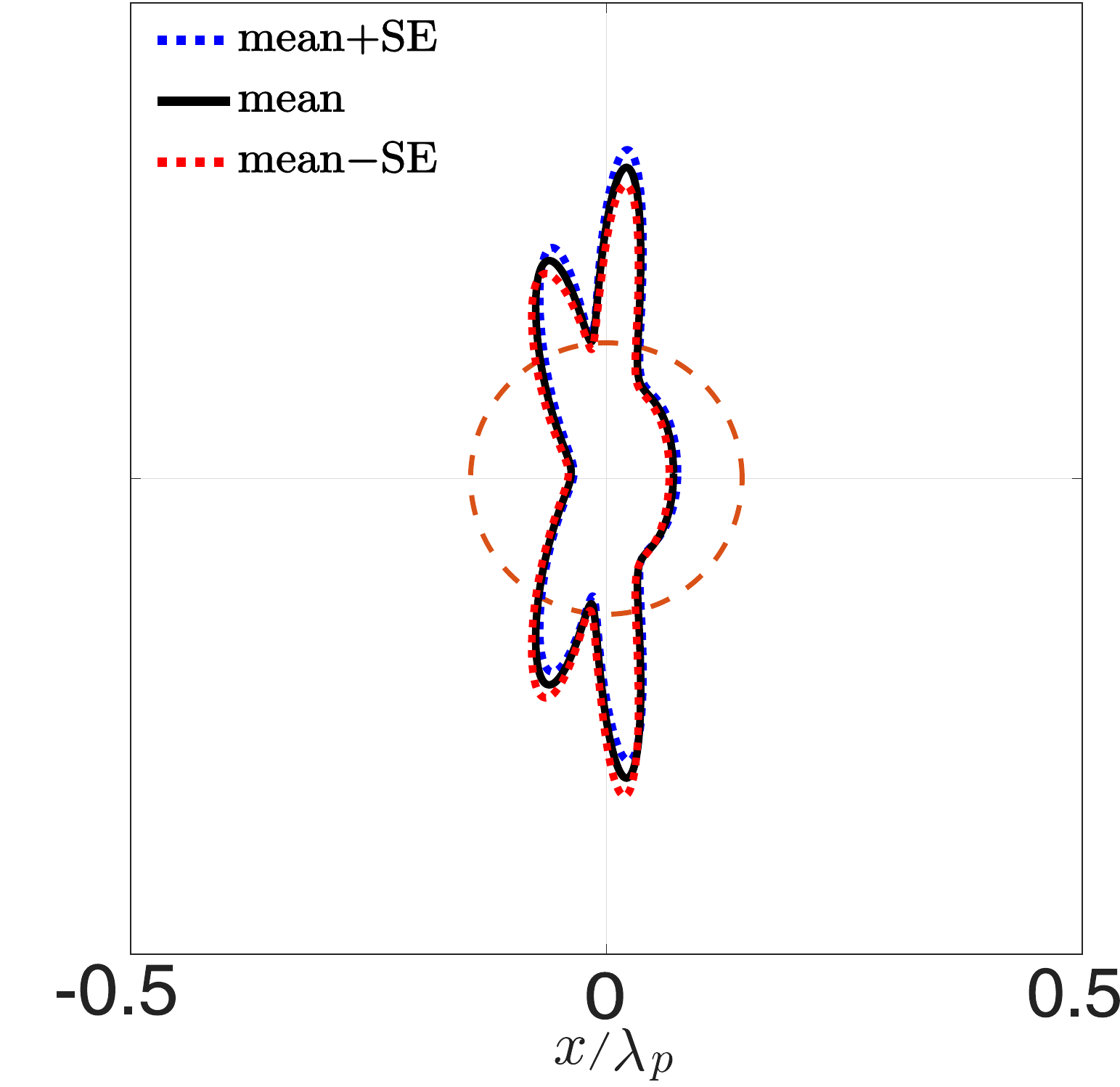}
                \caption{}
                \label{fig_sym_dir_all_d7}
        \end{subfigure}%
        \begin{subfigure}[b]{\dd}
                \includegraphics[width=\linewidth]{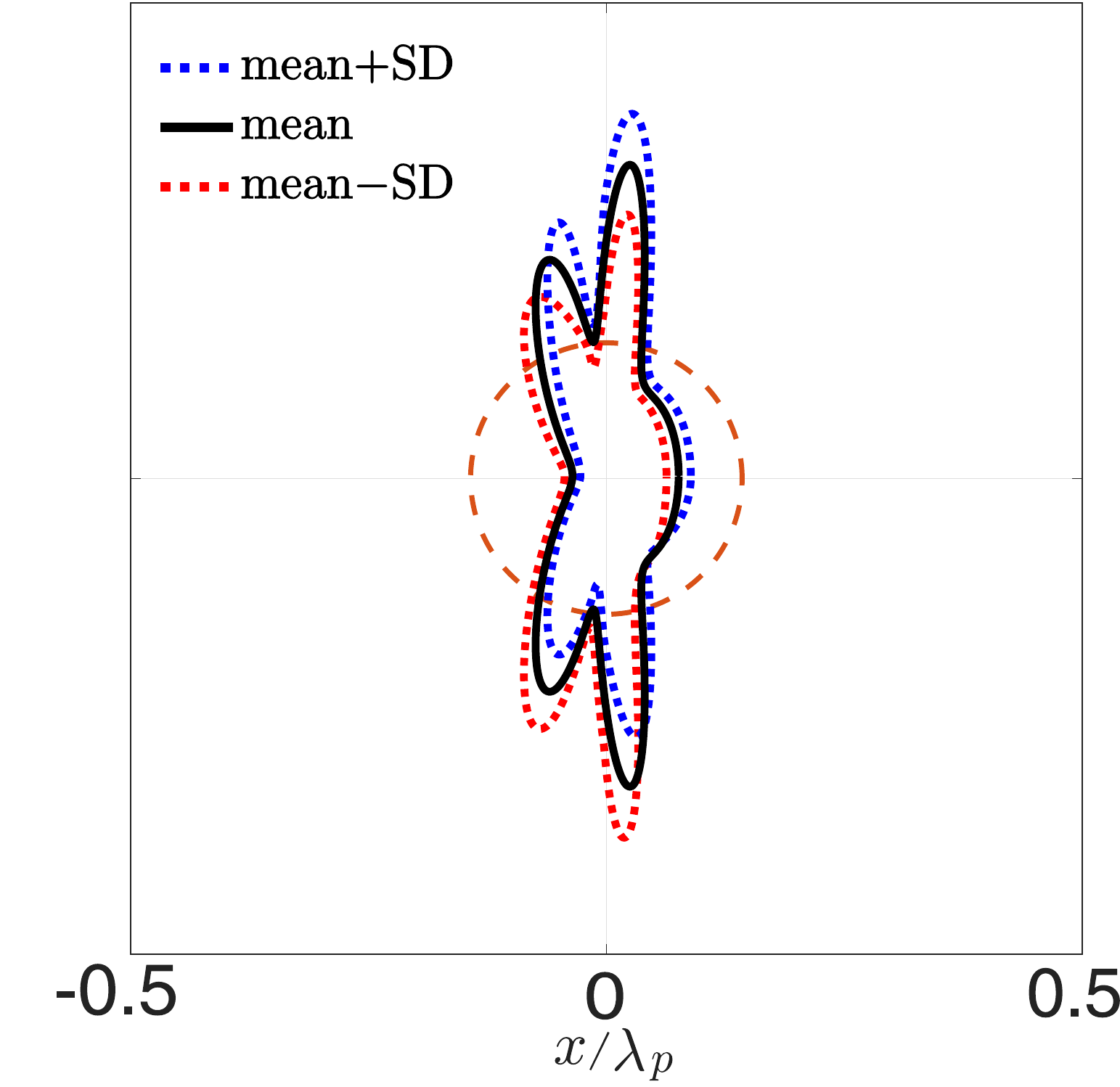}
                \caption{}
                \label{fig_sym_dir_all_ccc}
        \end{subfigure}%
        \begin{subfigure}[b]{\dd}
								\includegraphics[width=\linewidth]{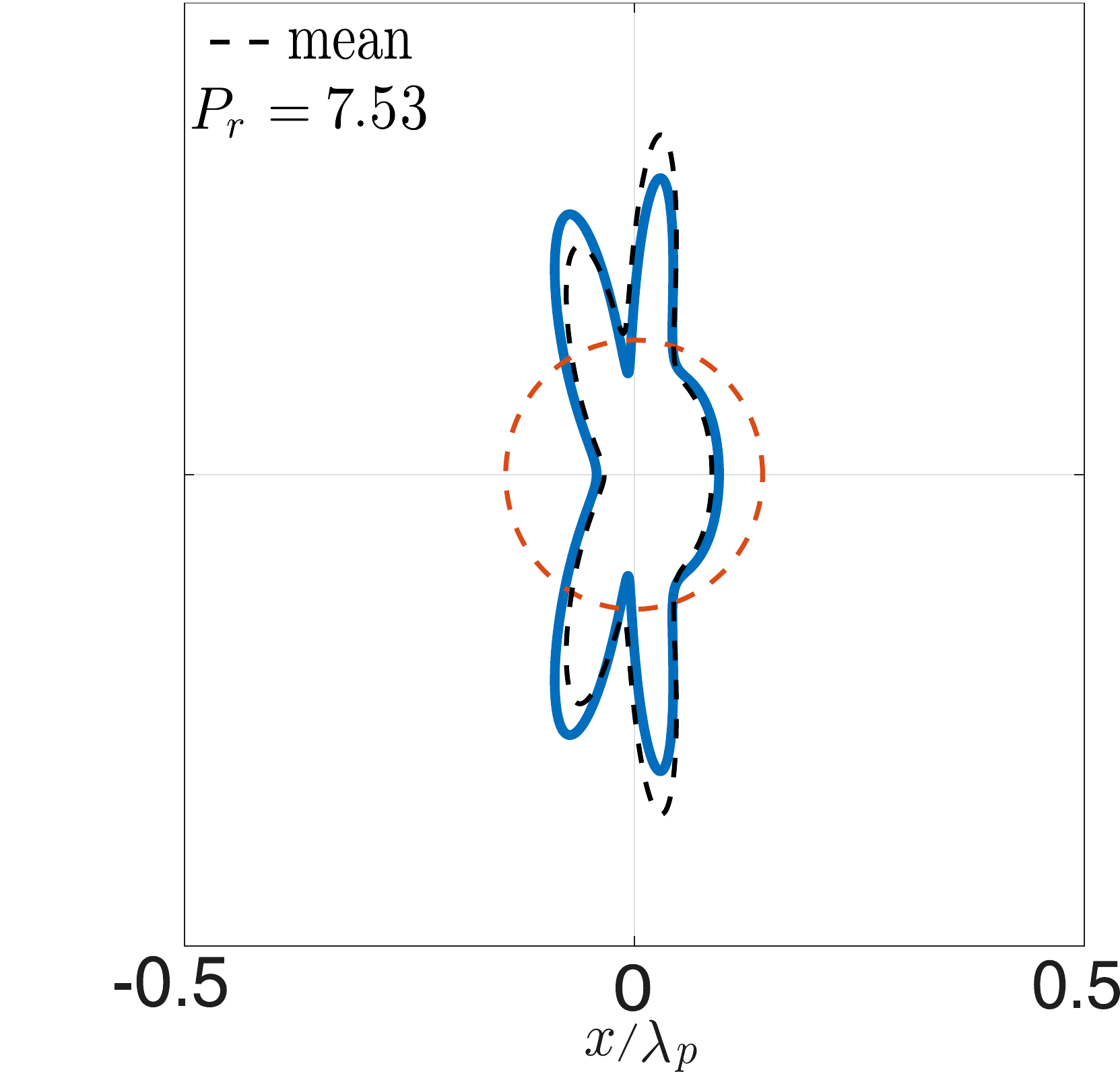}
                \caption{}
                \label{fig_sym_dir_a}
        \end{subfigure}%

        \caption{Optimization of the absorber shape under the action of polychromatic (broadband) directional incident waves. (a-h) optimum shape with eight different sets of random phases, (i) the averaged profile, (j) standard error of averaging, (k) standard deviation of the shape distribution and (l)  optimum shape under the assumption of the entire spectrum being in phase (all phases set to zero). Parameters are $k_ph_{opr}=1.13$, $k_pd=0.11$, $0.2$ [Hz] $<f<4$ [Hz], $|\theta|<80^\circ$ and coefficients of shape generation Fourier modes are given in Appendix II.}  \label{specdir}
\end{figure}

To demonstrate this case, we consider an incident JONSWAP spectrum of sea-state five given by equations \eqref{Eq_Somega} and \eqref{Eq_S_theta} with $H_s=3.25$ meters and $T_p=9.7$ seconds, and spreading parameters of $\mu=0$, $s=2$, $\Theta=160^\circ$. For an absorber of area $A/\lambda_p^2=0.064$, figure \ref{specdir}a-h shows optimum shapes for eight different sets of random phases. The obtained optimum shapes, similar to the case of monochromatic directional waves in figure \ref{monodirphase}, are different from one case to the other, with the two wings more highlighted in almost all figures when compared to the cases shown in figure \ref{monodirphase}a-h. The enhancement in power absorption ranges from 208\% to 637\%, that is two to more than six times higher than a circular absorber of the same area. The average absorption enhancement is 371\%.

We also show the average profile of the absorber (figure \ref{specdir}i), standard error which is less than 1\% (figure \ref{specdir}j), and standard deviation (figure \ref{specdir}k). The average absorber geometry of figure\ref{specdir}i, under each of incident waves of figure\ref{specdir}a-h obtains respectively $P_r=$ 4.94, 1.91, 5.21, 1.86, 3.87, 1.80, 1.94 and 3.89, whose average is $\bar P_r$=3.17. This means that the average profile shown in figure\ref{specdir}i has more than 300\% higher absorption capability than a circular shape of the same area. For the sake of comparison, we also calculate the optimum shape under the assumption of all waves being in phase (i.e. we set all phases equal to zero). Figure \ref{specdir}l shows the optimum shape obtained (blue solid line) and compares it with the average shape (black dashed line) and the same-area circle (red dashed line). The relative power in this case has the surprisingly high value of $P_r=7.53$, and it is interesting to note that the obtained shape under in-phase assumption is very close to the mean shape.

\begin{figure}[h!]
\centering
\includegraphics[width=0.35\textwidth]{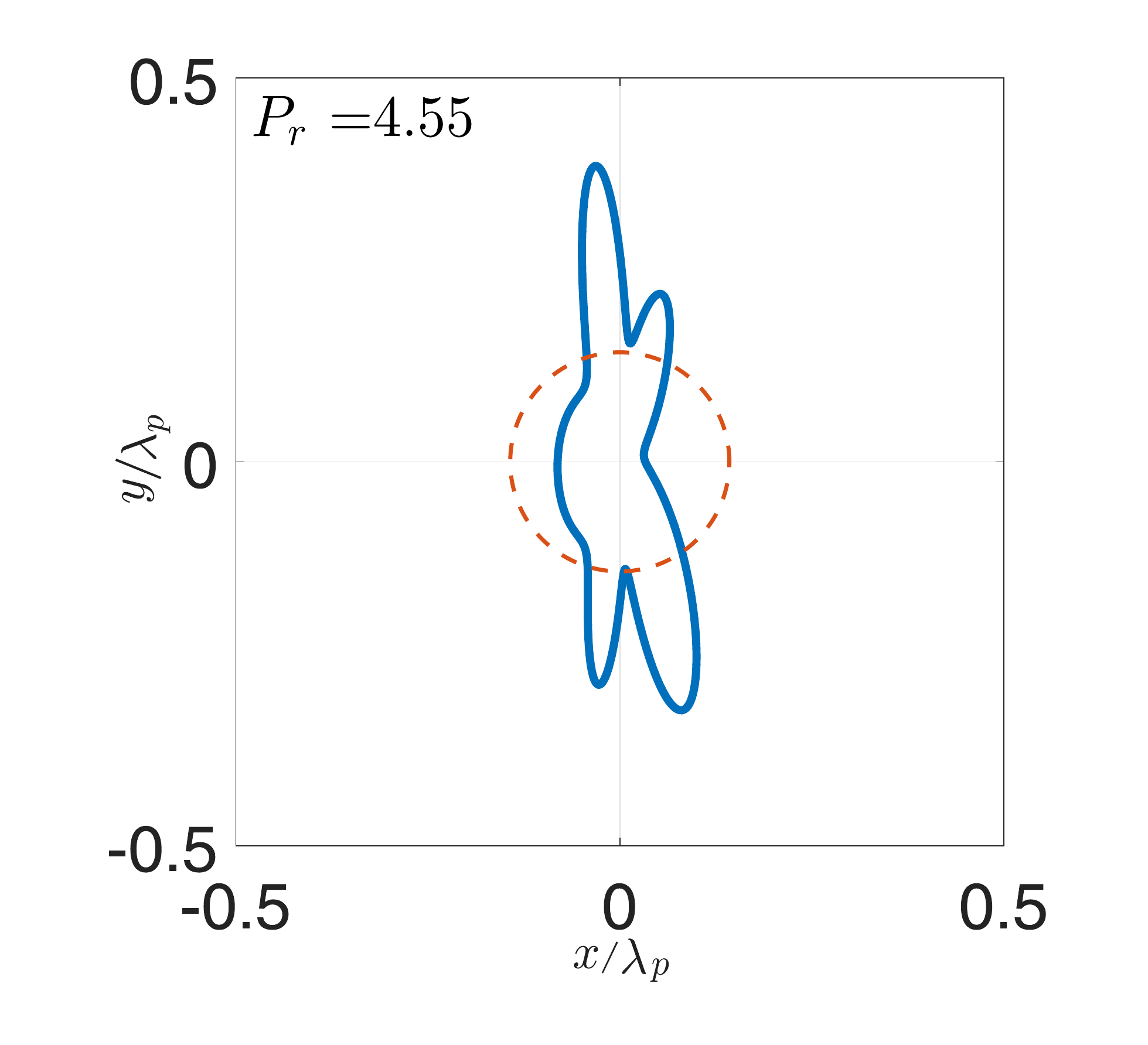}
	\caption{Optimum geometry of the absorber plate under the action of a polychromatic directional incident wave whose directionality is not symmetric (cf. figure \ref{figspr}c). Chosen parameters are $k_ph_{opr}=1.13$, $k_pd=0.11$,  $0.2$ [Hz] $<f<4$ [Hz], $|\theta|<80^\circ$ and all waves are assumed to be in-phase (cf. figure \ref{specdir}l) and coefficients of shape generation Fourier modes are given in Appendix II.}
	\label{fig_dir_optt_asym}
\end{figure}

As discussed before, the methodology developed here does not assume the incident waves having angular symmetry with respect to $\theta= 0^\circ$. It is, therefore, worth investigating a case in which the incident wave spectrum is not symmetric. We consider an asymmetric directional JONSWAP spectrum of sea state five (rough sea conditions) with $H_s=3.25$ meters and $T_p=9.7$ seconds, and spreading parameters of $\mu=0.5$, $\Theta=160^\circ$, and $s=75$ (figure \ref{fig_amp_dis_nonsym_dir_spec}). The procedure is similar as before, and we consider in-phase incident waves (as in figure\ref{specdir}l). The optimum shape is shown in figure \ref{fig_dir_optt_asym} for which $P_r=$ 4.55. Clearly the shape is asymmetric as a response to an asymmetric incident spectrum. Note that the incident spectrum (figure \ref{fig_amp_dis_nonsym_dir_spec}) has an inclination angle of $\theta\simeq10^\circ \sim 5^\circ$, and accordingly the optimum shape has almost the same inclination angle in its vertical axis in order to position itself nearly perpendicular to the direction of waves in order to maximize the absorbed power. 

\section{Conclusion}
Harnessing energy of water waves is based on conversion of the energy within the waves to drive a power take-off unit by means of an intermediate absorber or interface. However, for  having efficient energy conversion, the wave energy converter must be optimized in the design stage. We presented here a robust and systematic method for optimizing the absorber shape of a submerged planar pressure differential wave energy converter, using the Genetic Algorithm aiming to improve its wave power absorption level. A new parametric description of absorber shape based on Fourier decomposition of geometrical shapes was introduced. For each shape configuration we optimized the shape parameters and power take-off characteristics. The shape optimizations were done for different sea states as well as a general directional wave spectrum with asymmetric angular distribution. An analytic expression for calculation of absorbed power by wave energy converters subject to directional spectrum of waves was presented. Optimum shapes were found to be elongated perpendicular to the mean direction of incident waves with round corners. Having symmetry with respect to the mean wave direction for waves with symmetric angular distribution, the optimum shapes became asymmetric for an asymmetric angular distribution of incident waves. We showed that the optimum shape may have a significant higher energy capturing capability, sometimes nearly an order of magnitude, when compared to a circular absorber shape of the same area. 

The focus of the current manuscript is on a device with one degree of freedom (i.e. heave motion). A natural extension would be to relax other degrees of freedom and find the optimum shape and corresponding power capture. While this is straight forward and follows closely the optimization methodology presented here, it is computationally a lot more expensive than a single degree of freedom optimization.  It is to be noted that while relaxing more degrees of freedom may offer advantages in engineering and fabrication of the wave energy device, it may not necessarily result in enhancement of the efficiency. Current manuscript also considers small amplitude incident waves for which linear theory is valid. At such a limit, and for scales at which wave energy conversion devices work in high Reynolds numbers, viscous effects are known to be of a negligible significance \cite{Folley2016}. However, when incident wave amplitude is large (e.g. during a storm), nonlinear effects and viscous effects both become important and must be taken into account. Specifically, if waves are large enough that they break over the device, then certainly both nonlinearity and viscosity play major roles in the dynamics of wave-device interaction. In this case, direct simulation of full governing equation is inevitable.

\section*{Appendix I: Energy damping and the superposition principle}

The objective here is to present expressions to calculate generated (damped) energy in a damper of a linear mass-spring-damper system under the action of two sinusoidal forcing functions. In brief, if the two forcing functions have different frequencies, then generated energy is the linear superposition of energy generated under each of the excitation components. Nevertheless, if the two excitation forces have the same frequency (e.g. two same-frequency waves arriving from different directions), then the \textit{relative phases} of the two waves play a critical role in the overall energy production. Overall energy production in this case may become zero in the special case of the two phases being $\pi$-Radian different (at the location of the device). The energy production is maximum if the two phases are the same. While this is a classic subject, we repeatedly find mistakes and confusion in the work of researchers, and therefore decided to discuss this here. \\
\begin{figure}[h!]
\captionsetup{justification=centering,margin=2cm} 
	\centering		
	\includegraphics[width=0.7\textwidth]{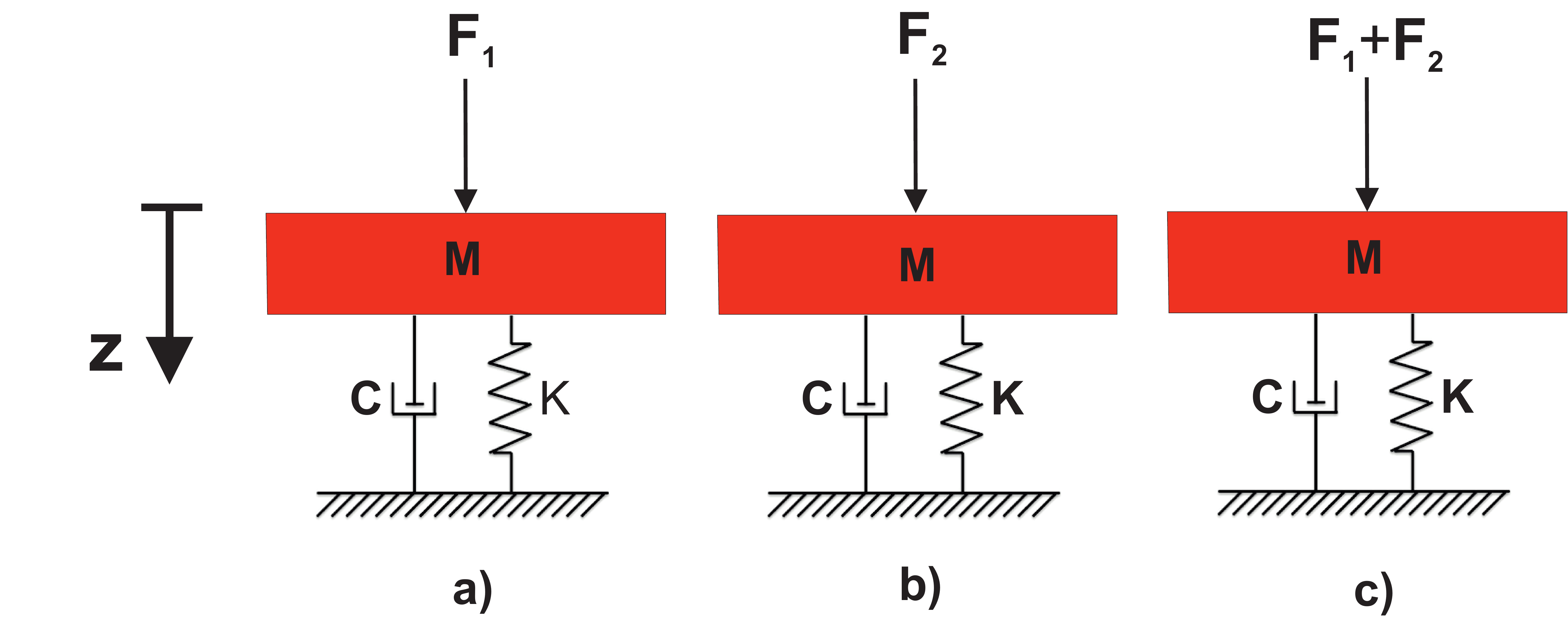}
	\caption{Mass-spring-damper system subject to \textbf{(a)} external force with frequency $\omega_1$; \textbf{(b)} external force with frequency $\omega_2$; and \textbf{(c)} superposition of cases \textbf{(a)} and \textbf{(b)} }
	\label{msd}
\end{figure}

Consider a linear mass-spring-damper system under the action of sinusoidal forces $F_1(\omega_j)=\hat{F}_1\,e^{i\,(\omega_1 t + \phi_{F,1})}$, $F_2(\omega_2)=\hat{F}_2\,e^{i\,(\omega_2 t + \phi_{F,2})}$, and the linear superposition of the two (figure \ref{msd}a-c), where $\omega_j,\phi_{F,j}$ are frequency and phase of each excitation. Under the action of each of the single forcing, corresponding system response is $z_j=\hat{z}_j e^{i(\omega_j t + \phi_{z,j})}$ with
\begin{subequations}  \label{Eq_amp_and_phase}
\begin{align}
\hat{z}_j=\frac{\hat{F}_j}{\sqrt{(K-M\omega^2_j)^2+(C\,\omega_j)^2}},~~~~~~~~~~~~\\
\phi_{z,j}=\phi_{F,j}-\phi_{K,C}=\phi_{F,j}-\arctan\Bigg( \frac{C\,\omega_j}{K-M\omega^2_j} \Bigg).
\end{align}
\end{subequations}
Therefore, the absorbed power is 
\begin{equation} \label{Eq_power}
P_j(t)=Re \lbrace F_{c,j}(t) \rbrace \, Re \lbrace V_j(t) \rbrace=
C\,(Re \lbrace \dot{z}_j(t) \rbrace)^2=C\, [ -\omega_j \hat{z}_j \sin(\omega_j\,t+\phi_{z,j}) ]^2 ,
\end{equation}
where $F_{c,j}$ is the force of damper  and $V_j$ is the vertical velocity of the mass $M$ due to external force $F_j$. Therefore, if the system of mass-spring-damper is subjected to two forcings, then
\begin{equation}\label{Eq_3-28}
P(t)=C\,[-\omega_1\hat{z}_1\sin(\omega_1 t+\phi_{z,1})-
\omega_2\hat{z}_2\sin(\omega_2 t+\phi_{z,2})]^2.
\end{equation}

The quantity of interest, however, is the time-averaged power
\begin{equation}
P^T_{ave}=\frac{1}{T}\int_{0}^{T}P(t)\,dt,
\end{equation}
where $T$ is integration time, which for single frequency cases is the period of the external forcing, and for multiple frequency cases is integer multiples of \textit{Least common multiple} of periods (if does not exist, then it is infinity). Therefore, it is always accurate to calculate $P^\infty_{ave}$ in the limit of $T\rightarrow\infty$. 

For cases of interest discussed above, assuming that $\omega_1 \neq \omega_2$ the average power becomes
\begin{equation}\label{exp1}
P^T_{ave}=\,\frac{1}{2}C\omega^2_1\,\hat{z}_1+\frac{1}{2}C\omega^2_2\,\hat{z}_2+\frac{C}{T}\text{(bounded value)},
\end{equation}
and if $\omega_1 = \omega_2$
\begin{equation}\label{exp2} 
P^T_{ave}=\,\frac{1}{2}C\omega^2_1\,(\hat{z}^2_1+\hat{z}^2_2)+C\omega^2_1\,\hat{z}_1 \hat{z}_2 \cos(\phi_1-\phi_2)+\frac{C}{T}\text{(bounded value)}.
\end{equation}
Clearly the last term in both expressions go to zero as $T\rightarrow\infty$. Therefore, effect of relative phases of the two forcing excitations with \textit{different frequencies} does not affect the overall power production, but if they have the same frequency then these phases play an important role. Expressions \eqref{exp1} and \eqref{exp2} (without the last term) are used to calculate total energy in different cases studied in this manuscript.

\section*{Appendix II: Optimum coefficients}
In the following, $a_{n}, \phi_n, p, q$ are shape coefficients (cf. equations 9, 10) where $a_{n,r}=a_n/r_0$, $r_0$ is the base circle radius, and $\zeta, \omega/\omega_n$ are non-dimensional parameters.
{\renewcommand{\arraystretch}{1.35}
\begin{table} [!h]
\centering
\resizebox{0.9\columnwidth}{!}{
\begin{tabular}{ c c c c c c c c c c c c c c c c c | c c}
&$a_{1,r}$ & $a_{2,r}$ & $a_{3,r}$ & $a_{4,r}$ & $a_{5,r}$ & $a_{6,r}$ & $a_{7,r}$ & $p$ & $q$ & 
$\phi_1$ & $\phi_2$ & $\phi_3$ & $\phi_4$ & $\phi_5$ & $\phi_6$ & $\phi_7$ & $\zeta$ & $\omega / \omega_n$ \\ 
 \hline
a)&0.397 & 0.158 & 0.120 & - & - & - & - & 0.5 & 2 
& $\frac{\pi}{2}$ & $\frac{3\pi}{2}$ & $\frac{3\pi}{2}$ & - & - & - & - & 0.11 & 1 \\

b)&0.377 & 0.156 & 0.160 & 0.08 & - & - & - & 0.5 & 2
& $\frac{\pi}{2}$ & $\frac{3\pi}{2}$ & $\frac{3\pi}{2}$ & $\frac{3\pi}{2}$ & - & - & - & 0.106 & 1\\

c)&0.351 & 0.159 & 0.158 & 0.156 & 0.002 & - & - & 0.5 & 2
& $\frac{\pi}{2}$ & $\frac{3\pi}{2}$ & $\frac{3\pi}{2}$ & $\frac{3\pi}{2}$ & $\frac{\pi}{2}$ & - & - & 0.103 & 1\\

d)&0.312 & 0.157 & 0.156 & 0.158 & 0.159 & 0 & - & 0.5 & 2
& $\frac{\pi}{2}$ & $\frac{3\pi}{2}$ & $\frac{3\pi}{2}$ & $\frac{\pi}{2}$ & $\frac{3\pi}{2}$ & $\frac{\pi}{2}$ & - & 0.104 & 1\\

e)&0.313 & 0.160 & 0.159 & 0.157 & 0 & 0 & 0.152 & 0.5 & 2
& $\frac{\pi}{2}$ & $\frac{3\pi}{2}$ & $\frac{3\pi}{2}$ & $\frac{3\pi}{2}$ & $\frac{\pi}{2}$ & $\frac{\pi}{2}$ & $\frac{\pi}{2}$ & 0.228 & 1 
\end{tabular}
}
\caption{ Optimum shape coefficients for Fig.\,\,\eqref{mono}; $\zeta={C}/[2\sqrt{K(m+A)}]$, $\omega_n=\sqrt{K/(m+A)}$ where $m,A,C,K$ are absorber's mass and added mass, and PTO's optimal damping and restoring coefficients respectively}
\end{table}
{\renewcommand{\arraystretch}{1.35}
\begin{table} [!h]
\centering
\resizebox{0.7\columnwidth}{!}{
\begin{tabular}{ c c c c c c c c c c c c c | c c}
&$a_{1,r}$ & $a_{2,r}$ & $a_{3,r}$ & $a_{4,r}$ & $a_{5,r}$ & $p$ & $q$ & 
$\phi_1$ & $\phi_2$ & $\phi_3$ & $\phi_4$ & $\phi_5$ & $\zeta$ & $\omega/\omega_n$\\ 
 \hline
a)&0.351 & 0.156 & 0.158 & 0.160 & 0.011 & 0.5 & 2 
& $\frac{\pi}{2}$ & $\frac{3\pi}{2}$ & $\frac{3\pi}{2}$ & $\frac{3\pi}{2}$ & $\frac{\pi}{2}$ &0.103 & 1 \\

b)&0.155 & 0.160 & 0.351 & 0.158 & 0 & 0.5 & 2
& $\frac{3\pi}{2}$ & $\frac{3\pi}{2}$ & $\frac{\pi}{2}$ & $\frac{3\pi}{2}$ & $\frac{\pi}{2}$ & 0.097 & 1\\

c)&0.154 & 0.159 & 0.312 & 0.158 & 0.160 & 0.5 & 2
& $\frac{3\pi}{2}$ & $\frac{3\pi}{2}$ & $\frac{\pi}{2}$ & $\frac{3\pi}{2}$ & $\frac{\pi}{2}$ & 0.093 & 1\\

d)&0.158 & 0.158 & 0 & 0.159 & 0.3511 & 0.5 & 2
& $\frac{3\pi}{2}$ & $\frac{3\pi}{2}$ & $\frac{\pi}{2}$ & $\frac{3\pi}{2}$ & $\frac{\pi}{2}$ & 0.095 & 1 \\

e)&0.160 & 0.159 & 0.157 & 0 & 0.3513 & 0.5 & 2
& $\frac{3\pi}{2}$ & $\frac{3\pi}{2}$ & $\frac{\pi}{2}$ & $\frac{\pi}{2}$ & $\frac{\pi}{2}$ & 0.101 & 1 \\
\end{tabular}
}
\caption{Optimum shape coefficients for Fig.\,\,\eqref{monodir}; $\zeta={C}/[2\sqrt{K(m+A)}]$, $\omega_n=\sqrt{K/(m+A)}$ where $m,A,C,K$ are absorber's mass and added mass, and PTO's optimal damping and restoring coefficients respectively}
\end{table}
{\renewcommand{\arraystretch}{1.35}
\begin{table}[h!]
\centering
\resizebox{0.7\columnwidth}{!}{
\begin{tabular}{ c c c c c c c c c c c c c | c c }
&$a_{1,r}$ & $a_{2,r}$ & $a_{3,r}$ & $a_{4,r}$ & $a_{5,r}$ & $p$ & $q$ & $\phi_1$ & $\phi_2$ & $\phi_3$ & $\phi_4$ & $\phi_5$ & $\zeta$ & $\omega/\omega_n$\\ 
 \hline
a)& 0.157 & 0.001 & 0.156 & 0.161 & 0.3513 & 0.5 & 2 
& $\frac{\pi}{2}$ & $\frac{\pi}{2}$ & $\frac{\pi}{2}$ & $\frac{3\pi}{2}$ & $\frac{\pi}{2}$ & 0.143 & 1 \\

b)&0.3122 & 0.159 & 0.162 & 0.160 & 0.157 & 0.5 & 2
& $\frac{\pi}{2}$ & $\frac{3\pi}{2}$ & $\frac{3\pi}{2}$ & $\frac{3\pi}{2}$ & $\frac{\pi}{2}$ & 0.103 & 1 \\

c)& 0.3511 & 0.162 & 0.161 & 0.003 & 0.157 & 0.5 & 2
& $\frac{\pi}{2}$ & $\frac{3\pi}{2}$ & $\frac{3\pi}{2}$ & $\frac{\pi}{2}$ & $\frac{\pi}{2}$ & 0.109 & 1\\

d)& 0.002 & 0.157 & 0.161 & 0.160 & 0.3511 & 0.5 & 2
& $\frac{\pi}{2}$ & $\frac{3\pi}{2}$ & $\frac{3\pi}{2}$ & $\frac{3\pi}{2}$ & $\frac{\pi}{2}$ & 0.009 & 1 \\

e)& 0.161 & 0.157 & 0.3122 & 0.159 & 0.160 & 0.5 & 2
& $\frac{3\pi}{2}$ & $\frac{3\pi}{2}$ & $\frac{\pi}{2}$ & $\frac{\pi}{2}$ & $\frac{\pi}{2}$ & 0.1 & 1 \\

f)&0.161 & 0.158 & 0.001 & 0.160 & 0.3513 & 0.5 & 2
& $\frac{3\pi}{2}$ & $\frac{3\pi}{2}$ & $\frac{\pi}{2}$ & $\frac{3\pi}{2}$ & $\frac{\pi}{2}$  & 0.099 & 1\\

g)& 0.158 & 0.162 & 0.161 & 0.002 & 0.3512 & 0.5 & 2
& $\frac{3\pi}{2}$ & $\frac{3\pi}{2}$ & $\frac{\pi}{2}$ & $\frac{\pi}{2}$ & $\frac{\pi}{2}$ & 0.101 & 1\\

h)&0.161 & 0.159 & 0.158 & 0.161 & 0.3122 & 0.5 & 2
& $\frac{3\pi}{2}$ & $\frac{\pi}{2}$ & $\frac{\pi}{2}$ & $\frac{3\pi}{2}$ & $\frac{\pi}{2}$ & 0.145 & 1\\
\end{tabular}
}
\caption{Optimum shape coefficients for Fig.\,\,\eqref{monodirphase}; $\zeta={C}/[2\sqrt{K(m+A)}]$, $\omega_n=\sqrt{K/(m+A)}$ where $m,A,C,K$ are absorber's mass and added mass, and PTO's optimal damping and restoring coefficients respectively}
\end{table}
%
%
%
{\renewcommand{\arraystretch}{1.35}
\begin{table} [!h]
\centering
\resizebox{0.7\columnwidth}{!}{
\begin{tabular}{ c c c c c c c c c c c c c | c c }
&$a_{1,r}$ & $a_{2,r}$ & $a_{3,r}$ & $a_{4,r}$ & $a_{5,r}$ & $p$ & $q$ & 
$\phi_1$ & $\phi_2$ & $\phi_3$ & $\phi_4$ & $\phi_5$ & $\zeta$ &  $\omega_p/\omega_n$\\ 
 \hline
a)&0.156 & 0.159 & 0.161 & 0.011 & 0.356 & 0.5 & 2 
& $\frac{\pi}{2}$ & $\frac{3\pi}{2}$ & $\frac{3\pi}{2}$ & $\frac{\pi}{2}$ & $\frac{\pi}{2}$ & 6.09 & 6.67 \\

b)&0 & 0.157 & 0.013 & 0.160 & 0.388 & 0.5 & 2
& $\frac{\pi}{2}$ & $\frac{\pi}{2}$ & $\frac{\pi}{2}$ & $\frac{3\pi}{2}$ & $\frac{\pi}{2}$  & 0.196 & 0.0519\\

c)&0.159 & 0.005 & 0 & 0.155 & 0.391 & 0.5 & 2
& $\frac{\pi}{2}$ & $\frac{\pi}{2}$ & $\frac{\pi}{2}$ & $\frac{3\pi}{2}$ & $\frac{\pi}{2}$ & 0.0136 & 0.0517\\

d)&0.007 & 0.159 & 0.156 & 0.161 & 0.353 & 0.5 & 2
& $\frac{\pi}{2}$ & $\frac{\pi}{2}$ & $\frac{\pi}{2}$ & $\frac{3\pi}{2}$ & $\frac{\pi}{2}$ & 0.429 & 0.164\\

e)&0.159 & 0.161 & 0.156 & 0.157 & 0.313 & 0.5 & 2
& $\frac{\pi}{2}$ & $\frac{3\pi}{2}$ & $\frac{3\pi}{2}$ & $\frac{3\pi}{2}$ & $\frac{\pi}{2}$  & 0.308 & 0.1714\\

f)&0.158 & 0.162 & 0.007 & 0.160 & 0.351 & 0.5 & 2
& $\frac{\pi}{2}$ & $\frac{3\pi}{2}$ & $\frac{\pi}{2}$ & $\frac{\pi}{2}$ & $\frac{\pi}{2}$ & 1.04 & 0.2736\\

g)&0.156 & 0.153 & 0.001 & 0.158 & 0.351 & 0.5 & 2
& $\frac{3\pi}{2}$ & $\frac{\pi}{2}$ & $\frac{\pi}{2}$ & $\frac{3\pi}{2}$ & $\frac{\pi}{2}$ & 0.196 & 0.052 \\

h)&0.156 & 0.159 & 0.160 & 0.155 & 0.31241 & 0.5 & 2
& $\frac{\pi}{2}$ & $\frac{3\pi}{2}$ & $\frac{3\pi}{2}$ & $\frac{3\pi}{2}$ & $\frac{\pi}{2}$  & 0.308 & 0.1714\\

l)&0 & 0.157 & 0.013 & 0.160 & 0.388 & 0.5 & 2
& $\frac{\pi}{2}$ & $\frac{\pi}{2}$ & $\frac{\pi}{2}$ & $\frac{3\pi}{2}$ & $\frac{\pi}{2}$  & 1.04 & 0.274 \\

\end{tabular}
}
\caption{Optimum shape coefficients for Fig.\,\,\eqref{specdir}; $\zeta={C}/[2\sqrt{K(m)}]$, $\omega_n=\sqrt{K/(m)}$ where $m,C,K$ are absorber's mass, and PTO's optimal damping and restoring coefficients respectively, $\omega_p$ is the angular frequency of spectrum's corresponding peak amplitude}
\end{table}
{\renewcommand{\arraystretch}{1.35}
\begin{table}[h!]
\centering
\resizebox{0.7\columnwidth}{!}{
\begin{tabular}{ c c c c c c c c c c c c c | c c}
&$a_{1,r}$ & $a_{2,r}$ & $a_{3,r}$ & $a_{4,r}$ & $a_{5,r}$ & $p$ & $q$ & 
$\phi_1$ & $\phi_2$ & $\phi_3$ & $\phi_4$ & $\phi_5$& $\zeta$ & $\omega_p/\omega_n$\\ 
 \hline
a)&0.05 & 0.151 & 0.154 & 0.159 & 0.338 & 0.5 & 2 
& $0$ & $\frac{2.006\pi}{2}$ & $\frac{-2.002\pi}{2}$ & $\frac{3.014\pi}{2}$ & $\frac{3.007\pi}{2}$ & 0.199 & 0.051 \\
\end{tabular}
}
\caption{Optimum shape coefficients for Fig.\,\,\eqref{fig_dir_optt_asym}; $\zeta={c}/[2\sqrt{k(m)}]$, $\omega_n=\sqrt{k/(m)}$ where $m,c,k$ are absorber's mass, and PTO's optimal damping and restoring coefficients respectively, $\omega_p$ is the angular frequency of spectrum's corresponding peak amplitude}
\end{table}

\section*{Appendix III: Derivation of expression for the area of the optimization candidate shapes}
{For the shapes built using equations \eqref{Eq_radius_fouriermodes} and \eqref{Eq_xy} it was mentioned in section \ref{sec_ShapeoptimizationMethodology} that the area of the shapes can be found by equation \eqref{Eq_sinModes_Area} which  by using Green's theorem \citep{Odzijewicz2013} here we derive it analytically.\\
Green's theorem asserts that if 
\begin{equation}
\gamma:~~~~~\theta\mapsto (x(\theta),y(\theta))~~~~~(0 \leq \theta\leq 2\pi)
\end{equation}
is a closed curve bounding counter-clockwise a region $B\subset \R^2$, then 
\begin{equation}\label{Eq_green_area}
Area\,(B)=\frac{1}{2}\int_0^{2\pi}[x(\theta)\,\dot{y}(\theta)-y(\theta)\,\dot{x}(\theta)]\,d\theta.
\end{equation}
Now, for a shape generated using the equations below
\begin{equation}
r(\theta)=r_0+\sum_{n=1}^{N_c}a_n\,\sin\,(n\,\theta+\phi_n),
\end{equation}
\ba 
x=p\,r(\theta)\,\cos\,(\theta), \qquad y=q\,r(\theta)\,\sin\,(\theta), 
\ea
by replacing the expressions of $x,y$ into equation \eqref{Eq_green_area} and calculating the integral by hand or a software the area of the shape becomes
\ba
Area\,(B)=pq\lb\pi{r_0}^2+\frac{1}{2}\left(\sum_{n=1}^{N_c} {\pi\,a_n}^2\right)\rb.
\ea
As a check one can put $p=q=1$ and do not add Fourier modes to the circle (i.e. $a_n=0$) in order to recover the known formula for area of a circle with radius $r_0$ as $\pi r_0^2$.}

\clearpage
\bibliographystyle{ieeetr}
\small{\bibliography{fdg}}

\end{document}